\documentclass[journal]{IEEEtran}
\usepackage{cite}
\usepackage{float}
\usepackage{xcolor}
\usepackage{url}
\usepackage{graphics}
\usepackage{graphicx}
\usepackage{epstopdf}
\usepackage{amsmath,amssymb,amsfonts}
\usepackage[linesnumbered,ruled,vlined]{algorithm2e}
\usepackage{algorithmic}
\usepackage{subcaption}

\newcommand{\gf}[1]{\textcolor{black}{{#1}}}

\begin{document}

\title{Channel Estimation in \textcolor{black}{RIS-Assisted MIMO} Systems Operating Under \textcolor{black}{Imperfections}} 

\author{Paulo R. B. Gomes,~Gilderlan~T.~de~Ara\'{u}jo,~Bruno~Sokal,~Andr\'{e}~L.~F.~de~Almeida,~\IEEEmembership{Senior~Member,~IEEE,}~Behrooz~Makki,~\IEEEmembership{Senior~Member,~IEEE,}~and~G\'{a}bor~Fodor,~\IEEEmembership{Senior~Member,~IEEE}% <-this % stops a space
\thanks{Paulo R. B. Gomes, Gilderlan T. de Ara\'{u}jo, Bruno Sokal, and Andr\'{e} L. F. de Almeida are with the Wireless Telecom Research Group (GTEL), Department of Teleinformatics Engineering, Federal University of Cear\'{a}, Fortaleza-CE. E-mails: \{paulo,gilderlan,brunosokal,andre\}@gtel.ufc.br.}% <-this % stops a space
\thanks{Behrooz Makki is with Ericsson Research, G$\ddot{\text{o}}$teborg, Sweden. E-mail: behrooz.makki@ericsson.com.}% <-this % stops a space
\thanks{G\'{a}bor Fodor is with Ericsson Research and KTH Royal Institute of Technology, Stockholm, Sweden. E-mail: gabor.fodor@ericsson.com.}% <-this % stops a space
\thanks{This work was supported by the Ericsson Research, Sweden, and Ericsson Innovation Center, Brazil, under UFC.48 Technical Cooperation Contract Ericsson/UFC. This study was financed in part by the Coordena\c{c}\~{a}o de Aperfei\c{c}oamento de Pessoal de N\'{i}vel Superior - Brasil (CAPES)-Finance Code 001, and CAPES/PRINT Proc. 88887.311965/2018-00. Andr\'{e} L. F. de Almeida acknowledges CNPq for its financial support under the grant 312491/2020-4. G\'{a}bor Fodor was partially supported by the Digital Futures project PERCy.}
\thanks{\textcolor{black}{Part of this work has been submitted for possible presentation in IEEE GLOBECOM 2022 \cite{ourGlobecom}.}
}}



\maketitle

\begin{abstract}
Reconfigurable intelligent surface (RIS) is a potential technology component of future wireless networks due to its capability of shaping the wireless environment. The promising gains of RIS-assisted multiple-input multiple-output (MIMO) systems in terms of extended coverage and enhanced capacity are, however, critically dependent
on the accuracy of the channel state information. However, traditional channel estimation schemes are not applicable in RIS-assisted MIMO networks, since passive RISs typically lack the signal processing capabilities that are assumed by channel estimation algorithms. This becomes most problematic when physical
imperfections or electronic impairments affect the RIS due to
its exposition to different environmental effects or caused by
hardware limitations from the circuitry. While these real-world effects are typically ignored in the literature, in this paper we propose efficient channel estimation schemes for RIS-assisted MIMO systems taking different imperfections into account. 
Specifically, we propose two sets of tensor-based algorithms, based on the parallel factor analysis decomposition schemes. First, by assuming a long-term model -- in which the RIS imperfections, modeled as unknown phase shifts, are static within the channel coherence time -- we formulate an iterative alternating least squares (ALS)-based algorithm for the joint estimation of the communication channels and the unknown phase deviations. 
Next, we develop the short-term imperfection model, which allows both amplitude and phase RIS imperfections to be non-static with respect to the channel coherence time. We  propose two iterative ALS-based and closed-form higher order singular value decomposition-based algorithms for the joint estimation of the channels and the unknown impairments. Moreover, we analyze the identifiability and computational complexity of the proposed algorithms and study the effects of various imperfections on the channel estimation quality. Simulation results demonstrate the effectiveness of the proposed tensor-based algorithms in terms of  estimation accuracy and computational complexity.
\end{abstract}


\IEEEpeerreviewmaketitle

\section{Introduction}\label{intro}
\vspace{-0.1cm}
Wireless communications have become a necessity in our daily lives providing significant benefits to individuals, businesses and the society at large. The steadily increasing demands for ubiquitous 
wireless services drive the efforts by the research and standardization communities 
to improve coverage, system capacity as well as the reliability and quality of a growing number of applications \cite{SCisco,RCITU}. 
The continuous growth of the number of mobile subscriptions, devices and traffic increases the number of deployed infrastructure nodes, which makes capital and operational expenditures as well as energy consumption \textcolor{black}{challenging} for mobile network operators. 
\cite{SZhang2017}. Therefore, to guarantee green sustainable wireless networks, one needs to carefully take the energy consumption and the hardware costs into account \cite{CYou,QWu}. 

Thanks to the recent development of meta-materials, reconfigurable intelligent surface (RIS) 
has emerged as a potential technology applicable in future wireless networks through the novel \textit{smart and programming environment} paradigm \cite{Zhou:22}. 
Unlike conventional networks, RIS-assisted networks enable the system to shape the
wireless environment to be more suitable for wireless communication \cite{DiRenzo2019,YLiang2019}. 
In conventional networks, the wireless transmission medium is seen as an uncontrollable element in the system due to the randomness in the radio environment. Therefore, the propagation of electromagnetic waves through the wireless channel cannot be controlled after they are emitted from the transmitters and before they reach the receivers \cite{DiRenzo2020}. 
An RIS is a planar meta-surface that has a large number of low-cost passive reflecting elements with adjustable capability of the reflected signal parameters such as amplitude, phase, frequency and polarization. With a well-configured RIS, the propagation conditions are improved by controlling the scattering characteristics to create passive and active beamforming at the RIS and the wireless transceivers, \textcolor{black}{respectively,} to achieve high beamforming gains \cite{BehroozSug}. \textcolor{black}{In addition}, a passive RIS does not require active radio frequency 
chains for signal transmission, reception and processing, since it simply relies on passive signal reflection, which makes it a power-efficient 
%cost-effective, 
and \textcolor{black}{less-complexity} technology \textcolor{black}{\cite{EBasar2019,QWuRZhang20192,arxivBehrooz,ivd01}.}

%%%%%%%%%%%%%%%%%%%%%%%%%%%%%%
%In terms of standardization, RIS has not yet been considered by the 3rd Generation Partnership Project (3GPP). 
%During the initial discussions on 3GPP Release 18, RIS-based communications were suggested as a part of the discussions around network-controlled
%repeaters \cite{ivd01,ivd02}. \textcolor{black}{However, it was decided not to continue with RIS, and leave it} 
%for studies in future 3GPP releases.
%%%%%%%%%%%%%%%%%%%%%%%%%%%%%%%%%%%%%

Despite its potential benefits,
the performance gains achieved in RIS-assisted systems are strongly dependent on, among others factors, the quality of the channel state information (CSI), where the channel between the wireless transceivers is effectively split into two-hop RIS-assisted channels.
This is because CSI is required to jointly design the passive beamforming at the RIS and the active beamforming at the transceivers \cite{AZappone2021}. 
However, due to the passive nature of the RIS, \textcolor{black}{channel estimation is not performed at the RIS but only on gNB or UE side.}
%no baseband processing capability may be available at the RIS, and channel estimation should be performed only at the intended receiver or at the transmitter. 

\gf{Recognizing this issue}, recent works propose
different strategies to tackle the channel estimation problem in RIS-assisted wireless communications under various configurations (see, e.g., \textcolor{black}{\cite{p20221,GilWCL,GilJournal}} and references therein). The most popular approach in the literature, known as the on/off method, consists of switching groups of RIS elements (or each individual element) to estimate the associated channel coefficients sequentially \cite{YYang2020}. More specifically, each element of the cascaded channel is estimated at each time slot from pilots sent by the transmitter and reflected by the corresponding RIS element \textcolor{black}{e.g., \cite{Mishra2019}}.

In the context of RIS-assisted multiple-input multiple-output (MIMO) systems, 
%the pioneer work 
\cite{ZQing} proposes a two-stage algorithm including sparse matrix factorization and matrix completion to estimate the channels in massive array setups. In \cite{Taha2021}, the authors leverage compressed sensing (CS) and deep learning tools to reduce the training overhead. In a similar way, \cite{JChen2019} also \textcolor{black}{uses} CS techniques and \textcolor{black}{proposes} a two-step iterative solution to solve the problem in a multi-user scenario. 
%Unlike previous works, 
\textcolor{black}{Then,} \cite{JHe} proposes an iterative method to estimate the channel parameters (angle of departure, angle of arrival and propagation path gains) in RIS-assisted millimeter wave (mmWave) MIMO systems, \textcolor{black}{resulting in} limited training overhead.  

Over the past two decades, tensor algebra and higher-order tensor decompositions have been successfully applied to signal modeling and processing in wireless communications. Tensor-based signal processing exploits the intrinsic multidimensional structure of wireless channels and signals to achieve \textcolor{black}{an} improved estimation accuracy under more relaxed and flexible system parameter \textcolor{black}{settings,} compared to conventional matrix-based approaches \cite{CDMA,tensorOverview,sug01}.
%\cite{CDMA,Parafac,Tucker,btd,paratuck,confac,tensorOverview,sug01}.

%For more details, \cite{tensorOverview} provides a complete and comprehensive overview of tensor decomposition applications within the wireless communications and radar areas, while \cite{sug01} presents an overview of tensor decompositions with applications in signal processing and machine learning.
\textcolor{black}{Recently,} tensor approaches have been proposed in the context of RIS-assisted wireless communications. \textcolor{black}{Particularly,} \cite{RIS1} \textcolor{black}{capitalizes} on the parallel factor (PARAFAC) decomposition to formulate an efficient iterative algorithm based on the alternating least squares (ALS) concept to solve the channel estimation problem in the downlink of a multi-user multiple input single output (MISO) network. \textcolor{black}{Also,} in \cite{gg1} and \cite{gg2}, the authors \textcolor{black}{develop} simple \textcolor{black}{iterative and closed-form} channel estimation algorithms based on a PARAFAC modeling \textcolor{black}{of} the single-user MIMO scenario. 

\begin{figure}[t]
	\centering
	\includegraphics[width=0.4\textwidth]{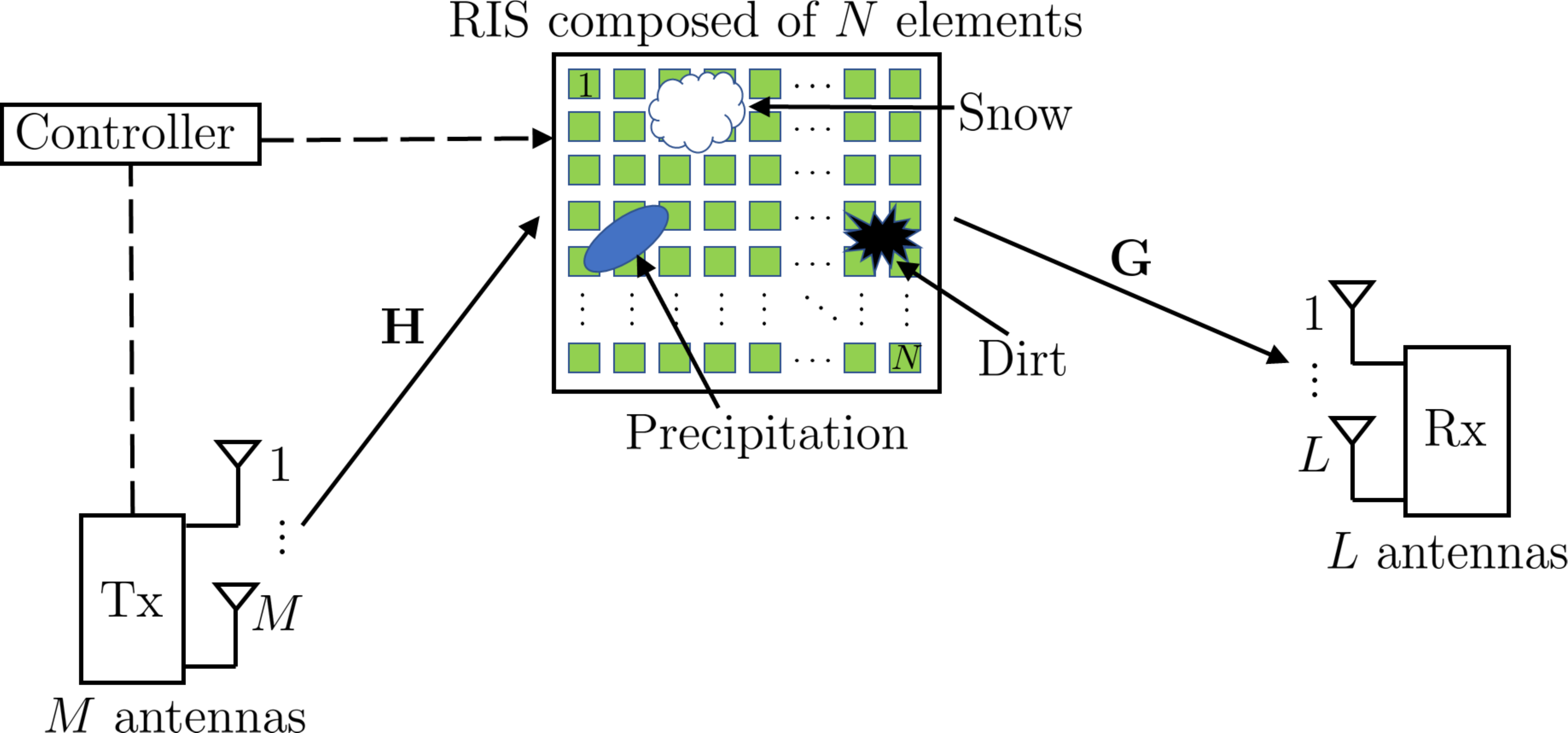}
	\caption{\small{An illustration of a RIS-assisted MIMO wireless communication system operating under imperfections.}}
	\label{figure1}
	%\label{figure2}
\end{figure}

%\begin{figure}[t]
%	\centering
%	\includegraphics[width=0.3\textwidth]{figs/ilu02.pdf}
%	\caption{Illustration of an outdoor RIS subject to different types of \textcolor{blue}{imperfections} in its passive reflecting elements. The \textcolor{blue}{imperfections} induce unwanted attenuation and phase shift on the reflected signals that introduce errors in the channel estimation \cite{mainblock}.}
%	\label{figure1}
%\end{figure}

\textcolor{black}{
The main motivation for RIS is \textcolor{black}{complexity}/energy reduction, otherwise
integrated access and backhaul (IAB) nodes, relays, and/or repeaters \textcolor{black}{may be alternatives.} 
%\textcolor{black}{may} show better performance. 
Then, \textcolor{black}{complexity} reduction may affect the hardware capability and, consequently, its performance. Therefore, to have a realistic view on the performance and usefulness of RIS, one needs to carefully take the hardware impairments as well as the environmental effects, such as water precipitations, flying debris, air particles, snowflakes, \gf{freezing rain, sleet}, dry/damp sand particles and dirt into account, as shown in Fig. \ref{figure1}.}

In practice, such blocking objects as well as the hardware impairments induce unwanted attenuation and phase shift on the reflected signals by the RIS introducing static or time-varying distortions in the received signal, which directly affect the channel estimation accuracy and, \textcolor{black}{consequently,} the system performance. However, the effects of hardware impairments and environmental imperfections have been rarely studied in the literature. For instance, \cite{imp1,imp3,imp4,imp5}, consider RISs operating under finite resolution of the phase shifts or phase estimation errors from imperfect channel estimation. \textcolor{black}{Then,}
%More completely,
\textcolor{black}{\cite{mainblock} and \cite{BiLi02}} consider different environmental effects on the RIS and propose methods to jointly estimate the channel and array blockage parameters in mmWave RIS-assisted systems. 

%Despite the strong attention received recently, the studies on RIS are still in their \textcolor{blue}{infancy,} and some research directions become crucial to reveal the real benefits and limitations of this technology. \textcolor{blue}{Particularly, the} previous works \textcolor{blue}{mainly} assume an ideal hardware at the RIS, neglecting the impact of inevitable hardware \textcolor{blue}{impairments} or physical \textcolor{blue}{imperfections} which introduce uncertainties, or unknown perturbations, in the cascaded channel model. 
%that directly affects the system performance. 
%Most of the existing works 
%\textcolor{blue}{Here,} e.g., \cite{imp1,imp2,imp3,imp4,imp5}, consider RIS operating under finite resolution of the phase shifts or phase estimation errors from imperfect channel estimation. However, in addition to the phase noise, \textcolor{blue}{different} types of physical \textcolor{blue}{imperfections} may exist due to the exposition of the RIS to different environmental \textcolor{blue}{effects}. For instance, in outdoor locations the RIS elements are subject to partial or complete blockages due to objects such as water precipitations, flying debris, air particles, snowflakes, ice stones, dry/damp sand particles and dirt, as shown in \textcolor{blue}{Fig.} \ref{figure1}. In practice, the existence of \textcolor{blue}{such blocking} objects induce unwanted attenuation and phase shift on the reflected signals introducing static or time-varying distortions in the received signal.

As we explain \textcolor{black}{in the following,} typical
%\textcolor{blue}{Because on this,} the aforementioned 
channel estimation methods may not be able to deal with \textcolor{black}{different} \textcolor{black}{imperfections} and \textcolor{black}{may} fail to properly estimate the channel. Therefore, it is necessary to continuously monitor the channel and compensate \textcolor{black}{for imperfections} in order to maintain \textcolor{black}{robust} system operation. 

In this paper, 
%by considering long-timescale and short-timescale impairment models for an RIS-assisted MIMO system, 
we propose tensor-based algorithms for the joint estimation of the involved channels and \textcolor{black}{imperfections in RIS-assisted MIMO systems. We take both the long- and short-term imperfections into account.} First, we show that the received signal under the \textcolor{black}{long- and short-term} 
%long- and short-timescale 
\textcolor{black}{imperfection} models can be recast as tensors following trilinear and quadrilinear PARAFAC models, respectively. Exploiting the \textcolor{black}{multi-linear} structure of these models, we derive two sets of tensor-based algorithms. For the \textcolor{black}{long-term} 
%long-timescale 
\textcolor{black}{imperfection (LTI)} model, where the RIS \textcolor{black}{imperfections}, modeled as unknown phase shifts, are static within the channel coherence time, we formulate an iterative trilinear ALS-based algorithm, named TALS-LTI, for the joint estimation of the involved channels and the unknown RIS phase deviations. Next, we generalize the \textcolor{black}{imperfections} behavior to be non-static with respect to the channel coherence time, referred to as the \textcolor{black}{short-term} 
%short-timescale 
\textcolor{black}{imperfection (STI)} model. For \textcolor{black}{such a} more challenging scenario, we propose iterative and closed-form tensor decomposition-based algorithms named TALS-STI and HOSVD-STI, respectively, to solve the joint channel and \textcolor{black}{RIS imperfections} estimation. We also study the identifiability of the proposed \textcolor{black}{estimators,} discuss their computational complexity \textcolor{black}{and investigate the effect of imperfections on the network performance.} \textcolor{black}{The key features of the proposed tensor-based algorithms are \gf{their} ability to properly estimate the channel and \gf{their} robustness to different kinds of real-world imperfections at the RIS.}

The simulation results show that, compared to the state-of-the-art methods, 
%competing method in \cite{GilJournal}, 
the proposed algorithms properly estimate the involved channels when different kinds of imperfections are takes into account. As a example, in the high signal-to-noise ratio (SNR) regime, our proposed algorithms improve the channel estimation by approximately $100$x compared to the method of \cite{GilJournal}, while present performance close to the lower-bound least squares (LS) estimator. Also, our proposed algorithms reduce considerably the overall \textcolor{black}{computational complexity}, compared to the related state-of-the-art method. %Then, the TALS-STI and HOSVD-STI algorithms are shown to be approximately $33$x and $21$x faster, respectively. 
Finally, the proposed TALS-LTI and TALS-STI algorithms are more flexible for the choices of training parameters compared to the proposed HOSVD-STI algorithm. Thus, the TALS-LTI and TALS-STI algorithms are preferable when more flexible
choices for training parameters are required, while the HOSVD-STI is preferred when low processing delay is desired.

The rest of this paper is organized as follows. In Section \ref{sModel}, the signal model of the RIS-assisted MIMO communication system \textcolor{black}{operating under imperfections} is introduced. We distinguish between two different types of \textcolor{black}{imperfections} under the RIS operation, \textcolor{black}{namely,} \textcolor{black}{LTI and STI.} 
%\textcolor{black}{long- and short-term imperfections}. 
Then, the channel estimation problem is discussed for the \textcolor{black}{LTI and STI scenarios.} 
%this more realistic scenario. 
These signal models are reformulated as higher-order tensors that \textcolor{black}{follow} trilinear and quadrilinear PARAFAC models, from which \textcolor{black}{two sets of} iterative and closed-form \textcolor{black}{tensor-based} algorithms for the joint estimation of the involved channels and \textcolor{black}{the} RIS \textcolor{black}{imperfections} are developed in Section \ref{CEAlg}. A detailed identifiability analysis and \textcolor{black}{its} link to the system design recommendations \textcolor{black}{as well as} \textcolor{black}{the computational complexity of the proposed tensor-based algorithms} are provided in Section \ref{Iden}. Simulation results are presented in Section \ref{Resul}. Finally, conclusions 
%and perspectives for the sequence of this work 
are drawn in Section \ref{Conc}.

\vspace{-0.3cm}
\subsection{Notations and Properties}

\textcolor{black}{The} notation conventions and \textcolor{black}{the} properties that will be used throughout this \textcolor{black}{paper} are defined in the following. Scalars are denoted by lower-case letters ($a, b, \ldots$), column vectors by bold lower-case letters ($\mathbf{a}, \mathbf{b}, \ldots$), matrices by bold upper-case letters ($\mathbf{A}, \mathbf{B}, \ldots$) and tensors are represented by upper-case calligraphic letters ($\mathcal{A}, \mathcal{B}, \ldots$). \textcolor{black}{Then,} $\mathbf{A}^{\text{T}}$ and $\mathbf{A}^{\dag}$ stand for the transpose and Moore-Penrose pseudo-inverse of $\mathbf{A}$, respectively. The operator $\text{vec}(\cdot)$ vectorizes its matrix argument by stacking its columns on top of each other, while $\text{vecd}(\cdot)$ forms a vector out of the diagonal of its matrix argument. \textcolor{black}{Also,} $\| \cdot \|_{\text{F}}$ represents the Frobenius norm \textcolor{black}{of a matrix or a tensor}, which is defined as the square root of the sum of the squared of its elements. $\lceil x \rceil$ is equal to the smallest integer that is greater than or equal to $x$. \textcolor{black}{Moreover,} $\mathbf{I}_{M}$ is the $M \times M$ identity matrix and $j = \sqrt{-1}$ is the imaginary unit. The operator $\mathbf{D}_{i}\left(\mathbf{A}\right)$ forms a diagonal matrix from the $i$-th row of its matrix argument $\mathbf{A}$, while the operator $\text{diag}(\mathbf{a})$ forms a diagonal matrix out of its vector argument $\mathbf{a}$. \textcolor{black}{We} define the Kronecker, Hadamard (element-wise product) and the outer product operators \textcolor{black}{by} $\otimes$, $\odot$ and $\circ$, respectively. The Khatri-Rao product (column-wise Kronecker product) between two matrices is defined as
\vspace{-0.1cm}
\begin{equation}
\mathbf{A} \diamond \mathbf{B} = \left[\mathbf{a}_{1} \otimes \mathbf{b}_{1}, \ldots, \mathbf{a}_{Q} \otimes \mathbf{b}_{Q}\right] \in \mathbb{C}^{IJ \times Q},
\label{pp1}
\end{equation}
or, equivalently, 
\vspace{-0.1cm}
\begin{equation}
\mathbf{A} \diamond \mathbf{B} = \left[\mathbf{D}_{1}\left(\mathbf{A}\right)\mathbf{B}^{\text{T}}, \ldots, \mathbf{D}_{Q}\left(\mathbf{A}\right)\mathbf{B}^{\text{T}}\right]^{\text{T}},
\label{pp2}
\end{equation}
where $\mathbf{A} = \left[\mathbf{a}_{1}, \ldots, \mathbf{a}_{Q}\right] \in \mathbb{C}^{I \times Q}$ and $\mathbf{B} = \left[\mathbf{b}_{1}, \ldots, \mathbf{b}_{Q}\right] \in \mathbb{C}^{J \times Q}$.

We shall make use of the following properties of the Khatri-Rao and Kronecker products
\textcolor{black}{
\begin{eqnarray}
\text{vec}\left(\mathbf{A}\text{diag}\left(\mathbf{c}\right)\mathbf{B}^{\text{T}}\right) &=& \left(\mathbf{B} \diamond \mathbf{A}\right)\mathbf{c}, \space \forall \space \mathbf{A}, \mathbf{B}, \mathbf{c},\label{pp3} \\
\mathbf{a} \otimes \mathbf{b} \otimes \mathbf{c} &=& \text{vec}\left(\mathbf{c} \circ \mathbf{b} \circ \mathbf{a}\right), \space \forall \space \mathbf{a}, \mathbf{b}, \mathbf{c}. \label{pp4}
\end{eqnarray}
}

Furthermore, the definitions and operations involving tensors are in accordance \textcolor{black}{with} \cite{Kolda} and \cite{hosvd}. The $n$-mode unfolding matrix of $\mathcal{A}$ along its $n$-th mode (or dimension) is represented by $\left[\mathcal{A}\right]_{(n)}$. The $n$-mode product between $\mathcal{A}$ and $\mathbf{B}$, returns a tensor $\mathcal{C} = \mathcal{A} \times_{n} \mathbf{B}$ such that $\left[\mathcal{C}\right]_{(n)} = \mathbf{B}\left[\mathcal{A}\right]_{(n)}$.

%\begin{figure}[t]
%	\centering
%	\includegraphics[width=0.35\textwidth]{figs/ff01.pdf}
%	\caption{Illustration of the considered channel estimation protocol.
%	%, proposed in \cite{gg1}. 
%	The training time is divided into $K$ time-blocks of duration $T$ symbol periods each. The RIS activation pattern $\mathbf{s}[k]$ is fixed during the $k$-th time-block and varies between different time-blocks while the \textcolor{red}{pilot symbols $\mathbf{x}[1], \ldots, \mathbf{x}[T]$} are reused from block-to-block.}
%	\label{protocol}
%\end{figure}

\begin{figure}[t]
	\centering
	\includegraphics[width=0.35\textwidth]{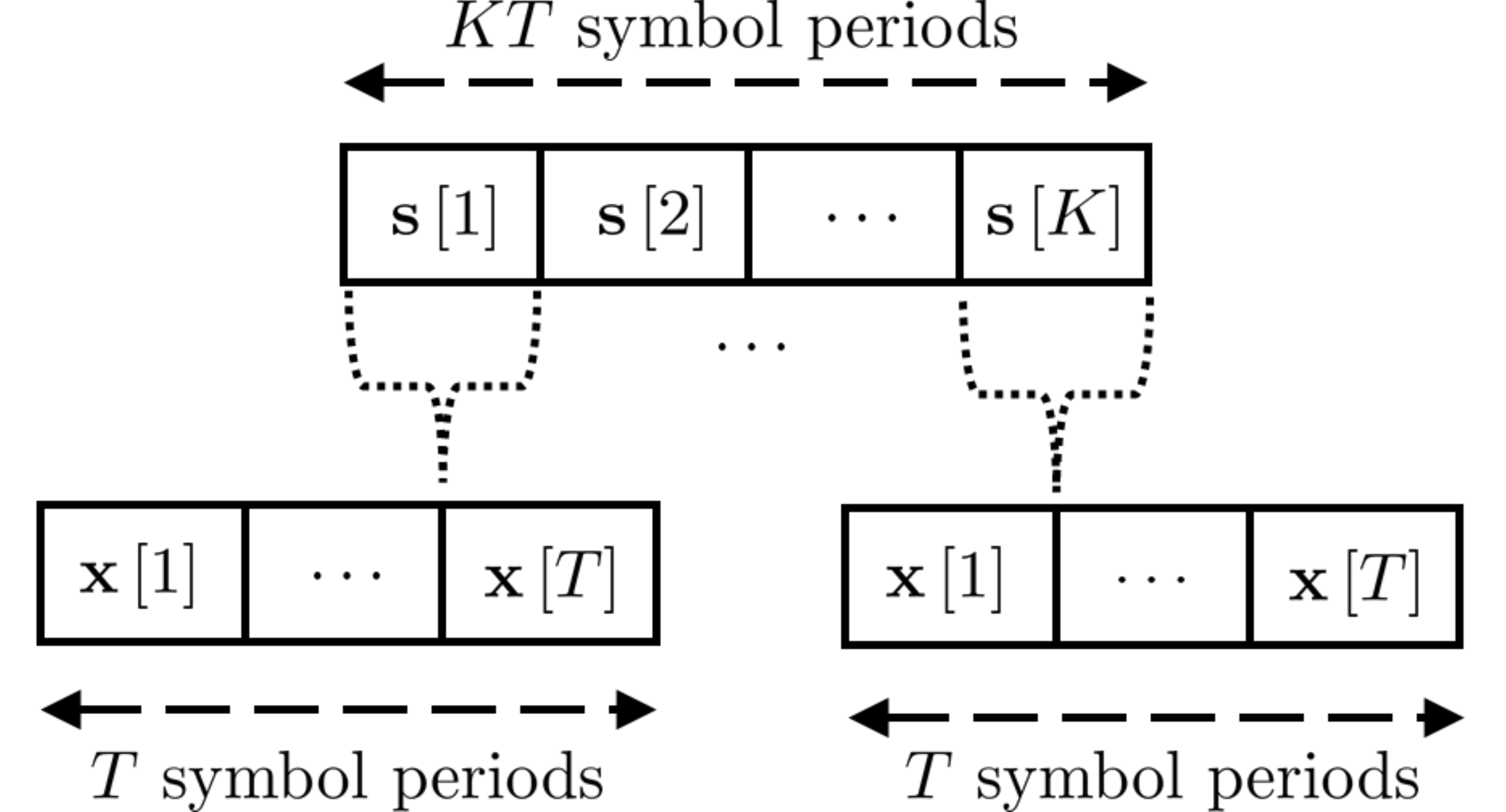}

	\caption{\small{An illustration of the considered channel estimation protocol.
	The training time is divided into $K$ time-blocks of duration $T$ symbol periods each. The RIS activation pattern $\mathbf{s}[k]$ is fixed during the $k$-th time-block and varies between different time-blocks while the pilot symbols $\mathbf{x}[1], \ldots, \mathbf{x}[T]$ are reused from block-to-block.}}
	\label{protocol}
\end{figure}
\vspace{-0.2cm}
\section{Signal Model and Problem Description}\label{sModel}

In this section, we first introduce the signal model and describe in detail the two considered 
%\textcolor{blue}{long- and short-term imperfection}
\textcolor{black}{LTI and STI} models. 
%Then, the proposed tensor-based channel estimation algorithms are presented.

\vspace{-0.1cm}
\subsection{RIS Operating \textcolor{black}{under} \textcolor{black}{LTI}}\label{longtime}

We consider a single-user narrowband RIS-assisted MIMO communication system, in which the transmitter (Tx) and the receiver (Rx) are \gf{equipped} with arrays composed of $M$ and $L$ antennas, respectively. Without loss of generality, although this work assumes a downlink communication, our signal models also \textcolor{black}{apply} to the uplink case by inverting the roles of the transmitter and the receiver. 
To improve the communication performance, an RIS with $N$ individually adjustable passive reflecting elements is deployed in a proper place that creates an alternative Tx-RIS-Rx link. The direct Tx-Rx link is assumed to be too weak or unavailable due to unfavorable propagation conditions. The considered system setup is illustrated in \textcolor{black}{Fig.} \ref{figure1}. 
%In addition, 
\textcolor{black}{We} also assume a block-fading channel where the Tx-RIS and RIS-Rx channels remain constant for at least $k = 1, \ldots, K,$ time-blocks, each with \textcolor{black}{duration of $T$ symbols}, representing a total duration of \textcolor{black}{$KT$} symbol periods dedicated for channel estimation within the channel coherence time.

%\begin{figure}[t]
%	\centering
%	\includegraphics[width=0.45\textwidth]{figs/ilu03.pdf}
%	\caption{Illustration of the RIS-assisted MIMO wireless communication system \textcolor{magenta}{operating under imperfections}.}
%	\label{figure2}
%\end{figure}

Let us define the RIS activation pattern $\mathbf{s}\left[k\right] \in \mathbb{C}^{N \times 1}$ configured at the $k$-th time-block as
\vspace{-0.1cm}
\begin{equation}
\mathbf{s}\left[k\right]= \left[\beta_{1,k}e^{j\phi_{1,k}}, \ldots, \beta_{N,k}e^{j\phi_{N,k}}\right]^{\text{T}} \in \mathbb{C}^{N \times 1},
\label{eleimp}
\end{equation}
where $0 \leq \phi_{n,k} \leq 2\pi$ and $0 \leq \beta_{n,k} \leq 1$ denote the phase shift and the amplitude reflection coefficient of the $n$-th RIS element tuned at the $k$-th time-block $\forall n = 1, \ldots, N$, and $\forall k = 1, \ldots, K$, respectively. 
As a protocol for channel estimation we assume the following \textcolor{black}{(see Fig. \ref{protocol})}:

\noindent 1) At each time-block $k = 1, \ldots, K$, of duration $T$, the elements of $\mathbf{s}\left[k\right]$ are dynamically tuned in a passive way via the smart controller; 

\noindent 2) The activation pattern $\mathbf{s}\left[k\right]$ remains constant within the $k$-th time-block but \textcolor{black}{may vary} between different time-blocks, yielding a total of $\mathbf{s}\left[1\right], \ldots, \mathbf{s}\left[K\right]$ different adjustable patterns to the RIS during the channel estimation stage; 
	
\noindent 3) The \textcolor{black}{pilot symbol} $\mathbf{x}\left[t,k\right] \in \mathbb{C}^{M \times 1}$ transmitted at the $t$-th symbol period within the $k$-th time-block is reused for each $k = 1, \ldots, K$, i.e., $\mathbf{x}\left[t,k\right] = \mathbf{x}\left[t\right]$ $\forall k = 1, \ldots, K$.

%\begin{enumerate}
%	\item At each time-block $k = 1, \ldots, K$, of duration $T$, the elements of $\mathbf{s}\left[k\right]$ are dynamically tuned in a passive way via the smart controller; 
%	\item The activation pattern $\mathbf{s}\left[k\right]$ remains constant within the $k$-th time-block but \textcolor{black}{may vary} between different time-blocks, yielding a total of $\mathbf{s}\left[1\right], \ldots, \mathbf{s}\left[K\right]$ different adjustable patterns to the RIS during the channel estimation stage; 
%	\item The \textcolor{black}{pilot symbol} $\mathbf{x}\left[t,k\right] \in \mathbb{C}^{M \times 1}$ transmitted at the $t$-th symbol period within the $k$-th time-block is reused for each $k = 1, \ldots, K$, i.e., $\mathbf{x}\left[t,k\right] = \mathbf{x}\left[t\right]$ $\forall k = 1, \ldots, K$.
%\end{enumerate} 

%For completeness, the considered channel estimation protocol detailed above is illustrated in \textcolor{black}{Fig.} \ref{protocol}. 

The baseband received \textcolor{black}{pilot} signal $\mathbf{y}\left[t,k\right] \in \mathbb{C}^{L \times 1}$ associated with the $t$-th symbol period at the $k$-th time-block can be expressed as
\vspace{-0.1cm}
\begin{equation}
\mathbf{y}\left[t,k\right] = \mathbf{G}\text{diag}\left(\mathbf{s}\left[k\right]\right)\mathbf{H}^{\text{T}}\mathbf{x}\left[t\right] + \mathbf{v}\left[t,k\right].
\label{simples}
\end{equation}
Collecting the received signals during the $T$ symbol periods at the $k$-th time-block, the model in (\ref{simples}) can be rewritten as
\vspace{-0.1cm}
\begin{equation}
\mathbf{Y}\left[k\right]= \mathbf{G}\text{diag}\left(\mathbf{s}\left[k\right]\right)\mathbf{H}^{\text{T}}\mathbf{X} + \mathbf{V}\left[k\right] \in \mathbb{C}^{L \times T},
\label{pmod1}
\end{equation} 
where $\mathbf{Y}\left[k\right] = \left[\mathbf{y}\left[1,k\right], \ldots, \mathbf{y}\left[T,k\right]\right] \in \mathbb{C}^{L \times T}$. The matrices $\mathbf{H} \in \mathbb{C}^{M \times N}$ and $\mathbf{G} \in \mathbb{C}^{L \times N}$ denote the Tx-RIS and RIS-Rx channels, respectively, while $\mathbf{X} = \left[\mathbf{x}\left[1\right], \ldots, \mathbf{x}\left[T\right]\right] \in \mathbb{C}^{M \times T}$ collects the pilot signals transmitted within the $k$-th time-block, and $\mathbf{V}\left[k\right] = \left[\mathbf{v}\left[1,k\right], \ldots, \mathbf{v}\left[T,k\right]\right] \in \mathbb{C}^{L \times T}$ is the additive white Gaussian noise (AWGN) matrix with zero mean and unit variance elements. In order to simplify our formulation and analysis, without loss of \textcolor{black}{generality,} we assume the transmission of the pilot signal $\mathbf{X} = \mathbf{I}_{M}$.
%during the channel estimation stage. 

%\begin{figure}[t]
%	\centering
%	\includegraphics[width=0.35\textwidth]{figs/ff01.pdf}
%	\caption{Illustration of the considered channel estimation protocol, proposed in \cite{gg1}. The training time is divided into $K$ time-blocks of duration $T$ symbol periods each. The RIS activation pattern $\mathbf{s}[k]$ is fixed during the $k$-th time-block and varies between different time-blocks while the \textcolor{red}{pilot symbols $\mathbf{x}[1], \ldots, \mathbf{x}[T]$} are reused from block-to-block.}
%	\label{protocol}
%\end{figure}

In this work, a special attention is given to the structure of the RIS activation pattern. 
%As alluded to in the introduction in Section \ref{intro}, 
\textcolor{black}{In practice,} some \textcolor{black}{imperfections} at the RIS elements are common to occur. Initially, we assume the case in which such \textcolor{black}{imperfections} induce \textcolor{black}{long-term} static phase shift perturbations at the RIS response. \textcolor{black}{Such imperfections may come from, e.g., phase noise due to the finite resolution of the phase shifts or by phase estimation errors from imperfect channel estimation.}
%This is a valid assumption, for instance, when the RIS elements are impaired by phase noise caused due to the finite resolution of the phase shifts or by phase estimation errors from imperfect channel estimation. 
In the presence of these \textcolor{black}{imperfections,} the structure of the RIS activation pattern in (\ref{eleimp}) is modified in an \textcolor{black}{undesired} 
%unwanted 
manner leading to the following resulting RIS reflection pattern that incorporates the \textcolor{black}{imperfection} contributions:
\vspace{-0.1cm}
\begin{equation}
\bar{\mathbf{s}}\left[k\right] = \left[\beta_{1,k}e^{j(\phi_{1,k} + \theta_{1})}, \ldots, \beta_{N,k}e^{j(\phi_{N,k} + \theta_{N})}\right]^{\text{T}}. 
%\in \mathbb{C}^{N \times 1},
\label{uyg}
\end{equation}
\textcolor{black}{Here,} $0 \leq \theta_{n} \leq 2\pi$ $\forall n = 1, \ldots, N$ denotes the phase shift perturbation that affects the $n$-th RIS element. \textcolor{black}{Also,}
%Equation 
(\ref{uyg}) can 
%also 
be alternatively represented in a more attractive form for our formulation as 
\vspace{-0.2cm}
\begin{equation}
\bar{\mathbf{s}}\left[k\right] = \mathbf{e} \odot \mathbf{s}\left[k\right] \in \mathbb{C}^{N \times 1},
\label{contlr}
\end{equation}
where the entries of the random vector $\mathbf{e} \in \mathbb{C}^{N \times 1}$ that collects \textcolor{black}{all} unknown existing phase perturbations are defined as
\vspace{-0.1cm}
\begin{equation}
e_{n} = \left\{\begin{array}{ll}
%1, & \quad \text{\textcolor{blue}{ideal case}} \\
1, & \quad \text{\textcolor{black}{non-impaired case}} \\
e^{j\theta_{n}}, & \quad \text{\textcolor{black}{otherwise}},
\end{array}
\right.
\label{condsN}
\end{equation}
%\begin{equation}
%e_{n} = \left\{\begin{array}{ll}
%e^{j\theta_{n}}, & \quad \text{if the $n$-th RIS element is impaired} \\
%1, & \quad \text{otherwise},
%\end{array}
%\right.
%\label{condsN}
%\end{equation}
\noindent for $n = 1, \ldots, N$. Making use of definitions in (\ref{contlr}) and (\ref{condsN}), the impaired version of the received signal \textcolor{black}{at the Rx node} in (\ref{pmod1}) can be expressed as
\vspace{-0.1cm}
\begin{equation}
\mathbf{Y}\left[k\right]= \mathbf{G}\text{diag}\left(\underbrace{\mathbf{e} \odot \mathbf{s}\left[k\right]}_{\bar{\mathbf{s}}\left[k\right]}\right)\mathbf{H}^{\text{T}} + \mathbf{V}\left[k\right] \in \mathbb{C}^{L \times M}, %\space \textcolor{blue}{\forall k = 1, \ldots, K.}
\label{improot1}
\end{equation}   
$\forall k = 1, \ldots, K$. \noindent Equivalently, in matrix form we have
\vspace{-0.1cm}
\begin{equation}
\mathbf{Y}\left[k\right]= \mathbf{G}\text{diag}\left(\mathbf{e}\right)\mathbf{D}_{k}\left(\mathbf{S}\right)\mathbf{H}^{\text{T}} + \mathbf{V}\left[k\right],
\label{improot}
\end{equation}
where $\mathbf{S} = \left[\mathbf{s}\left[1\right], \ldots, \mathbf{s}\left[K\right]\right]^{\text{T}} \in \mathbb{C}^{K \times N}$ collects in its rows the RIS activation patterns used accross $K$ time-blocks specifically configured for the channel estimation. 

\begin{figure}
\centering
\begin{subfigure}[b]{0.45\textwidth}
   \includegraphics[width=1\linewidth]{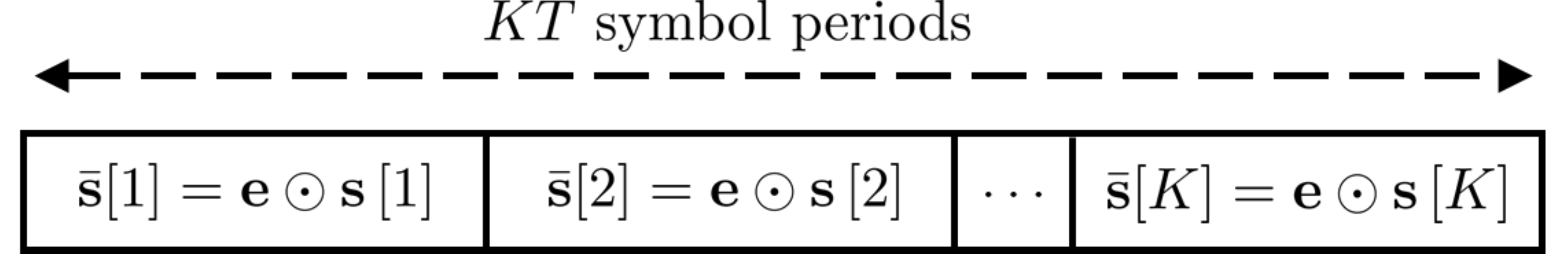}
   \caption{\textcolor{black}{An} illustration of the \textcolor{black}{LTI model.} 
   %\textcolor{blue}{long-term} RIS \textcolor{blue}{imperfection} model. 
   The vector $\mathbf{e}$ with phase perturbations is static during the \textcolor{black}{$KT$} symbol periods.}
   \label{lti} 
\end{subfigure}

\vspace{+0.1cm}

\begin{subfigure}[b]{0.5\textwidth}
   \includegraphics[width=1\linewidth]{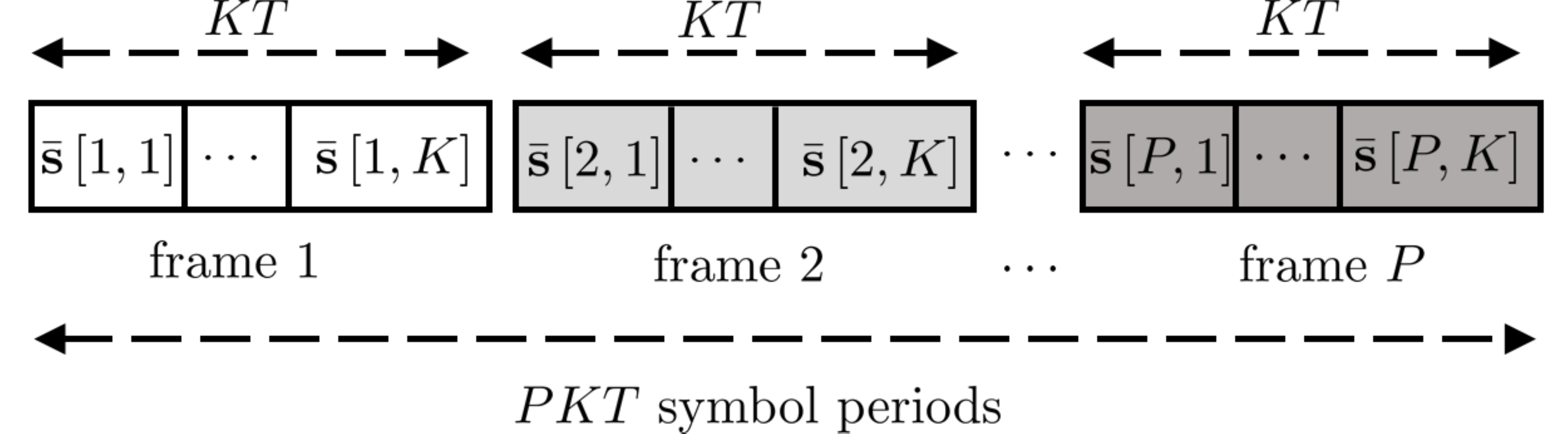}
   \caption{\textcolor{black}{An} illustration of the generalized \textcolor{black}{STI model with time-varying RIS fluctuations during the channel coherence time.}} 
   %\textcolor{blue}{short-tem} RIS \textcolor{blue}{imperfection} model.}
   \label{sti}
\end{subfigure}

%\caption[]{Illustration of the long- and short-term imperfection models.}
\caption[]{\small{Illustration of the LTI and STI models considered in the paper.}} 
%{long- and short-term imperfection models.}
\end{figure}

Particularly, the impaired received signal in (\ref{improot}) \textcolor{black}{considers} an RIS operating under \textcolor{black}{LTI} 
%\textcolor{blue}{long-term} \textcolor{blue}{imperfections} 
that \textcolor{black}{induces} phase shift perturbations in the reflected signals. In other words, in this model \textcolor{black}{we assume} that the vector $\mathbf{e} \in \mathbb{C}^{N \times 1}$ is formed only by phase components that \textcolor{black}{remain} static within the \textcolor{black}{$KT$} symbol periods. Figure \ref{lti} illustrates \textcolor{black}{the considered LTI model.} 
%this RIS \textcolor{blue}{imperfection} model. 
This occurs, for instance, when the behavior of the \textcolor{black}{imperfections} at the RIS elements \textcolor{black}{are static} 
%vary more slowly 
compared to the channel coherence time. 
%In the following, we formulate a second RIS impairment model by considering a dynamic scenario in which the RIS impairments induce both amplitude and phase perturbations and also vary quickly, i.e. are non-static during the training time.

%\begin{figure}[t]
%	\centering
%	\includegraphics[width=0.5\textwidth]{figs/ff02.pdf}
%	\caption{Illustration of the \textcolor{blue}{long-term} RIS \textcolor{blue}{imperfection} model. The vector $\mathbf{e}$ with phase perturbations is static during the $K \cdot T$ symbol periods.}
%	\label{lti}
%\end{figure}

\vspace{-0.3cm}
\subsection{RIS Operating \textcolor{black}{under} \textcolor{black}{STI}}\label{shorttime}

\textcolor{black}{Here, we} assume that the channel estimation occurs after the receiver collects $p = 1, \ldots, P$, frames composed of \textcolor{black}{$KT$} symbol periods each. The division of the reception time into $P$ frames is motivated by possible \textcolor{black}{short-term} variations caused by the RIS \textcolor{black}{imperfections,} \textcolor{black}{i.e., the imperfections changing more frequently.} In this approach, the behavior of the \textcolor{black}{imperfections} induces both amplitude and phase perturbations in the RIS elements and have a non-static nature with respect to channel coherence time but they present stationary characteristics within \textcolor{black}{each} 
%the $p$-th 
frame. \textcolor{black}{Figure} \ref{sti} illustrates the considered STI model.
%short-term RIS imperfection model.} 
In contrast to the received signal model formulated in \textcolor{black}{(\ref{improot})}, this assumption implies that such \textcolor{black}{imperfection} model takes into account rapid \textcolor{black}{amplitude and phase} fluctuations \textcolor{black}{at the RIS elements} during the channel coherence time. Note that the \textcolor{black}{STI} 
%\textcolor{blue}{short-term} \textcolor{blue}{imperfections} 
\textcolor{black}{induces} a block-fading effect in the signals reflected by the RIS during the \textcolor{black}{$PKT$} symbol periods for channel estimation. Therefore, by considering an RIS operating under this \textcolor{black}{STI model} 
%\textcolor{blue}{short-term imperfection} model 
the resulting reflection pattern related to the $k$-th time-block at the $p$-th frame can be written as %follows
\vspace{-0.1cm}
\begin{equation}
\bar{\mathbf{s}}\left[p,k\right] = \left[e_{n,p}\beta_{1,k}e^{j\phi_{1,k}}, \ldots, e_{N,p}\beta_{N,k}e^{j\phi_{1,k}}                                       \right]^{\text{T}},
\end{equation}
or, equivalently,
\vspace{-0.2cm}
\begin{equation}
\bar{\mathbf{s}}\left[p,k\right] = \mathbf{e}\left[p\right] \odot \mathbf{s}\left[k\right] \mathbb{C}^{N \times 1},
\label{spk}
\end{equation}
where $\mathbf{e}\left[p\right] = \left[e_{1,p}, \ldots, e_{N,p}\right]^{\text{T}} \in \mathbb{C}^{N \times 1}$ $\forall p = 1, \ldots, P$, depends on the $p$-th received frame and models the unknown non-static amplitude and phase fluctuations along the training time. 
%Figure \ref{sti} illustrates the considered short-timescale RIS impairment model. 
The entries of the random vector $\mathbf{e}\left[p\right]$ associated with the $p$-th frame are defined as
\vspace{-0.1cm}
\begin{equation}
e_{n,p} = \left\{\begin{array}{ll}
%\alpha_{n,p}\cdot e^{j\theta_{n,p}}, & \quad \text{if the $n$-th RIS element is impaired \\ at the $p$-th frame} \\
%1, & \quad \textcolor{blue}{\text{ideal case}} \\
1, & \quad \textcolor{black}{\text{non-impaired case}} \\
\alpha_{n,p}\cdot e^{j\theta_{n,p}}, & \quad \textcolor{black}{\text{otherwise}},
%\alpha_{n,p}\cdot e^{j\theta_{n,p}}, & \quad \text{if the $n$-th RIS element is impaired} \\
%1, & \quad \text{otherwise},
\end{array}
\right.
\label{conds}
\end{equation}
where $0 \leq \alpha_{n,p} \leq 1$ and $0 \leq \theta_{n,p} \leq 2\pi$ $\forall n = 1, \ldots, N$, and $\forall p = 1, \ldots, P$, denote the unwanted amplitude attenuation and phase shift perturbations that \textcolor{black}{affect} the $n$-th RIS element at the $p$-th frame, respectively. It is important to note that the model in (\ref{conds}) captures different kinds of real-world \textcolor{black}{imperfections} at the RIS. For example, we can note the \textcolor{black}{follow} situations: 

%\begin{figure}[t]
%	\centering
%	\includegraphics[width=0.5\textwidth]{figs/ff03.pdf}
%	\caption{Illustration of the generalized \textcolor{blue}{short-term} RIS \textcolor{blue}{imperfection} model.}
%	\label{sti}
%\end{figure}

\noindent 1) $\alpha_{n,p} \neq 0$ and $\theta_{n,p} \neq 0$ represent the amplitude absortion and phase shift caused by an object suspended on the $n$-th RIS element \cite{mainblock}, or caused by hardware impairments in the electronic circuits that make up the RIS \cite{var1}. 

\noindent 2) $\alpha_{n,p} = 1$ and $\theta_{n,p} \neq 0$ represent the phase noise perturbations from low-resolution phase shifts or phase errors from imperfect channel estimation \cite{imp1}-\cite{imp5}. 

\noindent 3) $\alpha_{n,p} = 0$ represents the maximum absorption i.e., the $n$-th RIS element is completely blocked \cite{mainblock}.

\noindent 4) $\alpha_{n,p} = 1$ and $\theta_{n,p} = 0$ represents the \textcolor{black}{non-impaired}
	%unfeasible perfect 
	RIS in which no \textcolor{black}{imperfection} affects its $n$-th element. Note that in this ideal case $\bar{\mathbf{s}}\left[p,k\right] = \mathbf{s}\left[p,k\right]$ holds since $e_{n,p} = 1$ $\forall n = 1, \ldots, N$, and $\forall p = 1, \ldots, P$.

%\begin{itemize}
%	\item $\alpha_{n,p} \neq 0$ and $\theta_{n,p} \neq 0$ represent the amplitude absortion and phase shift caused by an object suspended on the $n$-th RIS element \cite{mainblock}, or caused by hardware impairments in the electronic circuits that make up the RIS \cite{var1,var2}. 
%	\item $\alpha_{n,p} = 1$ and $\theta_{n,p} \neq 0$ represent the phase noise perturbations from low-resolution phase shifts or phase errors from imperfect channel estimation \cite{imp1}-\cite{imp5}. 
%	\item $\alpha_{n,p} = 0$ represents the maximum absorption i.e., the $n$-th RIS element is completely blocked \cite{mainblock}.
%	\item $\alpha_{n,p} = 1$ and $\theta_{n,p} = 0$ represents the \textcolor{black}{non-impaired}
%	%unfeasible perfect 
%	RIS in which no \textcolor{black}{imperfection} affects its $n$-th element. Note that in this ideal case $\bar{\mathbf{s}}\left[p,k\right] = \mathbf{s}\left[p,k\right]$ holds since $e_{n,p} = 1$ $\forall n = 1, \ldots, N$, and $\forall p = 1, \ldots, P$.
%\end{itemize} 
\textcolor{black}{We also} observe that the \textcolor{black}{LTI} 
%\textcolor{blue}{long-term} RIS \textcolor{blue}{imperfection} 
model presented in Section \ref{longtime} is a particular case of the generalized \textcolor{black}{STI} 
%\textcolor{blue}{short-term} \textcolor{blue}{imperfection} 
model when $\alpha_{n,p} = 1$ for $P =1$ and $n = 1, \ldots, N$. \textcolor{black}{However, we study these two cases separately for a clearer description of the static and non-static imperfection scenarios.}

By considering \textcolor{black}{an RIS} operating under the \textcolor{black}{STI}
%\textcolor{blue}{short-term} \textcolor{blue}{imperfection} 
model, we can rewrite the received signal in (\ref{improot1}) as
\vspace{-0.1cm}
\begin{equation}
\mathbf{Y}\left[p,k\right] = \mathbf{G}\text{diag}\left(\underbrace{\mathbf{e}\left[p\right] \odot \mathbf{s}\left[k\right]}_{\bar{\mathbf{s}}\left[p,k\right]}\right)\mathbf{H}^{\text{T}} + \mathbf{V}\left[p,k\right],
\label{dec1}
\end{equation}
$\forall p = 1, \ldots, P$, and $\forall k = 1, \ldots, K$. 
%Note that the novel resulting RIS reflection pattern $\bar{\mathbf{s}}\left[p,k\right] = \mathbf{e}\left[p\right] \odot \mathbf{s}\left[k\right] \mathbb{C}^{N \times 1}$ is now a function of the $p$-th short-timescale sub-block and can be defined directly from Equation (\ref{conds}) by adding the temporal dependence. 
In a more convenient form for our formulation, the received signal (\ref{dec1}) can be written in its complete matrix and decoupled format as
\vspace{-0.1cm}
\begin{equation}
\mathbf{Y}\left[p,k\right] = \mathbf{G}\mathbf{D}_{p}\left(\mathbf{E}\right)\mathbf{D}_{k}\left(\mathbf{S}\right)\mathbf{H}^{\text{T}} + \mathbf{V}\left[p,k\right],
\label{dec2}
\end{equation}
where each row of the matrix $\mathbf{E} = \left[\mathbf{e}\left[1\right], \ldots, \mathbf{e}\left[P\right]\right]^{\text{T}} \in \mathbb{C}^{P \times N}$ collects the amplitude and phase parameters for the RIS elements impaired at the $p$-th frame. Throughout this work, for the two approaches formulated in Sections \ref{longtime} and \ref{shorttime}, we assume that a number of \textcolor{black}{$N_{B} = NR_{B}$} random elements at the RIS are subject to \textcolor{black}{imperfections,} where \textcolor{black}{$R_{B} \in [0,1]$} denotes its occurrence probability.
\vspace{-0.1cm}
\subsection{Problem Description}

\textcolor{black}{In (\ref{contlr}) and (\ref{spk}),}
%As discussed in the previous section, the
%As aforementioned, when realistic situation in which the \Gls{ris} elements are affected by some kind of impairments, for instance outdoor occurrences or inherent electronic limitations, the \Gls{ris} activation pattern matrix $\mathbf{S}$ may not be accurately known at the receiver due to amplitude and/or phase perturbations that can lead to distortions in the received signals inserting some degradation in the performance of the channel estimation processing. In other words, the resulting 
the RIS reflection pattern corrupted with errors $\bar{\mathbf{s}}\left[k\right]$ and $\bar{\mathbf{s}}\left[p,k\right]$ induce unwanted amplitude and/or phase shift responses in the reflected signal by the RIS creating a mismatch between the ideal reflection pattern and the one that is actually applied by the RIS. 
%intended signal generated in a controller way by using the ideal activation pattern $\mathbf{s}\left[k\right]$ and what is distorted by the impaired \Gls{ris}.
To deal with these \textcolor{black}{imperfections,} in this work we propose to jointly estimate the involved channels $\mathbf{G}$ and $\mathbf{H}$, as well as the unknown \textcolor{black}{imperfections} that \textcolor{black}{affect} the $N_{B}$ impaired elements at the RIS. Decoupled estimations of the channels are required, for instance, to optimize the phase shifts at the RIS, the transmit precoder at the \textcolor{black}{transmitter} and the receive combiner at the \textcolor{black}{receiver} in order to maximize the rate and energy efficiencies in the data transmission phase \cite{AZappone2021,Yigit2020}. However, the issues of how to utilize the estimated channels to \textcolor{black}{jointly optimize the RIS phase shifts, the transmitter and receiver active beamformers,} \textcolor{black}{as well as the control overhead problem of optimal phase shifts \cite{Sokal01, Sokal02}} 
%solve the aforementioned optimization problem 
are out of the scope of this work and will be addressed in a future work. 

\textcolor{black}{In the following,} we show that the received signal models in (\ref{improot}) and (\ref{dec2}) can be represented as \textcolor{black}{third-} and fourth-order tensor models, \textcolor{black}{respectively.} Then, we show how such higher-order representations serve as reference models for the development of efficient PARAFAC-based algorithms to solve the channel estimation problem for more realistic scenarios where the RIS operates under \textcolor{black}{LTI and STI.} 
%\textcolor{blue}{long- or short-term} 
%long-timescale or short-timescale 
%\textcolor{blue}{imperfections}.   

%In the following, we show that the both received signal models can be represented as higher-order tensors and how such representations allows the development of efficient tensor-based algorithms to solve the channel estimation problem to the more realistic scenarios where RIS are working under long-timescale and short-timescale impairments. 

\vspace{-0.1cm}
\section{Proposed Channel Estimation Algorithms}\label{CEAlg}

\textcolor{black}{In this section, we propose three different channel estimation algorithms for the cases with LTI and STI.} 
%short- and long-term imperfections.}

%In this section, we propose three different PARAFAC-based algorithms for channel estimation in MIMO systems assisted by RIS operating under impairments. The first one, named TALS-LTI, is an iterative algorithm that address the scenario with long-timescale impairments. The other two, named TALS-STI and HOSVD-STI, are respectively iterative and closed-form algorithms to tackle the additional case with short-timescale impairments. 

%\subsection{Trilinear ALS for Channel Estimation in RIS Working Under Long-Timescale Impairments (TALSCE-LTI)}\label{proposta1}
%\subsection{TALS Algorithm for RIS Operating Under \textcolor{blue}{Long-Term} \textcolor{blue}{Imperfections} (TALS-LTI)}\label{proposta1}
\vspace{-0.1cm}
\subsection{TALS Algorithm for RIS Operating \textcolor{black}{under} \textcolor{black}{LTI} (TALS-LTI)}\label{proposta1}

We initially consider the scenario with \textcolor{black}{$N_{B}$} \textcolor{black}{unknown} passive elements of the RIS being affected by \textcolor{black}{LTI} 
%\textcolor{blue}{long-term} \textcolor{blue}{imperfections} 
as presented in Section \ref{longtime}. The values of the phase perturbations as well as \textcolor{black}{their} positions are assumed to be unknown at the receiver. 

\textcolor{black}{For simplicity} of presentation and without loss of generality, we neglect the noise term in our formulations. The noiseless part of the received signal in (\ref{improot}) can be naturally identified as the $k$-th frontal slice of a third-order tensor $\mathcal{Y} \in \mathbb{C}^{L \times M \times K}$ that admits the following PARAFAC decomposition \cite{Kolda} 
\begin{equation}
\mathcal{Y} = \mathcal{I}_{3,N} \times_{1} \mathbf{G} \times_{2} \mathbf{H} \times_{3} \bar{\mathbf{S}}.
\label{para01}
\end{equation}
%where 
\textcolor{black}{Here,} $\mathcal{I}_{3,N}$ represents a third-order identity tensor of size $N \times N \times N$.
%in which its elements have value 1 when all indices are equal and value 0 elsewhere, and $N$ is the rank of the \Gls{parafac} decomposition. 
According to (\ref{para01}), the factor matrices related to 1-mode, 2-mode and 3-mode of $\mathcal{Y}$ are respectively $\mathbf{G} \in \mathbb{C}^{L \times N}$, $\mathbf{H} \in \mathbb{C}^{M \times N}$ and $\bar{\mathbf{S}} = \left[\bar{\mathbf{s}}\left[1\right], \ldots, \bar{\mathbf{s}}\left[K\right]\right]^{\text{T}} \in \mathbb{C}^{K \times N}$ where according to (\ref{contlr}), $\bar{\mathbf{s}}\left[k\right]= \mathbf{e} \odot \mathbf{s} \left[k\right]$ or, equivalently, in matrix notation $\mathbf{D}_{k}\left(\bar{\mathbf{S}}\right) = \text{diag}\left(\mathbf{e}\right)\mathbf{D}_{k}\left(\mathbf{S}\right)$, $\forall k = 1, \ldots, K$. 

Resorting to the multilinear structure of the PARAFAC decomposition in (\ref{para01}), higher degrees of freedom for signal processing can be achieved by exploiting the dimensions of interest of the received signal tensor. In this sense, $\mathcal{Y}$ can also be expressed with respect to its 1-mode and 2-mode unfoldings, which can be expressed as
\begin{eqnarray}
\left[\mathcal{Y}\right]_{(1)} &=& \mathbf{G}\text{diag}\left(\mathbf{e}\right)\left(\mathbf{S} \diamond \mathbf{H}\right)^{\text{T}}\in \mathbb{C}^{L \times MK},\label{forvec}\label{sub1}\\
%\left[\mathcal{Y}\right]_{(2)} &=& \mathbf{H}\left(\mathbf{S} \diamond \mathbf{G}\text{diag}\left(\mathbf{e}\right)\right)^{\text{T}}  \in \mathbb{C}^{M \times LK}\label{sub2},
\left[\mathcal{Y}\right]_{(2)} &=& \mathbf{H}\text{diag}\left(\mathbf{e}\right)\left(\mathbf{S} \diamond \mathbf{G}\right)^{\text{T}}  \in \mathbb{C}^{M \times LK}\label{sub2},
%\left[\mathcal{Y}\right]_{(3)} &=& \bar{\mathbf{S}}\left(\mathbf{H} \diamond \mathbf{G}\right)^{\text{T}} \in \mathbb{C}^{K \times LM},
\end{eqnarray}
%\begin{eqnarray}
%\left[\mathcal{Y}\right]_{(1)} &=& \mathbf{G}\left(\bar{\mathbf{S}} \diamond \mathbf{H}\right)^{\text{T}} \in \mathbb{C}^{L \times MK},\\
%\left[\mathcal{Y}\right]_{(2)} &=& \mathbf{H}\left(\bar{\mathbf{S}} \diamond \mathbf{G}\right)^{\text{T}}  \in \mathbb{C}^{M \times LK},\\
%\left[\mathcal{Y}\right]_{(3)} &=& \bar{\mathbf{S}}\left(\mathbf{H} \diamond \mathbf{G}\right)^{\text{T}} \in \mathbb{C}^{K \times LM},
%\end{eqnarray}
%where $\left[\mathcal{Y}\right]_{(1)} \in \mathbb{C}^{L \times MK}$, $\left[\mathcal{Y}\right]_{(2)} \in \mathbb{C}^{M \times LK}$ and $\left[\mathcal{Y}\right]_{(3)} \in \mathbb{C}^{K \times LM}$ denote the 1-mode, 2-mode and 3-mode unfolding matrices of $\mathcal{Y}$ that are obtained from Equation (\ref{improot}) by stacking its frontal slices into wide matrices as defined below
where $\left[\mathcal{Y}\right]_{(1)} \in \mathbb{C}^{L \times MK}$ and $\left[\mathcal{Y}\right]_{(2)} \in \mathbb{C}^{M \times LK}$ are obtained from (\ref{improot}) by stacking the frontal slices into wide matrices, as defined \textcolor{black}{by}
\vspace{-0.3cm}
\begin{eqnarray}
	\left[\mathcal{Y}\right]_{(1)} &=& \left[\mathbf{Y}\left[1\right], \ldots, \mathbf{Y}\left[K\right]\right],\\
	\left[\mathcal{Y}\right]_{(2)} &=& \left[\mathbf{Y}^{\text{T}}\left[1\right], \ldots, \mathbf{Y}^{\text{T}}\left[K\right]\right].
	%\left[\mathcal{Y}\right]_{(3)} &=& \left[\text{vec}\left(\mathbf{Y}(1)\right), \ldots, \text{vec}\left(\mathbf{Y}(K)\right)\right]^{\text{T}}.
\end{eqnarray}
Additionally, (\ref{forvec}) can also be represented in a covenient vectorized form. By applying the property (\ref{pp3}) to (\ref{forvec}) we obtain
\begin{equation}
\text{vec}\left(\left[\mathcal{Y}\right]_{(1)}\right) = \left(\mathbf{S} \diamond \mathbf{H} \diamond \mathbf{G}\right)\mathbf{e} \in \mathbb{C}^{LMK \times 1}.\label{sub3}
\end{equation}

In the following, we describe an iterative way to estimate the channel matrices $\mathbf{H}$ and $\mathbf{G}$ from the received signal tensor $\mathcal{Y}$ that \textcolor{black}{models} the RIS-assisted MIMO system with RIS operating under \textcolor{black}{LTI.} 
%\textcolor{blue}{long-term} \textcolor{blue}{imperfections.} 
The estimation problem can be solved by computing a rank-$N$ approximation to the PARAFAC decomposition of $\mathcal{Y}$, i.e, 
\vspace{-0.1cm}
\begin{equation}
%\underset{\mathbf{G},\mathbf{H},\mathbf{e}}{\text{min}}\left\|\mathcal{Y} - \mathcal{I}_{3,N} \times_{1} \mathbf{G} \times_{2} \mathbf{H} \times_{3} \bar{\mathbf{S}}\right\|_{\text{F}}^{2},
\underset{\mathbf{G},\mathbf{H},\mathbf{e}}{\text{min}}\sum_{k=1}^{K}\left\|\mathbf{Y}\left[k\right] - \mathbf{G}\text{diag}\left(\mathbf{e}\right)\mathbf{D}_{k}\left(\mathbf{S}\right)\mathbf{H}^{\text{T}}\right\|_{\text{F}}^{2}.
\label{optpro}
\end{equation}
%In this problem, the knowledge of rank-$N$ of the \Gls{parafac} decomposition is necessary. In the context of this work, this implies that the receiver knows the number of elements at the \Gls{ris}, which is a feasible assumption in practice.
\textcolor{black}{Problem (\ref{optpro})}
%The above optimization problem 
can be solved in an efficient form by an ALS algorithm \cite{Kolda,Bro98}. It is a well-known \textcolor{black}{iterative method for} 
%a commonly adopted approach for 
estimating the factor matrices of a tensor model 
%in an iterative way 
thanks to its implementation simplicity and monotonic convergence property in which the update of \textcolor{black}{every} given matrix at each iteration may either improve or maintain but cannot worsen the current fit, leading usually to global minimum solution \cite{Feng2011,Dong2018}. The decoupled estimates of $\mathbf{G}$, $\mathbf{H}$ and $\mathbf{e}$ can be obtained by converting the trilinear fitting problem in (\ref{optpro}) into the following three simplest linear least squares (LS) sub-problems formulated from (\ref{sub1}), (\ref{sub2}) and (\ref{sub3}), respectively
\vspace{-0.1cm}
\begin{eqnarray}
\hat{\mathbf{G}} &=& \underset{\textcolor{black}{\mathbf{G}\mid\mathbf{e},\mathbf{H}}}{\text{argmin}}\left\|\left[\mathcal{Y}\right]_{(1)} -  \mathbf{G}\text{diag}\left(\mathbf{e}\right)\left(\mathbf{S} \diamond \mathbf{H}\right)^{\text{T}}\right\|_{\text{F}}^{2}, \label{o1}\\ 
\hat{\mathbf{H}} &=& \underset{\textcolor{black}{\mathbf{H}\mid\mathbf{e},\mathbf{G}}}{\text{argmin}}\left\|\left[\mathcal{Y}\right]_{(2)} - \mathbf{H}\text{diag}\left(\mathbf{e}\right)\left(\mathbf{S} \diamond \mathbf{G}\right)^{\text{T}}\right\|_{\text{F}}^{2}, \label{o2}\\
%\hat{\mathbf{H}} &=& \underset{\mathbf{H}}{\text{argmin}}\left\|\left[\mathcal{Y}\right]_{(2)} - \mathbf{H}\left(\mathbf{S} \diamond \mathbf{G}\text{diag}\left(\mathbf{e}\right)\right)^{\text{T}}\right\|_{\text{F}}^{2}, \label{o2}\\
\hat{\mathbf{e}} &=& \underset{\textcolor{black}{\mathbf{e}\mid\textbf{H},\mathbf{G}}}{\text{argmin}}\left\|\text{vec}\left(\left[\mathcal{Y}\right]_{(1)}\right) - \left(\mathbf{S} \diamond \mathbf{H} \diamond \mathbf{G}\right)\mathbf{e}\right\|_{\text{F}}^{2} \label{03}.
\end{eqnarray}
%\begin{eqnarray}
%\hat{\mathbf{G}} &=& \underset{\mathbf{G}}{\text{argmin}}\left\|\left[\mathcal{Y}\right]_{(1)} -  \mathbf{G}\text{diag}\left(\mathbf{e}\right)\left(\mathbf{S} \diamond \mathbf{H}\right)^{\text{T}}\right\|_{\text{F}}^{2}, \label{o1}\\ 
%\hat{\mathbf{H}} &=& \underset{\mathbf{H}}{\text{argmin}}\left\|\left[\mathcal{Y}\right]_{(2)} - \mathbf{H}\text{diag}\left(\mathbf{e}\right)\left(\mathbf{S} \diamond \mathbf{G}\right)^{\text{T}}\right\|_{\text{F}}^{2}, \label{o2}\\
%\hat{\mathbf{H}} &=& \underset{\mathbf{H}}{\text{argmin}}\left\|\left[\mathcal{Y}\right]_{(2)} - \mathbf{H}\left(\mathbf{S} \diamond \mathbf{G}\text{diag}\left(\mathbf{e}\right)\right)^{\text{T}}\right\|_{\text{F}}^{2}, \label{o2}\\
%\hat{\mathbf{e}} &=& \underset{\mathbf{e}}{\text{argmin}}\left\|\text{vec}\left(\left[\mathcal{Y}\right]_{(1)}\right) - \left(\mathbf{S} \diamond \mathbf{H} \diamond \mathbf{G}\right)\mathbf{e}\right\|_{\text{F}}^{2} \label{03}.
%\end{eqnarray}
According to (\ref{o1}), the conditional LS update for $\hat{\mathbf{G}}$ is given by 
\vspace{-0.2cm}
\begin{equation}
\hat{\mathbf{G}} = \left[\mathcal{Y}\right]_{(1)}\left[\text{diag}\left(\mathbf{e}\right)\left(\mathbf{S} \diamond \mathbf{H}\right)^{\text{T}}\right]^{\dag}.\label{sol1}
\end{equation} 
Similarly, according to (\ref{o2}) and (\ref{03}), the conditional LS updates for $\hat{\mathbf{H}}$ and $\hat{\mathbf{e}}$ are respectively given by 
\vspace{-0.1cm}
\begin{eqnarray}
%\hat{\mathbf{G}} &=& \left[\mathcal{Y}\right]_{(1)}\left[\text{diag}\left(\mathbf{e}\right)\left(\mathbf{S} \diamond \mathbf{H}\right)^{\text{T}}\right]^{\dag},\label{sol1}\\
\hat{\mathbf{H}} &=& \left[\mathcal{Y}\right]_{(2)}\left[\text{diag}\left(\mathbf{e}\right)\left(\mathbf{S} \diamond \mathbf{G}\right)^{\text{T}}\right]^{\dag},\label{sol2}\\
%\hat{\mathbf{H}} &=& \left[\mathcal{Y}\right]_{(2)}\left[\left(\mathbf{S} \diamond \mathbf{G}\text{diag}\left(\mathbf{e}\right)\right)^{\text{T}}\right]^{\dag},\label{sol2}\\
\hat{\mathbf{e}} &=& \left(\mathbf{S} \diamond \mathbf{H} \diamond \mathbf{G}\right)^{\dag}\text{vec}\left(\left[\mathcal{Y}\right]_{(1)}\right)\label{sol3}.
\end{eqnarray}
%respectively.
%whose solutions are easily deduced, resulting in the following LS estimates:
%\begin{eqnarray}
%\hat{\mathbf{G}} &=& \left[\mathcal{Y}\right]_{(1)}\left[\text{diag}\left(\mathbf{e}\right)\left(\mathbf{S} \diamond \mathbf{H}\right)^{\text{T}}\right]^{\dag},\label{sol1}\\
%\hat{\mathbf{H}} &=& \left[\mathcal{Y}\right]_{(2)}\left[\left(\mathbf{S} \diamond \mathbf{G}\text{diag}\left(\mathbf{e}\right)\right)^{\text{T}}\right]^{\dag},\label{sol2}\\
%\hat{\mathbf{e}} &=& \left(\mathbf{S} \diamond \mathbf{H} \diamond \mathbf{G}\right)^{\dag}\text{vec}\left(\left[\mathcal{Y}\right]_{(1)}\right)\label{sol3}.
%\end{eqnarray}

\begin{algorithm}[t]
	\caption{\small{TALS-LTI Algorithm}}
	\begin{algorithmic}\label{Alg1}
		\STATE \small{\textbf{1.} \textit{Set} $i = 0$;
			\STATE \quad \textit{Keep} $\mathbf{S}$ \textit{fixed;} \textit{Initialize randomly the matrix} $\hat{\mathbf{H}}_{(i=0)}$\\
			\STATE \quad \textit{and the imperfections vector} $\hat{\mathbf{e}}_{(i=0)}$;
			\STATE \textbf{2.} $i \leftarrow i + 1$;
			\STATE \textbf{3.} \textit{According to} (\ref{sol1}), \textit{obtain an LS estimate of} $\hat{\mathbf{G}}_{(i)}$:
			\STATE \begin{displaymath}
			\hat{\mathbf{G}}_{(i)} = \left[\mathcal{Y}\right]_{(1)}\left[\text{diag}\left(\hat{\mathbf{e}}_{(i-1)}\right)\left(\mathbf{S} \diamond \hat{\mathbf{H}}_{(i-1)}\right)^{\text{T}}\right]^{\dag};
			\end{displaymath}
			\STATE \textbf{4.} \textit{According to} (\ref{sol2}), \textit{obtain an LS estimate of} $\hat{\mathbf{H}}_{(i)}$:
			\STATE \begin{displaymath}
			\hat{\mathbf{H}}_{(i)} = \left[\mathcal{Y}\right]_{(2)}\left[\text{diag}\left(\hat{\mathbf{e}}_{(i-1)}\right)\left(\mathbf{S} \diamond \hat{\mathbf{G}}_{(i)}\right)^{\text{T}}\right]^{\dag};
			\end{displaymath}
			%\STATE \begin{displaymath}
			%\hat{\mathbf{H}}_{(i)} = \left[\mathcal{Y}\right]_{(2)}\left[\left(\mathbf{S} \diamond \hat{\mathbf{G}}_{(i)}\text{diag}\left(\hat{\mathbf{e}}_{(i-1)}\right)\right)^{\text{T}}\right]^{\dag};
			%\end{displaymath}	
			\STATE \textbf{5.} \textit{According to} (\ref{sol3}), \textit{obtain an LS estimate of} $\hat{\mathbf{e}}_{(i)}$:
			\STATE \begin{displaymath}
			\hat{\mathbf{e}}_{(i)} = \left(\mathbf{S} \diamond \hat{\mathbf{H}}_{(i)} \diamond \hat{\mathbf{G}}_{(i)}\right)^{\dag}\text{vec}\left(\left[\mathcal{Y}\right]_{(1)}\right);
			\end{displaymath}
			\STATE \textbf{6.} \textit{Calculate the residual error} $\textcolor{black}{\epsilon}_{(i)} = \|\left[\mathcal{Y}\right]_{(1)} - \hat{\left[\mathcal{Y}\right]}_{(1)\textcolor{black}{(i)}}\|_{\text{F}}^{2}$\\
			\STATE \quad \textit{where}
			\STATE \begin{displaymath}
			\hat{\left[\mathcal{Y}\right]}_{(1)\textcolor{black}{(i)}} = \hat{\mathbf{G}}_{(i)}\text{diag}\left(\hat{\mathbf{e}}_{(i)}\right)\left(\mathbf{S} \diamond \hat{\mathbf{H}}_{(i)}\right)^{\text{T}};
			\end{displaymath}
			\STATE \textbf{7.} \textit{Repeat Steps} 2-6 \textit{until} $|\textcolor{black}{\epsilon}_{(i)} - \textcolor{black}{\epsilon}_{(i-1)}| \leq \textcolor{black}{\delta}$.}
	\end{algorithmic}
\end{algorithm}

The proposed TALS-STI algorithm consists of three iterative and alternating update steps formulated from the LS solutions in (\ref{sol1}), (\ref{sol2}) and (\ref{sol3}). At each step, the fitting error is minimized with respect to one given factor matrix by fixing the other matrices to their values obtained at previous updating steps. This procedure is repeated until the convergence of the algorithm at the $i$-th iteration \textcolor{black}{determined by the designer.}

\textcolor{black}{Define}
\vspace{-0.1cm}
\begin{equation}
\hat{\left[\mathcal{Y}\right]}_{(1)\textcolor{black}{(i)}} = \hat{\mathbf{G}}_{(i)}\text{diag}\left(\hat{\mathbf{e}}_{(i)}\right)\left(\mathbf{S} \diamond \hat{\mathbf{H}}_{(i)}\right)^{\text{T}} \in \mathbb{C}^{L \times MK},
\end{equation}
as the reconstructed version of $\left[\mathcal{Y}\right]_{(1)}$ obtained from the estimates of $\hat{\mathbf{G}}_{(i)}$, $\hat{\mathbf{H}}_{(i)}$, and $\hat{\mathbf{e}}_{(i)}$, and the residual error as  
\vspace{-0.1cm}
\begin{equation}
\textcolor{black}{\epsilon_{(i)}} = \left\|\left[\mathcal{Y}\right]_{(1)} - \hat{\left[\mathcal{Y}\right]}_{(1)\textcolor{black}{(i)}}\right\|_{\text{F}}^{2},
\end{equation} 
computed at the end of the $i$-th iteration. The convergence of the algorithm is declared when $|\textcolor{black}{\epsilon}_{(i)} - \textcolor{black}{\epsilon}_{(i-1)}| \leq \delta$, \textcolor{black}{with $\delta$ being a constant considered by the designer,} meaning that the reconstruction error does not significantly change between two successive iterations. In this work, we set $\delta = 10^{-6}$ as a convergence threshold. The implementation steps of the proposed iterative TALS-LTI algorithm are summarized in the pseudocode shown in Algorithm 1. \textcolor{black}{For the complexity analysis of Algorithm 1, see Section \ref{Iden}.} 

%The proposed Trilinear ALS procedure for Channel Estimation in RIS working under Long-Timescale Imperfections developed in Section \ref{proposta1} is referred as TALSCE-LTI algorithm. We summarize in detail the steps of the proposed TALSCE-LTI algorithm in the pseudocode shown in \textbf{Algorithm 1}.

%\subsection{Channel Estimation With RIS Under Short-Timescale Imperfections}

%\subsection{Trilinear ALS for Channel Estimation in RIS Working Under Short-Timescale Impairments (TALSCE-STI)}\label{alsst}

%\subsection{TALS Algorithm for RIS Operating Under \textcolor{blue}{Short-Term} \textcolor{blue}{Imperfections} (TALS-STI)}\label{alsst}

\vspace{-0.2cm}
\subsection{TALS Algorithm for RIS Operating \textcolor{black}{under} \textcolor{black}{STI} (TALS-STI)}\label{alsst}

%In the previous section, we have considered long-timescale RIS impairments. 

In order to derive proposed channel estimators for a scenario with \textcolor{black}{STI,}
%\textcolor{blue}{short-term} RIS \textcolor{blue}{imperfections,} 
let us first establish a link between the received signal in (\ref{dec2}) and the PARAFAC decomposition. 
%we first need recast the received signal in Equation (\ref{dec2}) in an equivalent tensor formulation. 
According to \cite{Kolda}, the noiseless signal part of (\ref{dec2}) expresses the ($p,k$)-th frontal slice of a fourth-order tensor $\mathcal{Y} \in \mathbb{C}^{L \times M \times K \times P}$ that follows the PARAFAC decomposition
\vspace{-0.1cm}
\begin{equation}
\mathcal{Y} = \mathcal{I}_{4,N} \times_{1} \mathbf{G} \times_{2} \mathbf{H} \times_{3} \mathbf{S} \times_{4} \mathbf{E}.
\label{t4}
\end{equation} 
%where 
\textcolor{black}{Here,} $\mathcal{I}_{4,N}$ denotes the fourth-order identity tensor of size $N \times N \times N \times N$, while $\mathbf{G}$, $\mathbf{H}$, $\mathbf{S}$ and $\mathbf{E}$ are the 1,2,3,4-mode factor matrices of the decomposition, respectively.

By stacking column-wise the noiseless received signal in (\ref{dec2}) for the $K$ time-blocks at frame $p$ as the matrix $\mathbf{Y}_{p} = \left[\mathbf{Y}\left[p,1\right], \ldots, \mathbf{Y}\left[p,K\right]\right] \in \mathbb{C}^{L \times MK}$, we have
\vspace{-0.1cm}
\begin{equation}
\mathbf{Y}_{p} = \mathbf{G}\mathbf{D}_{p}\left(\mathbf{E}\right)\left[\mathbf{D}_{1}\left(\mathbf{S}\right)\mathbf{H}^{\text{T}}, \ldots,\mathbf{D}_{K}\left(\mathbf{S}\right)\mathbf{H}^{\text{T}}\right], 
%\left[\mathbf{Y}\left[p,1\right], \ldots, \mathbf{Y}\left[p,K\right]\right] = \mathbf{G}\mathbf{D}_{p}\left(\mathbf{E}\right)\left[\mathbf{D}_{1}\left(\mathbf{S}\right)\mathbf{H}^{\text{T}}, \ldots,\mathbf{D}_{K}\left(\mathbf{S}\right)\mathbf{H}^{\text{T}}\right].
\label{f4}
\end{equation}
$\forall p = 1, \ldots, P$. Applying the property (\ref{pp2}) to the right-hand side of (\ref{f4}), a more compact form is obtained as
\vspace{-0.1cm}
\begin{equation}
\mathbf{Y}_{p} = \mathbf{G}\mathbf{D}_{p}\left(\mathbf{E}\right)\left(\mathbf{S} \diamond \mathbf{H}\right)^{\text{T}} \in \mathbb{C}^{L \times MK}.
\label{gg}
\end{equation}
From (\ref{gg}), we can define the new column-wise collection $\left[\mathcal{Y}\right]_{(1)} = \left[\mathbf{Y}_{1}, \ldots, \mathbf{Y}_{P}\right] \in \mathbb{C}^{L \times MKP}$ as the 1-mode matrix unfolding of the received signal tensor $\mathcal{Y} \in \mathbb{C}^{L \times M \times K \times P}$ in (\ref{t4}), \textcolor{black}{which is given by}
%Therefore, we \textcolor{blue}{obtain}
\vspace{-0.1cm}
\begin{equation}
\left[\mathcal{Y}\right]_{(1)} = \mathbf{G}\left[\mathbf{D}_{1}\left(\mathbf{E}\right)\left(\mathbf{S} \diamond \mathbf{H}      \right)^{\text{T}}, \ldots, \mathbf{D}_{P}\left(\mathbf{E}\right)\left(\mathbf{S} \diamond \mathbf{H}\right)^{\text{T}}\right].
\label{ll}
\end{equation}
By applying property (\ref{pp2}) to the right-hand side of (\ref{ll}), we finally \textcolor{black}{obtain}
\vspace{-0.1cm}
\begin{equation}
\left[\mathcal{Y}\right]_{(1)} = \mathbf{G}\left(\mathbf{E} \diamond \mathbf{S} \diamond \mathbf{H}\right)^{\text{T}} \in \mathbb{C}^{L \times MKP}.
\label{unf41}
\end{equation} 

Additionally, for our purpose, we also need to define the 2-mode, 3-mode and 4-mode matrix unfoldings of the fourth-order received signal tensor $\mathcal{Y}  \in \mathbb{C}^{L \times M \times K \times P}$ since they will be exploited to formulate our second set of channel estimation algorithms in the sequel. The remaining unfoldings can be deduced using a similar procedure by permuting the factor matrices in (\ref{dec2}). 
%\cite{Favier2015}. 
This leads to the following factorizations to the other unfoldings 
%of $\mathcal{Y}$ 
%\cite{Kolda,Favier2015}: 
\vspace{-0.1cm}
\begin{eqnarray}
\left[\mathcal{Y}\right]_{(2)} &=& \mathbf{H}\left(\mathbf{E} \diamond \mathbf{S} \diamond \mathbf{G}\right)^{\text{T}} \in \mathbb{C}^{M \times LKP},\label{unf42} \\
\left[\mathcal{Y}\right]_{(3)} &=& \mathbf{S}\left(\mathbf{E} \diamond \mathbf{H} \diamond \mathbf{G}\right)^{\text{T}} \in \mathbb{C}^{K \times LMP}, \label{hop}\\
\left[\mathcal{Y}\right]_{(4)} &=& \mathbf{E}\left(\mathbf{S} \diamond \mathbf{H} \diamond \mathbf{G}\right)^{\text{T}} \in \mathbb{C}^{P \times LMK}.\label{unf43}
\end{eqnarray}

In the following, we show that the channel matrices $\mathbf{H}$ and $\mathbf{G}$ can be also estimated when \textcolor{black}{STI} 
%\textcolor{blue}{short-term} RIS \textcolor{blue}{imperfections} 
are assumed. 
%To this end, we propose two alternative iterative and closed-form channel estimation algorithms.  
From the received signal tensor described in (\ref{t4}), the estimates of $\mathbf{G}$ and $\mathbf{H}$ can be obtained by minimizing the following quadrilinear LS fitting \textcolor{black}{problem}
\vspace{-0.1cm}
\begin{equation}
\underset{\mathbf{G},\mathbf{H},\mathbf{E}}{\text{min}}\sum_{p=1}^{P}\sum_{k=1}^{K}\left\|\mathbf{Y}\left[p,k\right] - \mathbf{G}\mathbf{D}_{p}\left(\mathbf{E}\right)\mathbf{D}_{k}\left(\mathbf{S}\right)\mathbf{H}^{\text{T}}\right\|_{\text{F}}^{2}.
\label{hhh}
\end{equation}

Similar to the TALS-LTI algorithm, we also propose to solve this optimization problem by means of the ALS algorithm. Since the matrix $\mathbf{S}$ is known at the receiver, the quadrilinear fitting problem in (\ref{hhh}) is simplified to a trilinear fitting problem that reduces to iteratively minimize the following linear LS sub-problems formulated from (\ref{unf41}), (\ref{unf42}) and (\ref{unf43}), respectively  
\vspace{-0.3cm}
\begin{eqnarray}
\hat{\mathbf{G}} &=& \underset{\textcolor{black}{\mathbf{G}\mid\mathbf{E},\mathbf{H}}}{\text{argmin}}\left\|\left[\mathcal{Y}\right]_{(1)} - \mathbf{G}\left(\mathbf{E} \diamond \mathbf{S} \diamond \mathbf{H}\right)^{\text{T}}\right\|_{\text{F}}^{2}, \label{w1}\\
\hat{\mathbf{H}} &=& \underset{\textcolor{black}{\mathbf{H}\mid\mathbf{E},\mathbf{G}}}{\text{argmin}}\left\|\left[\mathcal{Y}\right]_{(2)} - \mathbf{H}\left(\mathbf{E} \diamond \mathbf{S} \diamond \mathbf{G}\right)^{\text{T}}\right\|_{\text{F}}^{2}, \label{w3}\\
\hat{\mathbf{E}} &=& \underset{\textcolor{black}{\mathbf{E}\mid\mathbf{H},\mathbf{G}}}{\text{argmin}}\left\|\left[\mathcal{Y}\right]_{(4)} - \mathbf{E}\left(\mathbf{S} \diamond \mathbf{H} \diamond \mathbf{G}\right)^{\text{T}}\right\|_{\text{F}}^{2} \label{w2}.
\end{eqnarray}
It follows from (\ref{w1}), (\ref{w3}) and (\ref{w2}) that the conditional LS updates of $\hat{\mathbf{G}}$, $\hat{\mathbf{H}}$ and $\hat{\mathbf{E}}$ are given by
\vspace{-0.1cm}
\begin{eqnarray}
\hat{\mathbf{G}} &=& \left[\mathcal{Y}\right]_{(1)}\left[\left(\mathbf{E} \diamond \mathbf{S} \diamond \mathbf{H}\right)^{\text{T}}           \right]^{\dag}, \label{hj1}\\
\hat{\mathbf{H}} &=& \left[\mathcal{Y}\right]_{(2)}\left[\left(\mathbf{E} \diamond \mathbf{S} \diamond \mathbf{G}\right)^{\text{T}}           \right]^{\dag},\label{hj2}\\
\hat{\mathbf{E}} &=& \left[\mathcal{Y}\right]_{(4)}\left[\left(\mathbf{S} \diamond \mathbf{H} \diamond \mathbf{G}\right)^{\text{T}}           \right]^{\dag}\label{hj3}, 
\end{eqnarray}
respectively. In the same way as in Algorithm 1, the updates of $\hat{\mathbf{G}}$, $\hat{\mathbf{H}}$ and $\hat{\mathbf{E}}$ are obtained by iteratively performing (\ref{hj1}), (\ref{hj2}) and (\ref{hj3}) until the convergence. The proposed iterative TALS-STI algorithm is detailed in the pseudocode shown in Algorithm 2. \textcolor{black}{For the complexity analysis of Algorithm 2, see Section \ref{Iden}.}

\begin{algorithm}[t]
	\caption{\small{TALS-STI Algorithm}}
	\begin{algorithmic}\label{Alg2}
		\STATE \small{\textbf{1.} \textit{Set} $i = 0$;
			\STATE \quad \textit{Keep} $\mathbf{S}$ \textit{fixed;} 
			\STATE \quad \textit{Initialize randomly the matrices} $\hat{\mathbf{H}}_{(i=0)}$ \textit{and} $\hat{\mathbf{E}}_{(i=0)}$;\\
			\STATE \textbf{2.} $i \leftarrow i + 1$;
			\STATE \textbf{3.} \textit{According to} (\ref{hj1}), \textit{obtain an LS estimate of} $\hat{\mathbf{G}}_{(i)}$:
			\STATE \begin{displaymath}
			\hat{\mathbf{G}}_{(i)} = \left[\mathcal{Y}\right]_{(1)}\left[\left(\hat{\mathbf{E}}_{(i-1)} \diamond \mathbf{S} \diamond \hat{\mathbf{H}}_{(i-1)}\right)^{\text{T}}\right]^{\dag};
			\end{displaymath}
			\STATE \textbf{4.} \textit{According to} (\ref{hj2}), \textit{obtain an LS estimate of} $\hat{\mathbf{H}}_{(i)}$:
			\STATE \begin{displaymath}
			\hat{\mathbf{H}}_{(i)} = \left[\mathcal{Y}\right]_{(2)}\left[\left(\hat{\mathbf{E}}_{(i-1)} \diamond \mathbf{S} \diamond \hat{\mathbf{G}}_{(i)}\right)^{\text{T}}\right]^{\dag};
			\end{displaymath}
			\STATE \textbf{5.} \textit{According to} (\ref{hj3}), \textit{obtain an LS estimate of} $\hat{\mathbf{E}}_{(i)}$:
			\STATE \begin{displaymath}
			\hat{\mathbf{E}}_{(i)} = \left[\mathcal{Y}\right]_{(4)}\left[\left(\mathbf{S} \diamond \hat{\mathbf{H}}_{(i)} \diamond \hat{\mathbf{G}}_{(i)}\right)^{\text{T}}\right]^{\dag}; 
			\end{displaymath}
			\STATE \textbf{6.} \textit{Calculate the residual error} $\textcolor{black}{\epsilon}_{(i)} = \|\left[\mathcal{Y}\right]_{(1)} - \hat{\left[\mathcal{Y}\right]}_{(1)\textcolor{black}{(i)}}\|_{\text{F}}^{2}$\\
			\STATE \quad \textit{where}
			\STATE \begin{displaymath}
			\hat{\left[\mathcal{Y}\right]}_{(1)\textcolor{black}{(i)}}= \hat{\mathbf{G}}_{(i)}\left(\hat{\mathbf{E}}_{(i)} \diamond \mathbf{S} \diamond \hat{\mathbf{H}}_{(i)}\right)^{\text{T}};
			\end{displaymath}
			\STATE \textbf{7.} \textit{Repeat Steps} 2-6 \textit{until} $|\textcolor{black}{\epsilon}_{(i)} - \textcolor{black}{\epsilon}_{(i-1)}| \leq \textcolor{black}{\delta}$.}
	\end{algorithmic}
\end{algorithm}

\textit{Remark 1:} In the application context of this work, the non-impaired RIS activation pattern matrix $\mathbf{S}$ is assumed to be known at the receiver as indicated in the first step of \textcolor{black}{Algorithms 1-2,} respectively. This is a feasible assumption in accordance with the channel estimation protocol shown in \textcolor{black}{Fig.} \ref{protocol}. \textcolor{black}{Among different design possibilities, we set} $\mathbf{S}$ 
%is designed 
as a semi-unitary matrix satisfying \textcolor{black}{$\mathbf{S}^{\text{H}}\mathbf{S} = K\mathbf{I}_{N}$}. According to \cite{RIS1}, a good choice is to consider $\mathbf{S}$ as a deterministic truncated discrete Fourier transform (DFT) matrix. This choice guarantees a good performance of the proposed algorithms \textcolor{black}{since the correlation properties of the additive noise are not affected during the estimation processing.} More details on the optimal design of $\mathbf{S}$ are found in \cite{Optimal}. Moreover, despite the iterative nature of the proposed TALS-LTI and TALS-STI algorithms, the convergence to the global minimum is always achieved within a few iterations (usually less than $80$ iterations as verified in our simulation results) due to the knowledge of $\mathbf{S}$ that remains fixed during the iterations. 

%\subsection{HOSVD-Based Channel Estimation in RIS Working Under Short-Timescale Impairments (HOSVDCE-STI)}\label{hoce}

%\subsection{HOSVD Algorithm for RIS Operating Under \textcolor{blue}{Short-Term} \textcolor{blue}{Imperfections} (HOSVD-STI)}\label{hoce}

\vspace{-0.2cm}
\subsection{HOSVD Algorithm for RIS Operating \textcolor{black}{under} \textcolor{black}{STI} (HOSVD-STI)}\label{hoce}

%\textcolor{blue}{Here, we}
\textcolor{black}{We now} derive a closed-form solution \textcolor{black}{based on higher order singular value decomposition (HOSVD)} for channel estimation under the \textcolor{black}{STI} 
%\textcolor{blue}{short-term} RIS \textcolor{blue}{imperfection} 
model. According to (\ref{hop}), the transpose of the 3-mode unfolding of $\mathcal{Y}$ is denoted by
\vspace{-0.1cm}
\begin{equation}
\left[\mathcal{Y}\right]^{\text{T}}_{(3)} = \left(\mathbf{E} \diamond \mathbf{H} \diamond \mathbf{G}\right)\mathbf{S}^{\text{T}}.
\label{ho00}
\end{equation}
The first processing step at the receiver is to apply a bilinear time-domain matched-filtering by multiplying both sides in (\ref{ho00}) by the pseudo-inverse of $\mathbf{S}^{\text{T}}$, \textcolor{black}{resulting in}
%. After this preprocessing step, we have
\vspace{-0.1cm}
\begin{equation}
\tilde{\mathbf{Y}} = \mathbf{E} \diamond \mathbf{H} \diamond \mathbf{G} \in \mathbb{C}^{LMP \times N}, 
\label{ho01}
\end{equation} 
where $\tilde{\mathbf{Y}} = \left[\mathcal{Y}\right]^{\text{T}}_{(3)}\left(\mathbf{S}^{\text{T}}\right)^{\dag}$. From (\ref{ho01}), decoupled estimates of the channel matrices and RIS \textcolor{black}{imperfections} can be obtained by separating each factor matrix in the Khatri-Rao product. In this sense, the estimates can be obtained by minimizing the following cost function
\vspace{-0.2cm}
\begin{equation} 
\underset{\mathbf{G},\mathbf{H},\mathbf{E}}{\text{min}}\left\|\tilde{\mathbf{Y}} - \mathbf{E} \diamond \mathbf{H} \diamond \mathbf{G}\right\|_{\text{F}}^{2}.
\label{optho}
\end{equation}
Here, we propose to solve this problem by means of multiple rank-one tensor approximations via the HOSVD 
%the higher order singular value decomposition (HOSVD) 
\cite{hosvd}. \textcolor{black}{To this end, let} us define $\tilde{\mathbf{Y}} = \left[\tilde{\mathbf{y}}_{1}, \ldots, \tilde{\mathbf{y}}_{N}\right] \in \mathbb{C}^{LMP \times N}$, and note that
%Making use of the Khatri-Rao product definition, i.e., that it is equal to the column-wise Kronecker product 
according to (\ref{pp1}), the $n$-th column of $\tilde{\mathbf{Y}}$ can be \textcolor{black}{written} as
\vspace{-0.2cm}
\begin{equation}
\tilde{\mathbf{y}}_{n} = \mathbf{e}_{n} \otimes \mathbf{h}_{n} \otimes \mathbf{g}_{n} \in \mathbb{C}^{LMP \times 1},
\label{ho05}
\end{equation} 
where $\mathbf{e}_{n} \in \mathbb{C}^{P \times 1}$, $\mathbf{h}_{n} \in \mathbb{C}^{M \times 1}$ and $\mathbf{g}_{n} \in \mathbb{C}^{L \times 1}$ denote the $n$-th column of $\mathbf{E}$, $\mathbf{H}$ and $\mathbf{G}$, respectively. Using the equivalence property \textcolor{black}{in} (\ref{pp4}) that relates the Kronecker product to the outer product, we can rewrite (\ref{ho05}) as 
\vspace{-0.2cm}
\begin{equation}
\tilde{\mathbf{y}}_{n} = \text{vec}\left(\mathbf{g}_{n} \circ \mathbf{h}_{n} \circ \mathbf{e}_{n}\right) \in \mathbb{C}^{LMP \times 1},
\end{equation}
that represents the vectorized form of the following third-order rank-one tensor
\vspace{-0.1cm}
\begin{equation}
\tilde{\mathcal{Y}}_{n} = \mathbf{g}_{n} \circ \mathbf{h}_{n} \circ \mathbf{e}_{n} \in \mathbb{C}^{L \times M \times P}.
\label{r1ho}
\end{equation}
Thus, the optimization problem in (\ref{optho}) is equivalent to finding the estimates of $\mathbf{H}$, $\mathbf{G}$ and $\mathbf{E}$ that minimize a set of $N$ rank-one tensor approximations, i.e, 
\vspace{-0.1cm}
\begin{equation} 
\left(\hat{\mathbf{G}}, \hat{\mathbf{H}}, \hat{\mathbf{E}}\right) = \underset{\mathbf{G},\mathbf{H},\mathbf{E}}{\text{argmin}}\sum_{n=1}^{N}\left\|\tilde{\mathcal{Y}}_{n} - \mathbf{g}_{n} \circ \mathbf{h}_{n} \circ \mathbf{e}_{n}\right\|_{\text{F}}^{2}.
\label{gj}
\end{equation}

\begin{algorithm}[t]
	\caption{\small{HOSVD-STI Algorithm}}
	\begin{algorithmic}\label{Alg3}
		\STATE \small{\textbf{for} $n = 1, \ldots, N$
			\STATE \textbf{1.} \textit{Rearrange the $n$-th column of $\tilde{\mathbf{Y}}$ in Equation (\ref{ho01})} \\
			\STATE \quad \textit{as the rank-one tensor $\tilde{\mathcal{Y}}_{n}$ in Equation (\ref{r1ho})};  
			\STATE \textbf{2.} \textbf{HOSVD procedure}
			\STATE \quad \textbf{2.1} \textit{Compute} $\mathbf{U}_{n}^{(1)}$ \textit{as the} $L$ \textit{left singular vectors of} $\left[\tilde{\mathcal{Y}}_{n}\right]_{(1)}$:   
			\begin{displaymath}
			\left[\tilde{\mathcal{Y}}_{n}\right]_{(1)} = \mathbf{U}_{n}^{(1)} \cdot \mathbf{\Sigma}_{n}^{(1)} \cdot \mathbf{V}_{n}^{(1)\text{H}};
			\end{displaymath}
			\STATE \quad \textbf{2.2} \textit{Compute} $\mathbf{U}_{n}^{(2)}$ \textit{as the} $M$ \textit{left singular vectors of} $\left[\tilde{\mathcal{Y}}_{n}\right]_{(2)}$:   
			\begin{displaymath}
			\left[\tilde{\mathcal{Y}}_{n}\right]_{(2)} = \mathbf{U}_{n}^{(2)} \cdot \mathbf{\Sigma}_{n}^{(2)} \cdot \mathbf{V}_{n}^{(2)\text{H}};
			\end{displaymath}
			\STATE \quad \textbf{2.3} \textit{Compute} $\mathbf{U}_{n}^{(3)}$ \textit{as the} $P$ \textit{left singular vectors of} $\left[\tilde{\mathcal{Y}}_{n}\right]_{(3)}$:   
			\begin{displaymath}
			\left[\tilde{\mathcal{Y}}_{n}\right]_{(3)} = \mathbf{U}_{n}^{(3)} \cdot \mathbf{\Sigma}_{n}^{(3)} \cdot \mathbf{V}_{n}^{(3)\text{H}};
			\end{displaymath}
			\STATE \quad \textbf{2.4} \textit{Compute the HOSVD core tensor} $\mathcal{G}_{n}$ \textit{as}: 
			\begin{displaymath}
			\mathcal{G}_{n} = \tilde{\mathcal{Y}}_{n} \times_{1} \mathbf{U}_{n}^{(1)\text{H}} \times_{2} \mathbf{U}_{n}^{(2)\text{H}} \times_{3} \mathbf{U}_{n}^{(3)\text{H}};
			\end{displaymath}
			\STATE \quad \textbf{end procedure}
			\STATE \textbf{3.} \textit{Obtain the estimates for} $\hat{\mathbf{g}}_{n}$, $\hat{\mathbf{h}}_{n}$ and $\hat{\mathbf{e}}_{n}$ \textit{from Equations} \\
			\STATE \quad (\ref{zs1}), (\ref{zs2}) \textit{and} (\ref{zs3}), \textit{respectively;}  
			\STATE \textbf{end}
			\STATE \textbf{4.} \textit{Return the matrices} $\hat{\mathbf{G}} = \left[\hat{\mathbf{g}}_{1}, \ldots, \hat{\mathbf{g}}_{N}\right]$, $\hat{\mathbf{H}} = \left[\hat{\mathbf{h}}_{1}, \ldots, \hat{\mathbf{h}}_{N}\right]$ \\
			\STATE \quad \textit{and} $\hat{\mathbf{E}} = \left[\hat{\mathbf{e}}_{1}, \ldots, \hat{\mathbf{e}}_{N}\right]$.
		}
	\end{algorithmic}
\end{algorithm}
\vspace{-0.1cm}
Let us introduce the HOSVD of $\tilde{\mathcal{Y}}_{n}$ as 
\vspace{-0.1cm}
\begin{equation}
\tilde{\mathcal{Y}}_{n} = \mathcal{G}_{n} \times_{1} \mathbf{U}_{n}^{(1)} \times_{2} \mathbf{U}_{n}^{(2)} \times_{3} \mathbf{U}_{n}^{(3)} \in \mathbb{C}^{L \times M \times P},
\end{equation}
where $\mathbf{U}_{n}^{(1)} \in \mathbb{C}^{L \times L}$, $\mathbf{U}_{n}^{(2)} \in \mathbb{C}^{M \times M}$ and $\mathbf{U}_{n}^{(3)} \in \mathbb{C}^{P \times P}$ are unitary matrices, while $\mathcal{G}_{n} \in \mathbb{C}^{L \times M \times P}$ denotes the HOSVD core tensor.
%$\forall n = 1, \ldots N$. 
The estimates of the vectors $\mathbf{g}_{n}$, $\mathbf{h}_{n}$ and $\mathbf{e}_{n}$ that \textcolor{black}{solve} the LS problem in (\ref{gj}) can be obtained by truncating the HOSVD of $\tilde{\mathcal{Y}}_{n}$ to its dominant rank-one component, \textcolor{black}{yielding}   
\vspace{-0.2cm}
\begin{eqnarray}
\hat{\mathbf{g}}_{n} &=& \sqrt[3]{\left(\mathcal{G}_{n}\right)_{1,1,1}} \cdot \mathbf{u}^{(1)}_{1,n}, \label{zs1}\\
\hat{\mathbf{h}}_{n} &=& \sqrt[3]{\left(\mathcal{G}_{n}\right)_{1,1,1}} \cdot \mathbf{u}^{(2)}_{1,n}, \label{zs2}\\
\hat{\mathbf{e}}_{n} &=& \sqrt[3]{\left(\mathcal{G}_{n}\right)_{1,1,1}} \cdot \mathbf{u}^{(3)}_{1,n}, \label{zs3}
\end{eqnarray} 
where $\mathbf{u}^{(1)}_{1,n} \in \mathbb{C}^{L \times 1}$, $\mathbf{u}^{(2)}_{1,n} \in \mathbb{C}^{M \times 1}$ and $\mathbf{u}^{(3)}_{1,n} \in \mathbb{C}^{P \times 1}$ are the first higher order singular vectors, i.e., the first column of $\mathbf{U}_{n}^{(1)}$, $\mathbf{U}_{n}^{(2)}$ and $\mathbf{U}_{n}^{(3)}$, respectively. \textcolor{black}{Here,} $\left(\mathcal{G}_{n}\right)_{1,1,1}$ is the first element of the core tensor $\mathcal{G}_{n}$. The estimates of $\hat{\mathbf{G}}$, $\hat{\mathbf{H}}$ and $\hat{\mathbf{E}}$ are obtained by repeating the procedure \textcolor{black}{of (\ref{r1ho})-(\ref{zs3})} for the $N$ columns of $\tilde{\mathbf{Y}}$ in (\ref{ho01}). In other words, a total of $N$ rank-one tensor approximations via HOSVD are necessary to obtain the full estimates of the matrices $\hat{\mathbf{G}} = \left[\hat{\mathbf{g}}_{1}, \ldots, \hat{\mathbf{g}}_{N}\right]$, $\hat{\mathbf{H}} = \left[\hat{\mathbf{h}}_{1}, \ldots, \hat{\mathbf{h}}_{N}\right]$ and $\hat{\mathbf{E}} = \left[\hat{\mathbf{e}}_{1}, \ldots, \hat{\mathbf{e}}_{N}\right]$ in a closed-form manner. The implementation steps of the proposed closed-form HOSVD-STI algorithm are summarized in Algorithm 3. \textcolor{black}{For the complexity analysis of Algorithm 3, see Section \ref{Iden}.}
%In addition, the detailing of the \Gls{hosvd} computation steps has also been inserted in the pseudocode. 

%\vspace{+0.2cm}
%The closed-form HOSVD-based procedure for Channel Estimation in RIS working under Short-Timescale Imperfections is referred as HOSVDCE-STI algorithm. The pseudocode shown in \textbf{Algorithm 3} summarizes step-by-step the proposed HOSVDCE-STI algorithm. In addition, the detailing of the HOSVD computation steps has also been inserted in the pseudocode. 

\vspace{-0.1cm}
\section{Identifiability \textcolor{black}{and Computational Complexity}}\label{Iden}

%We now 

\textcolor{black}{In this section, we} examine the identifiability aspects \textcolor{black}{and the computational complexity} associated with the proposed tensor-based TALS-LTI, TALS-STI and HOSVD-STI algorithms.  

\textit{1) TALS-LTI algorithm:} According to (\ref{sol1}) and (\ref{sol2}), the uniqueness of the LS estimates of $\hat{\mathbf{G}}$ and $\hat{\mathbf{H}}$ requires that $\text{diag}\left(\mathbf{e}\right)\left(\mathbf{S} \diamond \mathbf{H}\right)^{\text{T}} \in \mathbb{C}^{N \times KM}$ and $\text{diag}\left(\mathbf{e}\right)\left(\mathbf{S} \diamond \mathbf{G}\right)^{\text{T}} \in \mathbb{C}^{N \times KL}$ \textcolor{black}{are} full row-rank to be right-invertible. Additionally, the uniqueness of the LS estimate of $\hat{\mathbf{e}}$ requires that $\left(\mathbf{S} \diamond \mathbf{H} \diamond \mathbf{G}\right) \in \mathbb{C}^{KLM \times N}$ be full column-rank to be left-invertible in (\ref{sol3}). This means that the conditions $N \leq KM$, $N \leq KL$ and $N\leq KLM$ must be jointly satisfied. By combining these three necessary and sufficient conditions, we obtain the lower bound on the number of time-blocks  $K$  necessary for channel estimation so that \textcolor{black}{Steps} 3, 4 and 5 in Algorithm 1 yield unique solutions:
\vspace{-0.1cm}
\begin{equation}
%N \leq K \cdot \text{min}\left(L,M\right).
K \geq \left\lceil \frac{N}{\text{min}\left(L,M\right)}\right\rceil.
\label{lk}
\end{equation}
%where $\lceil x \rceil$ is equal to the smallest integer that is greater than or equal to $x$.

\textit{2) TALS-STI algorithm:} We can note from (\ref{hj1}), (\ref{hj2}) and (\ref{hj3}) that unique estimates of $\hat{\mathbf{G}}$, $\hat{\mathbf{H}}$ and $\hat{\mathbf{E}}$ in the LS sense requires that $\left(\mathbf{E} \diamond \mathbf{S} \diamond \mathbf{H}\right)^{\text{T}} \in \mathbb{C}^{N \times KMP}$, $\left(\mathbf{E} \diamond \mathbf{S} \diamond \mathbf{G}\right)^{\text{T}} \in \mathbb{C}^{N \times KLP}$ and $\left(\mathbf{S} \diamond \mathbf{H} \diamond \mathbf{G}\right)^{\text{T}} \in \mathbb{C}^{N \times KLM}$ \textcolor{black}{are} full row-rank to be right-invertible. This means that the conditions $N \leq KMP$, $N \leq KLP$ and $N \leq KLM$ must be satisfied. The combination of these inequalities leads to the following necessary and sufficient condition to be satisfied
\vspace{-0.1cm}
\begin{equation}
K \geq \left\lceil \frac{N}{\text{min}\left(MP,LP,LM\right)}\right\rceil.
\label{df}
\end{equation}
The condition in (\ref{df}) establishes the lower-bound on the \textcolor{black}{required} number of time-blocks \textcolor{black}{$K$} so that \textcolor{black}{Steps} 3, 4 and 5 in Algorithm 2 provide unique solutions when \textcolor{black}{STI are assumed.} 
%\textcolor{blue}{short-term} RIS \textcolor{blue}{imperfections} are assumed.

\textit{3) HOSVD-STI algorithm:} In contrast to the iterative TALS-LTI and TALS-STI algorithms in which three LS conditions must be jointly satisfied, the proposed HOSVD-STI algorithm is a closed-form solution requiring only that the RIS activation pattern matrix $\mathbf{S} \in \mathbb{C}^{K \times N}$ \textcolor{black}{has} full column-rank in order to guarantee the uniqueness in the LS sense when the bilinear time-domain matched-filtering preprocessing is performed at the receiver side, as indicated in (\ref{ho01}). This leads to the following necessary and sufficient condition to the use of the HOSVD-STI algorithm
\vspace{-0.3cm}
\begin{equation}
K \geq N.
\label{hgf}
\end{equation}  
%\vspace{-0.1cm}
This condition is satisfied under the truncated DFT design discussed \textcolor{black}{in \textit{Remark 1},} 
%previously (\textit{Remark 1}), 
where the RIS activation pattern matrix has orthonormal columns.

According to (\ref{lk})-(\ref{hgf}), independent of the imperfection model, the required number of time-blocks $K$ scales with the number of RIS elements $N$ at least linearly. However, the proposed TALS-LTI and TALS-STI algorithms present more flexible operation conditions for channel estimation compared to our HOSVD-STI algorithm. On the other hand, the HOSVD-STI is a closed-form approach that presents lower computational complexity compared to the others one, as discussed \textcolor{black}{in the following.} 
%as follows.

\textcolor{black}{\textit{4) Computational complexity:} As can be observed in \textcolor{black}{Algorithms 1-2,} the computational complexity of the proposed TALS-LTI and TALS-STI algorithms is dominated by the cost associated with the computation of the matrix pseudo-inverses in three LS update steps that calculate the estimates of the channels and \textcolor{black}{imperfection} matrices in an iterative and alternating way. Therefore, the computational complexity of the TALS-LTI and TALS-STI algorithms are $\mathcal{O}\left(N^{2}K[M + L + ML]\right)$ and $\mathcal{O}\left(N^{2}K[PM + PL + ML]\right)$ per iteration, respectively. Regarding the proposed HOSVD-STI in Algorithm 3, its computational complexity is dominated by the HOSVD of the third-order rank-one tensor in (\ref{r1ho}), which is equivalent to compute the truncated SVDs of its 1-mode, 2-mode and 3-mode unfolding matrices to rank-one. These truncated SVDs are repeated $N$ times. Therefore, the HOSVD-STI algorithm has a complexity $\mathcal{O}\left(NMLP\right)$, which is clearly less \textcolor{black}{than} that of the TALS-LTI and TALS-STI algorithms.} \textcolor{black}{Specially, as opposed to Algorithms 1-2, where the computational complexity is proportional to $N^2,$ in Algorithm 3 \textcolor{black}{the complexity} scales with the RIS size $N$ linearly. As a result, compared to Algorithms 1-2, the relative gain of Algorithm 3, in terms of complexity increases rapidly as the number of RIS elements} \textcolor{black}{increases.}

\textit{Remark 2:} By comparing the conditions in (\ref{df}) and (\ref{hgf}), we \textcolor{black}{observe} that the TALS-STI algorithm presents a less restrictive requirement on the minimum number $K$ of time-blocks necessary for the channel estimation compared to HOSVD-STI algorithm that works well when $K \geq N$ is satisfied. On the other hand, in the HOSVD-STI algorithm a single HOSVD is computed independently for each column of the post-filtered signal in (\ref{ho01}). This \textcolor{black}{leads to a total of} $N$ independent rank-one tensor approximations via HOSVD that can be parallelized if more than one processor is available leading to a considerable reduction in the total processing time for channel estimation. Based on this, we can conclude that the TALS-STI algorithm may be preferred in situations where more flexible choices for the number of time-blocks, \textcolor{black}{$K$,} are required while the HOSVD-STI algorithm becomes more attractive when low processing delay is needed and the receiver is equipped with multiple processors. Hence, there is a tradeoff between computational complexity and operation conditions for the proposed tensor-based algorithms.

\textit{Remark 3:} Since $\mathbf{S}$ is known at the receiver,  
%Being the \Gls{ris} activation pattern matrix $\mathbf{S}$ knowledge at the receiver in the application context of this work, the identifiability study of the proposed approaches by means of the classical Kruskal's uniqueness theorem for the \Gls{parafac} decomposition \cite{Kruskal} can be simplified by the uniqueness analyzes of the pseudo-inverses computed in each conditional \Gls{ls} update steps in our algorithms as explained above. Therefore, thanks to the knowledge of $\mathbf{S}$, 
by satisfying the design recommendations given by (\ref{lk}), (\ref{df}) and (\ref{hgf}) the estimated factor matrices $\hat{\mathbf{G}}$, $\hat{\mathbf{H}}$ and $\hat{\mathbf{E}}$ \textcolor{black}{do} not suffer from column permutation ambiguity. 
%Consequently,  the estimated factor matrices are simply column scaled versions of their corresponding true factor matrices and can be recovered in an unique manner up to scaling factors. 
The scaling ambiguity affecting the columns of the estimated matrices can be eliminated with a simple normalization procedure, as performed in \cite{ZQing} and \cite{RIS1}. 

\begin{figure}[!h]
\centering
\subfloat[\textcolor{black}{NMSE of $\hat{\mathbf{H}}$ \textit{versus} SNR (dB).}]{\includegraphics[scale=0.57]{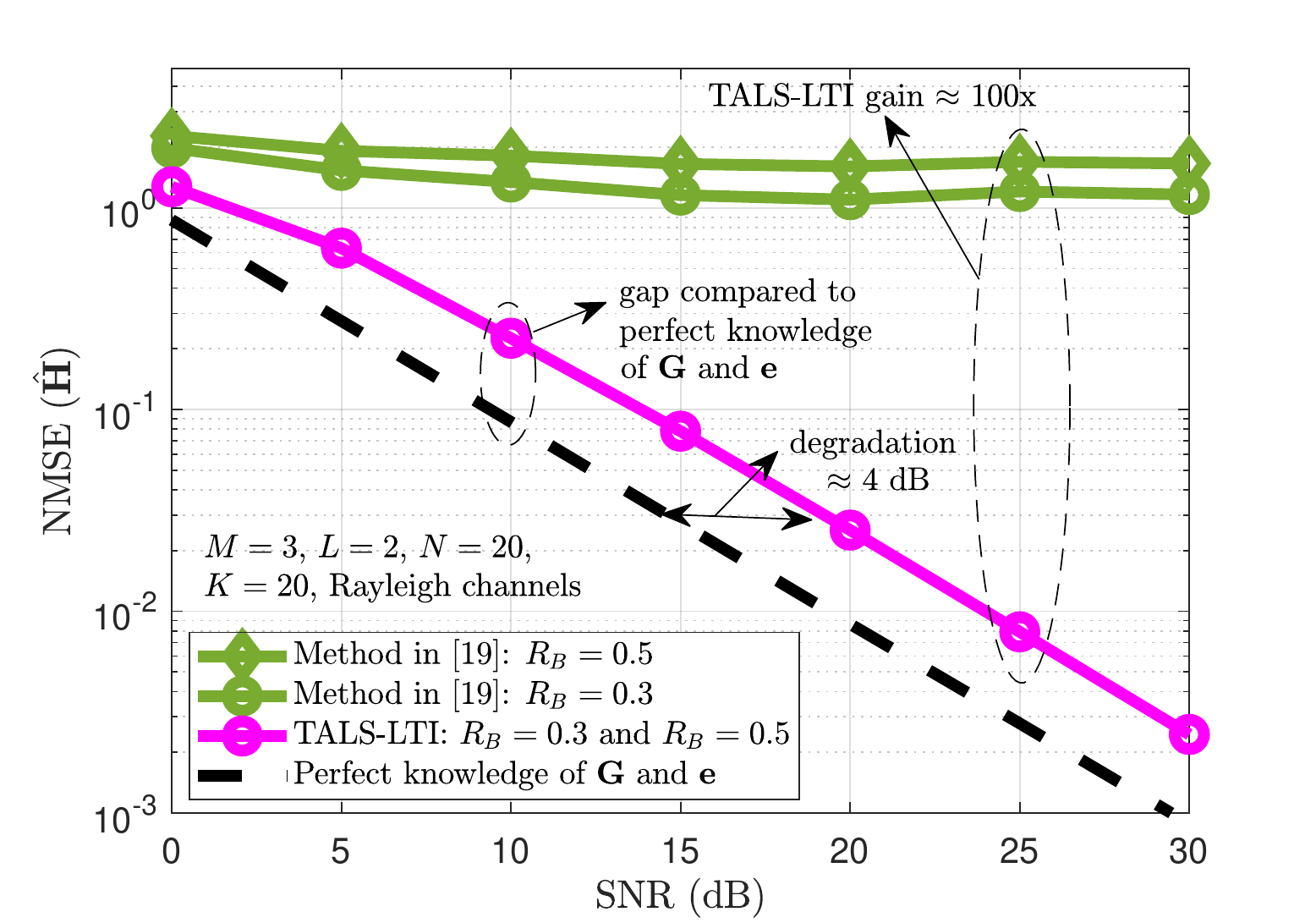}\label{Figura4a}}\\
\subfloat[\textcolor{black}{NMSE of $\hat{\mathbf{G}}$ \textit{versus} SNR (dB).}]{\includegraphics[scale=0.57]{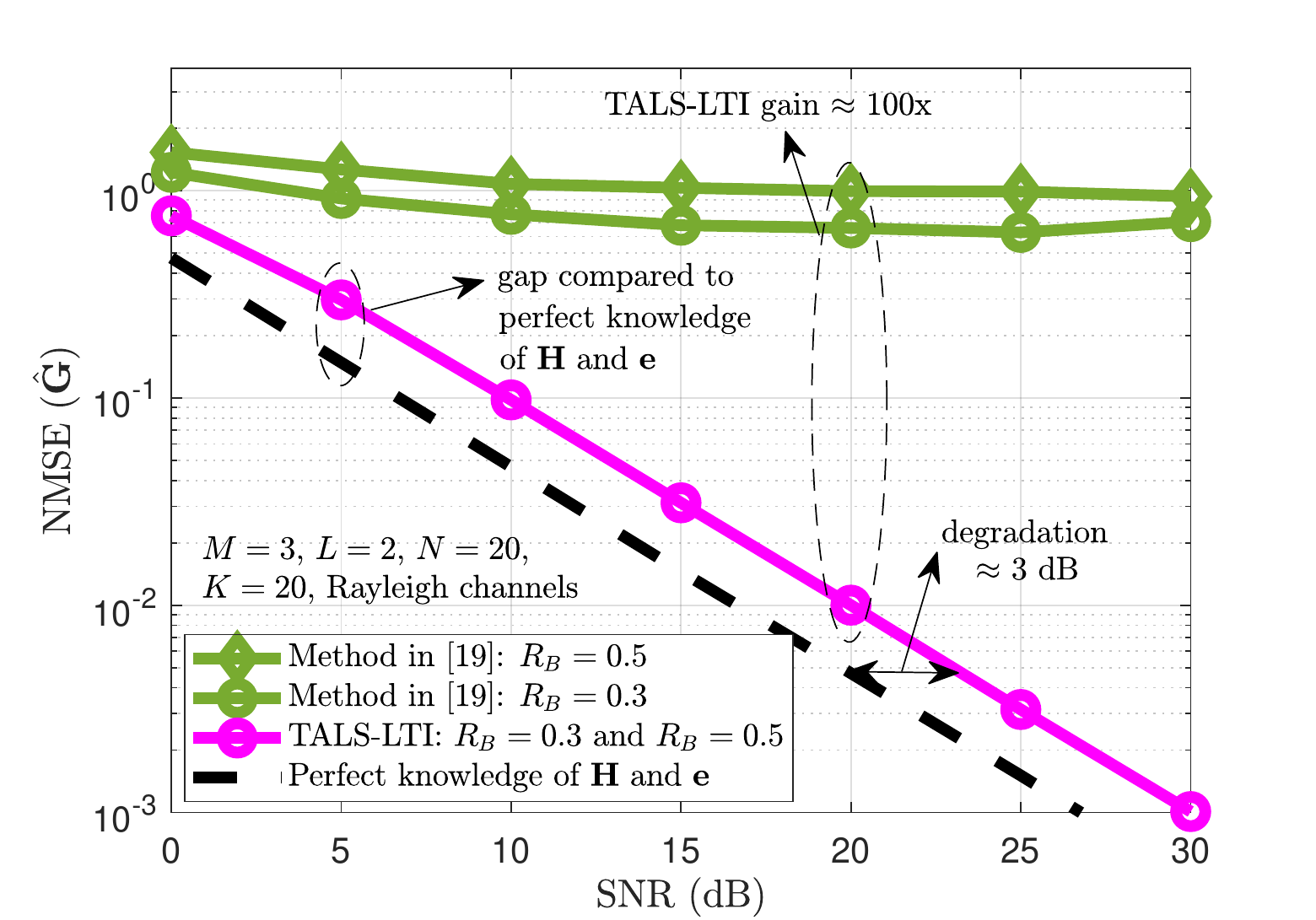}\label{Figura4b}}\\
\subfloat[\textcolor{black}{NMSE of $\hat{\mathbf{e}}$ \textit{versus} SNR (dB).}]{\includegraphics[scale=0.57]{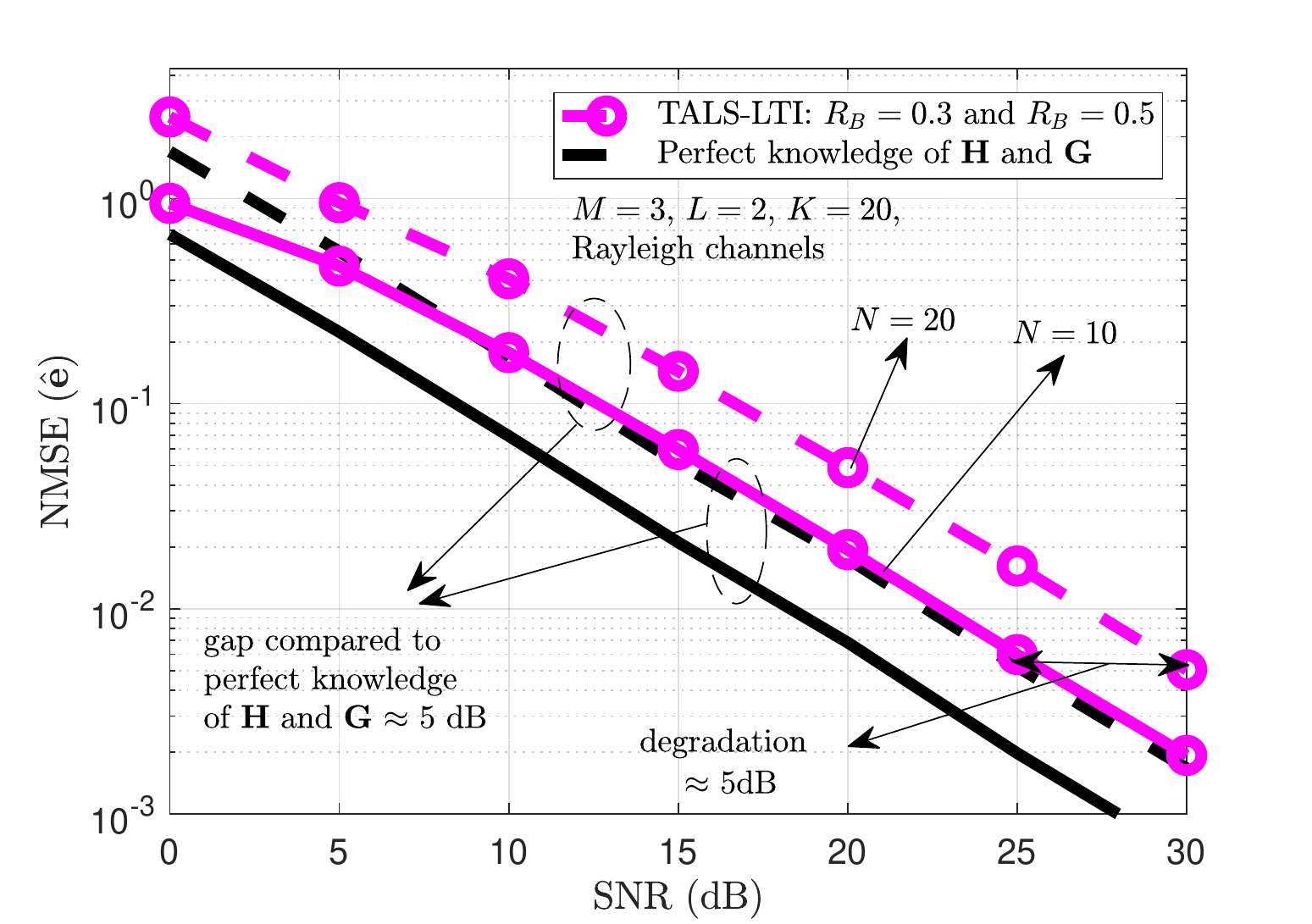}\label{Figura4c}}

\caption{\small{NMSE performance of the TALS-LTI algorithm \textit{versus} the SNR (dB) for $M=3$ transmit antennas, $L=2$ receive antennas, $K=20$ time-blocks and different RIS elements $N$ and impairments occurrence probability $R_{B}$ assuming i.i.d. Rayleigh fading channels.}}
\label{Figura4}
\end{figure}
\begin{figure}[!t]
	\centering
	\includegraphics[scale=0.57]{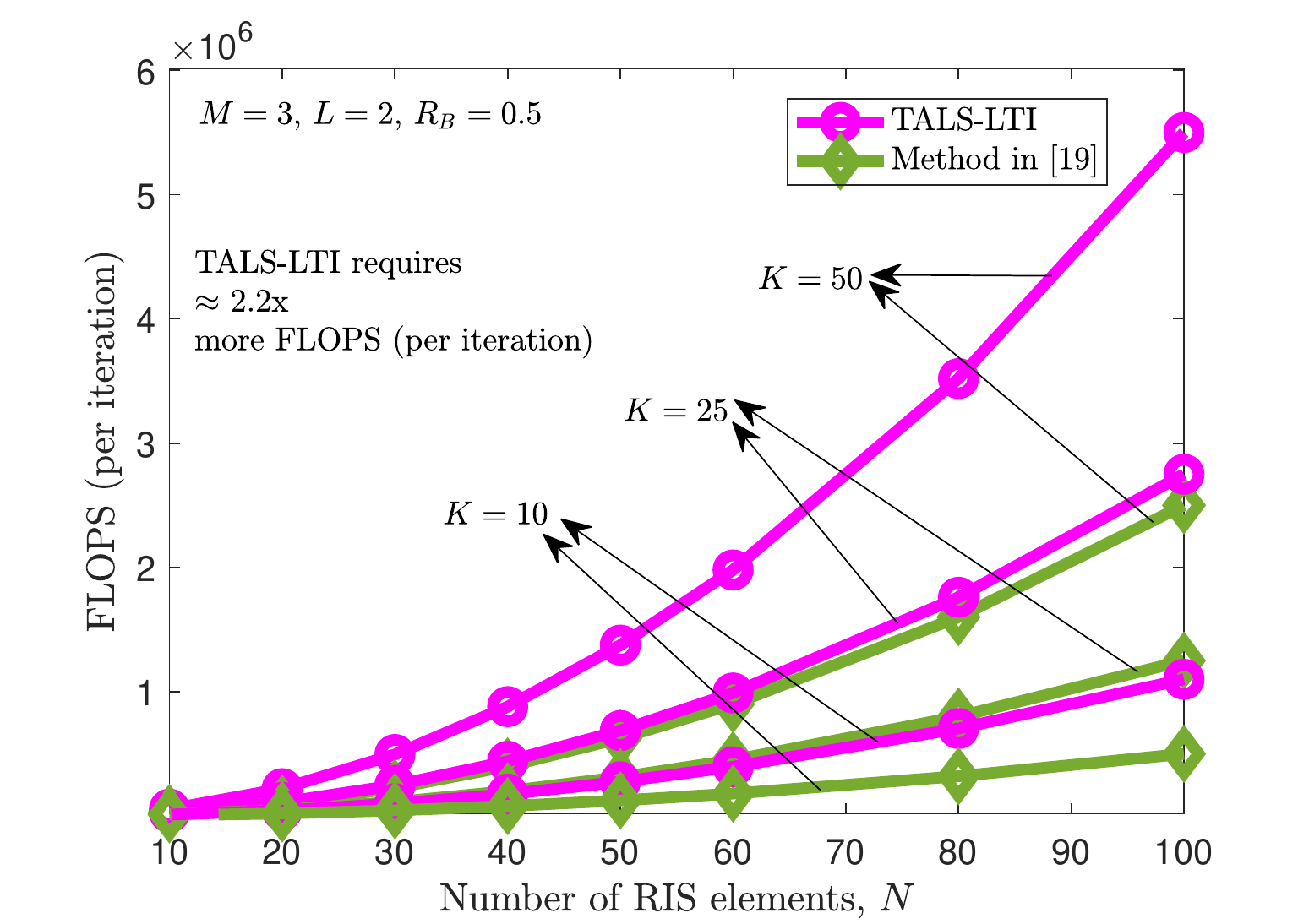}
	\caption{\small{Number of FLOPS required by the TALS-LTI algorithm \textit{versus} the number of the RIS elements $N$ for $M=3$ transmit antennas, $L=2$ receive antennas, impairments occurrence probability $R_{B}=0.5$ and different time-blocks $K$.}}
	\label{Figura5}
\end{figure}

\vspace{-0.5cm}
\section{Simulation Results}\label{Resul}

In this section, we present \textcolor{black}{the} simulation results for performance evaluation of our proposed tensor-based \textcolor{black}{algorithms,} in terms of computational complexity and estimation accuracy of the channels and imperfections, while comparing to benchmark approaches. The results presented here are averaged over \textcolor{black}{$\Omega = 3 \cdot 10^{3}$} independent Monte Carlo runs. Each run corresponds \textcolor{black}{to} a different realization of the involved communication channels, RIS patterns, impairment parameters and noise. We have designed the RIS pattern matrix $\mathbf{S}$ as a DFT matrix. The amplitude and phase impairments parameters embedded in the vector $\mathbf{e}$ and in the matrix $\mathbf{E}$ \textcolor{black}{follow a} uniform distribution between \textcolor{black}{$[0,1]$} and between \textcolor{black}{$[0, 2\pi]$,} respectively. \textcolor{black}{The location of the unknown $N_{B}$ impaired elements are assumed to be random with occurrence probability $R_{B}$, totalizing $N_{B} = NR_{B}$ impaired elements at the RIS in each simulated setup.} The metric used to evaluate the estimation accuracy is the normalized mean square error (NMSE) between the true and estimated matrices that provides a relative measure for the estimation error of the proposed algorithms. For the estimated channel \textcolor{black}{$\hat{\mathbf{H}}$,} we define
\vspace{-0.2cm}
\begin{equation}
\text{NMSE}(\hat{\mathbf{H}}) = \frac{1}{\Omega}\sum_{\omega=1}^{\Omega}\frac{\|\mathbf{H}^{(\omega)} - \hat{\mathbf{H}}^{(\omega)}\|_{\text{F}}^{2}}{\|\mathbf{H}^{(\omega)}\|_{\text{F}}^{2}},
\label{nmse}
\end{equation}
\noindent where \textcolor{black}{$\mathbf{H}^{(\omega)}$} and \textcolor{black}{$\hat{\mathbf{H}}^{(\omega)}$} denote the true channel and its estimate both related to the \textcolor{black}{$\omega$-th run,} respectively. Similar definitions \textcolor{black}{as in (\ref{nmse})} apply to the estimates of $\hat{\mathbf{G}}$, $\hat{\mathbf{E}}$ and $\hat{\mathbf{e}}$. \textcolor{black}{Finally, note that, while we present the simulation results for a set of parameter setting, we have test the results for a broad range of parameter settings and observed the same qualitative conclusions as those presented.}
\vspace{-0.3cm}
\subsection{\textcolor{black}{TALS-LTI Performance}}\label{perftals}

We first examine, in Figs. \ref{Figura4} and \ref{Figura5}, the performance of the proposed TALS-LTI algorithm. These plots show the NMSE and computational complexity in terms of floating-point operations (FLOPS) assuming i.i.d. Rayleigh fading channels. As a benchmark, we compare our TALS-LTI algorithm with the method proposed in \cite{GilJournal}, \textcolor{black}{which} is also a PARAFAC-based algorithm but formulated to the ideal case in which no impairments affect the RIS elements. Additionally, as a lower-bound for comparison, we also plot the performance of the \textit{clairvoyant} LS estimators of $\hat{\mathbf{H}}$, $\hat{\mathbf{G}}$ and $\hat{\mathbf{e}}$ in (\ref{sol1}), (\ref{sol2}) and (\ref{sol3}) obtained when the true factor matrices in the right-hand side of these equations are perfectly known. 

\textcolor{black}{From} Figs. \ref{Figura4a} and \ref{Figura4b}, we can observe that the method in \cite{GilJournal} is not suitable to tackle the channel estimation problem when RIS impairments are present. In contrast, \textcolor{black}{our} proposed TALS-LTI algorithm \textcolor{black}{provides} accurate estimates in which the NMSE of the estimated channels decreases linearly when the SNR increases, and \textcolor{black}{is} not sensitive to the number of impaired elements at the RIS. For instance, the TALS-LTI presents constant gaps when compared to the lower-bound LS estimator equal to $4$ dB and $3$ dB providing satisfactory performance in terms of the channel estimation for all simulated SNR range. Also, in contrast to method in \cite{GilJournal}, the TALS-LTI accurately \textcolor{black}{estimates} the RIS impairments that are treated as independent variables estimated beyond the involved channels, as illustrated in Fig. \ref{Figura4c}. 

In terms of computational complexity, we can see from Fig. \ref{Figura5} that the TALS-LTI presents greater complexity than competitor method requiring approximately $2.2\text{x}$ more FLOPS per iteration. This happens because the TALS-LTI has one more update step (per iteration) to \textcolor{black}{estimate} the RIS imperfections compared to \textcolor{black}{the} method in \cite{GilJournal} that does not perform the estimation of imperfections. Also, the complexity of both approaches increases with the number of RIS elements $N$ and time-blocks $K$. 

\begin{figure}[!h]
	\centering
	\subfloat[\textcolor{black}{NMSE of $\hat{\mathbf{H}}$ \textit{versus} SNR (dB).}]{\includegraphics[scale=0.57]{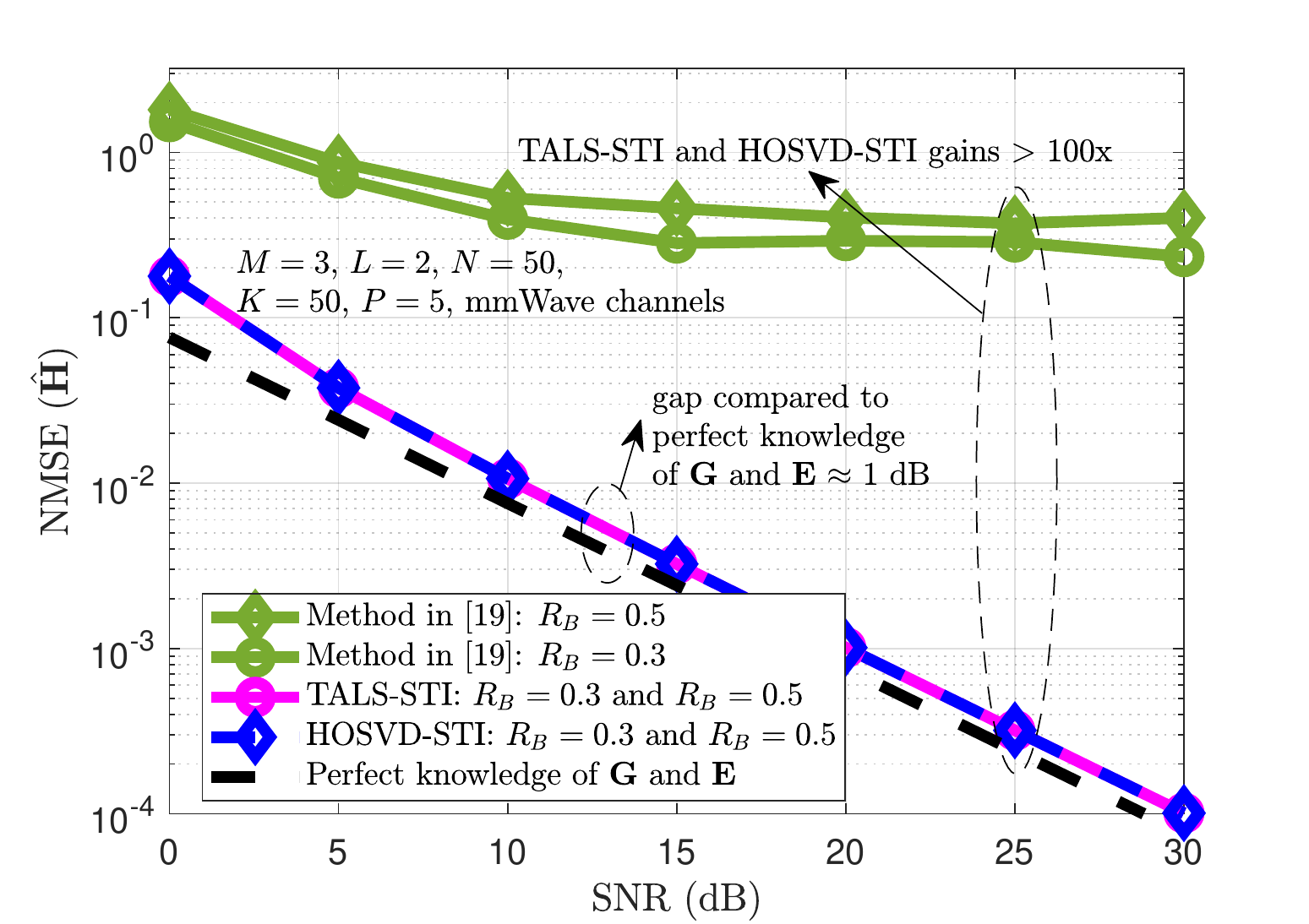}\label{Figura6a}}\\
	\subfloat[\textcolor{black}{NMSE of $\hat{\mathbf{G}}$ \textit{versus} SNR (dB).}]{\includegraphics[scale=0.57]{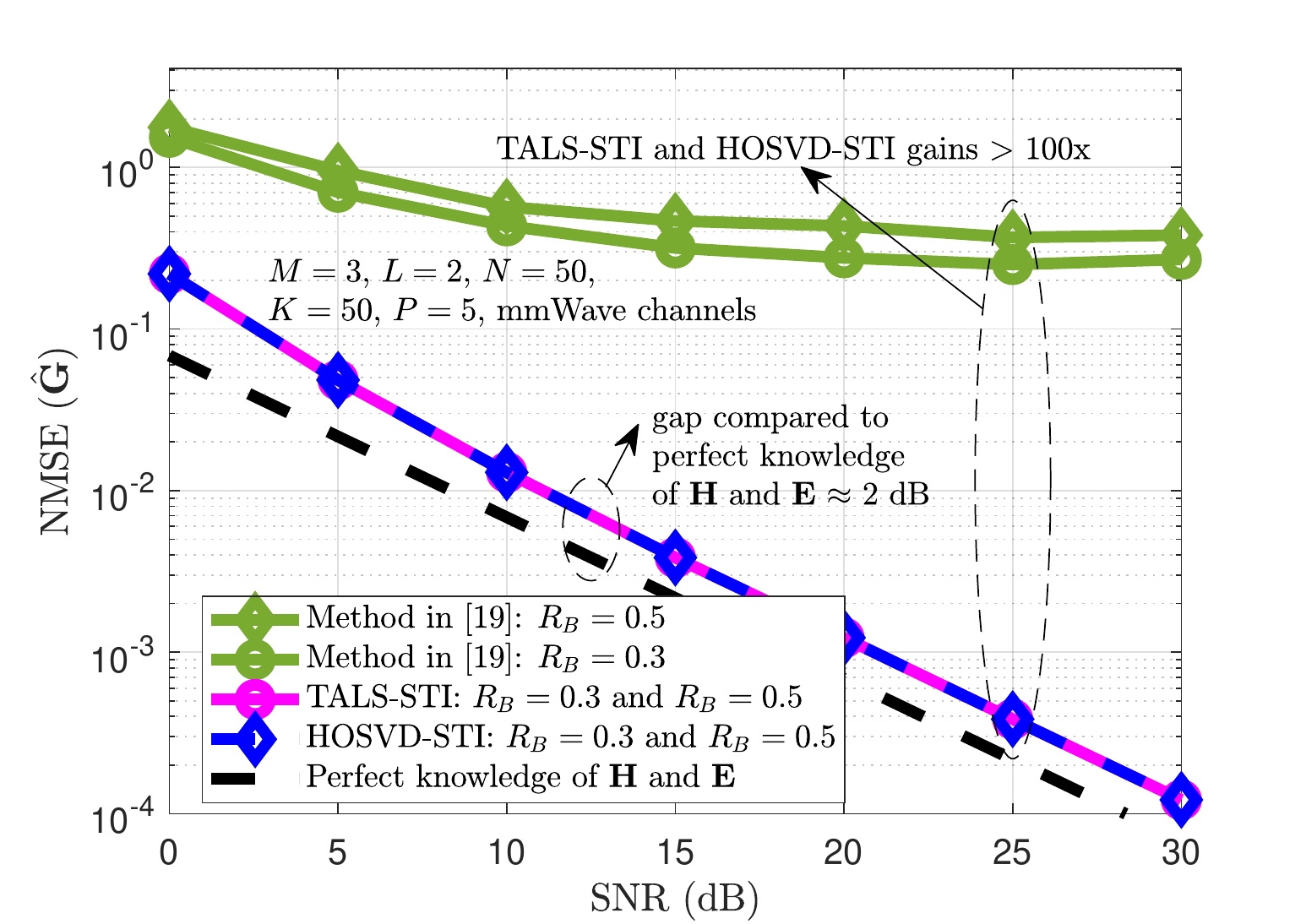}\label{Figura6b}}\\
	\subfloat[\textcolor{black}{NMSE of $\hat{\mathbf{E}}$ \textit{versus} SNR (dB).}]{\includegraphics[scale=0.57]{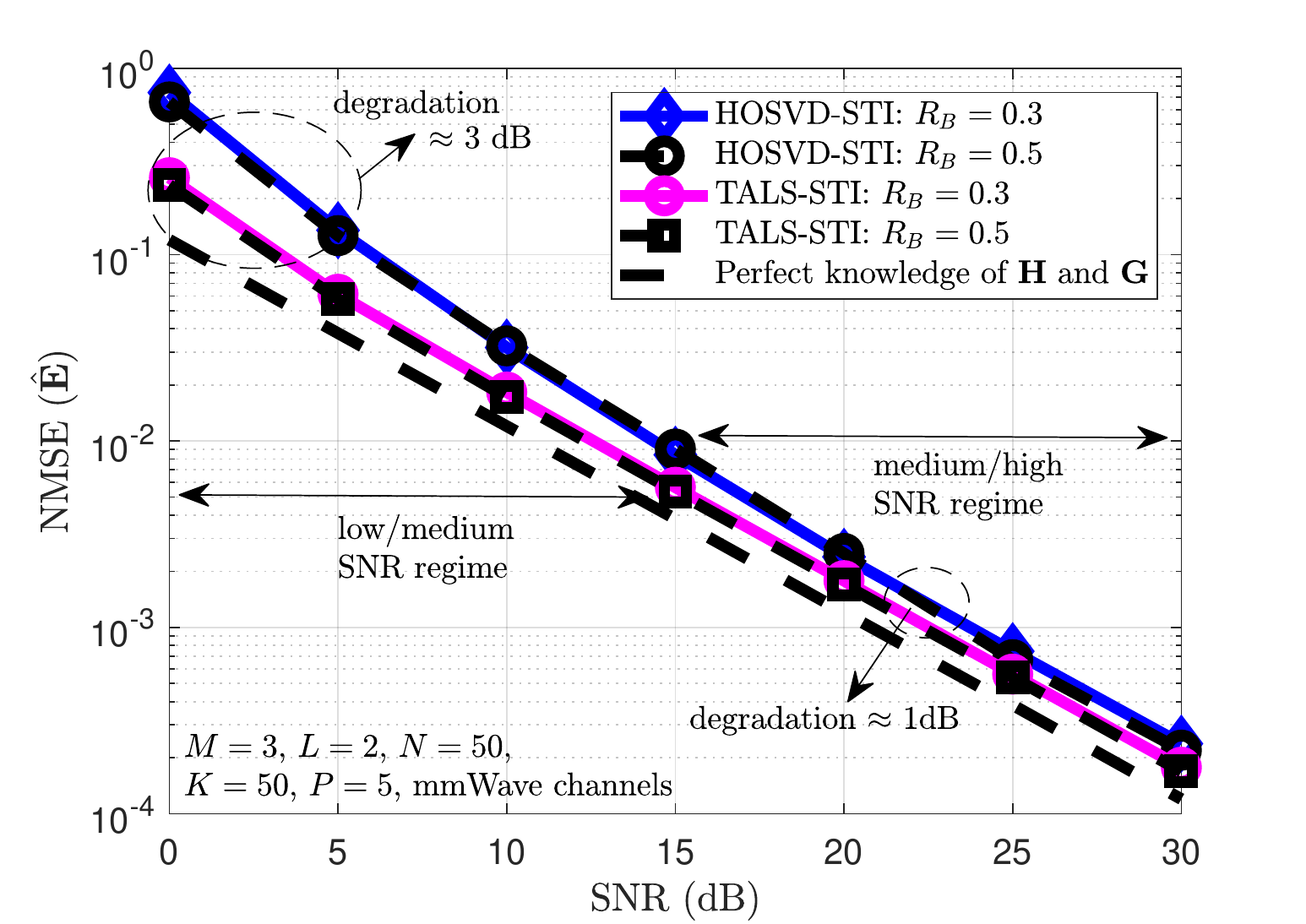}\label{Figura6c}}
	\caption{\small{NMSE performance of the TALS-STI and HOSVD-STI algorithms \textit{versus} the SNR (dB) for $M=3$ transmit antennas, $L=2$ receive antennas, $N=50$ RIS elements, $K=50$ time-blocks, $P=5$ frames and different impairments occurrence probability $R_{B}$ assuming a typical mmWave propagation environment.}}
	\label{Figura6}
\end{figure}
\begin{figure}[!t]
	\centering
	\includegraphics[scale=0.57]{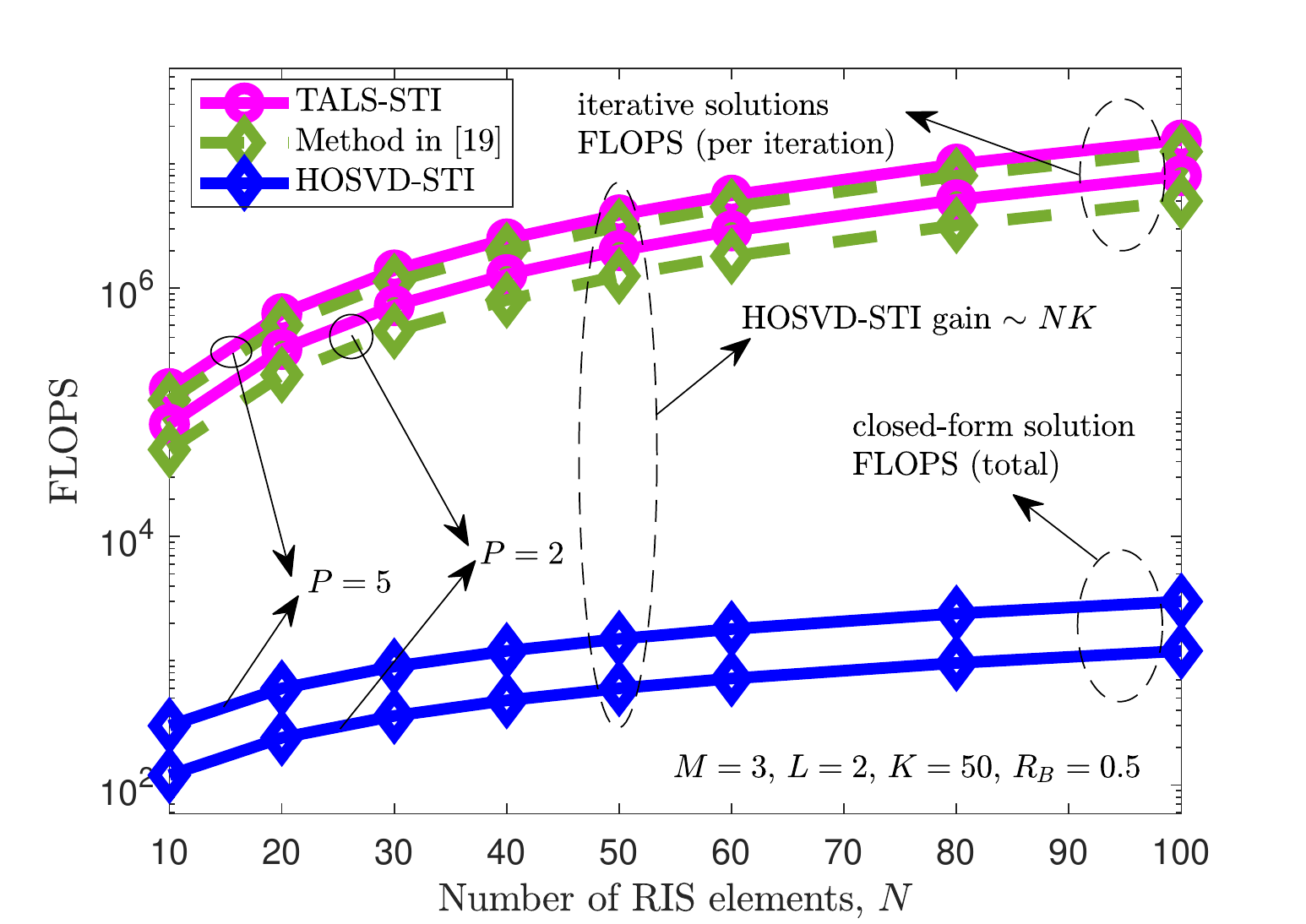}
	\caption{\small{Number of FLOPS required by the TALS-STI and HOSVD-STI algorithms \textit{versus} the number of the RIS elements $N$ for $M=3$ transmit antennas, $L=2$ receive antennas, $K=50$ time-blocks, impairments occurrence probability $R_{B}=0.5$ and different frames $P$.}}
	\label{Figura7}
\end{figure}
\vspace{-0.2cm}
\subsection{\textcolor{black}{TALS-STI and HOSVD-STI Performance}}

Here, we evaluate the performance of the proposed TALS-STI and HOSVD-STI algorithms assuming i.i.d. Rayleigh fading channels and a typical mmWave propagation environment. For the mmWave setup, we assume that the transmitter and the receiver are equipped with half-wavelength spaced uniform linear arrays while the RIS has half-wavelength spaced reflecting elements disposed in a uniform rectangular grid. In this case, the channel matrices are generated according to the widely used geometric channel model \textcolor{black}{\cite{Zhou:22}}, in which the angles of arrival and departure are randomly generated according to a uniform distribution. The azimuth and elevation angles are drawn between \textcolor{black}{$[-\pi/2,\pi/2]$} and \textcolor{black}{$[0,\pi/2]$}, 
%$-\pi/2$ and $\pi/2$ and between $0$ and $\pi/2$, 
respectively. The number of propagation paths \textcolor{black}{is} set to $1$ in the Tx-RIS link and to $2$ in the RIS-Rx link, respectively. The complex channel gains follow uniform distributions. The lower-bound LS estimators of $\hat{\mathbf{H}}$, $\hat{\mathbf{G}}$ and $\hat{\mathbf{E}}$ are obtained similarly to \textcolor{black}{Section \ref{perftals}}, but now from (\ref{hj1}), (\ref{hj2}) and (\ref{hj3}), respectively.

From Fig. \ref{Figura6}, similar conclusions to the results in \textcolor{black}{Fig. \ref{Figura4},} in terms of channel estimation \textcolor{black}{performance,} can be made by comparing the proposed TALS-STI and HOSVD-STI algorithms with the method in \cite{GilJournal}. However, more \textcolor{black}{accurate} estimates for the channels and impairments are obtained when the mmWave propagation scenario is considered. In this experiment, the TALS-STI and HOSVD-STI \textcolor{black}{present, respectively,} constant gaps of approximately $1$ dB and \textcolor{black}{$2$ dB,} compared to the lower-bound LS estimator for \textcolor{black}{the considered range of SNR,} 
%all the SNR simulated range,
confirming the effectiveness of the proposed tensor-based algorithms in terms of estimation accuracy. Similar results are obtained for the proposed TALS-LTI algorithm, but omitted here due to space limitation. Also, we see from Fig. \ref{Figura6c}, that in the \textcolor{black}{low/medium} SNR regime, the TALS-STI algorithm outperforms the HOSVD-STI one in terms of imperfections estimation, indicating that the \textcolor{black}{TALS-STI algorithm} becomes preferable with considerable gain in low SNR when complexity issues are not taken into account. The tradeoff between computational complexity and estimation performance is analyzed in the sequel. The \textcolor{black}{presented} result indicates that the proposed algorithms are able to work with high accuracy under different propagation environments and kinds of imperfections, i.e., our approaches are general techniques that provide accurate estimates for \textcolor{black}{different}
%non-specific 
channel models.

\begin{figure}[!t]
	\centering
	\includegraphics[scale=0.57]{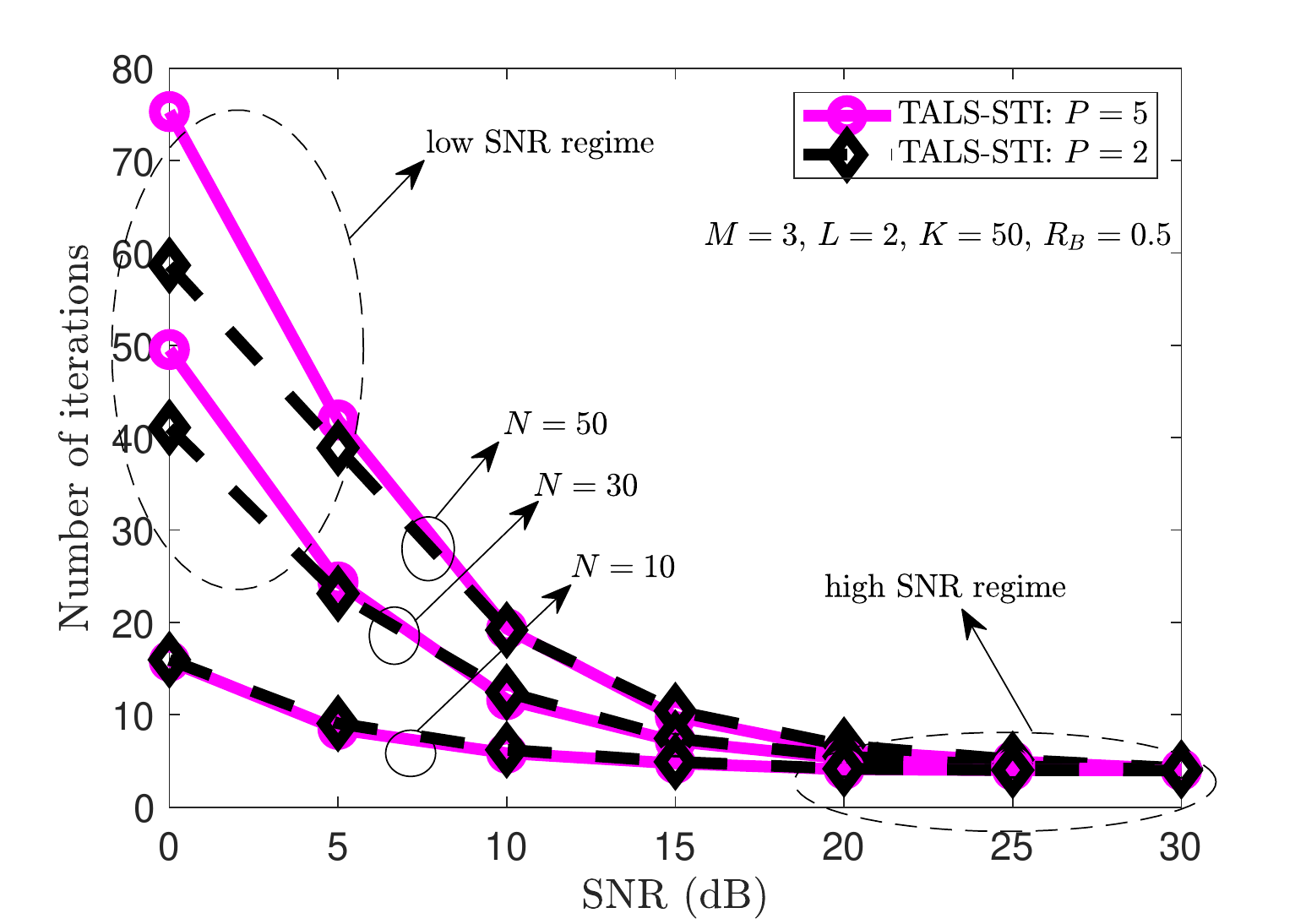}
	\caption{\small{Number of iterations required by the TALS-STI algorithm \textit{versus} the SNR (dB) for $M=3$ transmit antennas, $L=2$ receive antennas, $K=50$ time-blocks, impairments occurrence probability $R_{B}=0.5$ and different RIS elements $N$ and frames $P$.}}
	\label{Figura8}
\end{figure}

In Fig. \ref{Figura7}, we evaluate the computational complexity (in terms of FLOPS) of the TALS-STI and HOSVD-STI algorithms. We can see that the HOSVD-STI is less complex than TALS-STI for different values of RIS elements $N$ and frames $P$, since it is a closed-form solution for joint channel and imperfections estimation. The HOSVD-STI algorithm provides a remarkable gain of the order of $NK$ FLOPS compared to the TALS-STI one. It can be seen that the complexity of both methods grows when the number of RIS elements $N$ and frames $P$ increases, which is an expected result since the number of entries in $\mathbf{H}$, $\mathbf{G}$ and $\mathbf{E}$ also \textcolor{black}{increases} with $N$ and $P$. However, \textcolor{black}{in contrast} to the HOSVD-STI which is a closed-form solution, for a complete analysis on the overall complexity of the proposed techniques, Fig. \ref{Figura8} shows the number of iterations necessary for the convergence of the TALS-STI algorithm considering different values of $N$ and $P$. It can be seen that the TALS-STI algorithm rapidly converges thanks to the knowledge of the non-impaired matrix $\mathbf{S}$ that remains fixed during each iteration. \textcolor{black}{Also,} the number of iterations required for the convergence decreases as a function of the SNR. In the high SNR regime, the convergence of the proposed TALS-STI \textcolor{black}{algorithm} is no more sensitive to $N$, $P$ and SNR values. When the SNR is higher than $20$ dB, its convergence is quickly achieved within approximately $5$ iterations. 

\begin{figure}[!t]
	\centering
	\includegraphics[scale=0.57]{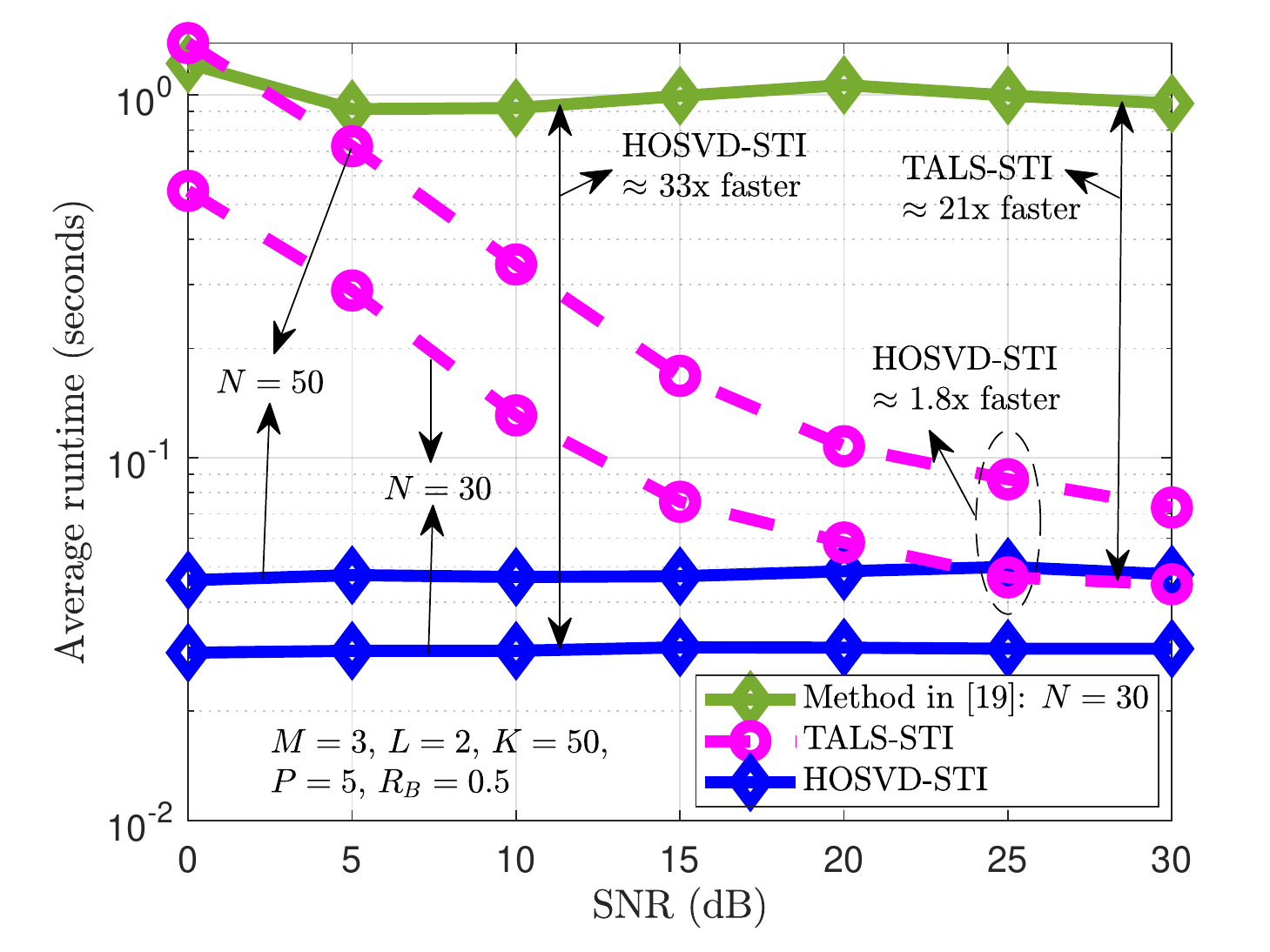}
	\caption{\small{Average runtime (in seconds) for the TALS-STI and HOSVD-STI algorithms \textit{versus} the SNR (dB) for $M=3$ transmit antennas, $L=2$ receive antennas, $K=50$ time-blocks, $P=5$ frames, impairments occurrence probability $R_{B}=0.5$ and different RIS elements $N$.}}
	\label{Figura9}
\end{figure}

\begin{figure}[!t]
	\centering
	\subfloat[\textcolor{black}{NMSE of $\hat{\mathbf{H}}$ \textit{versus} $N$.}]{\includegraphics[scale=0.57]{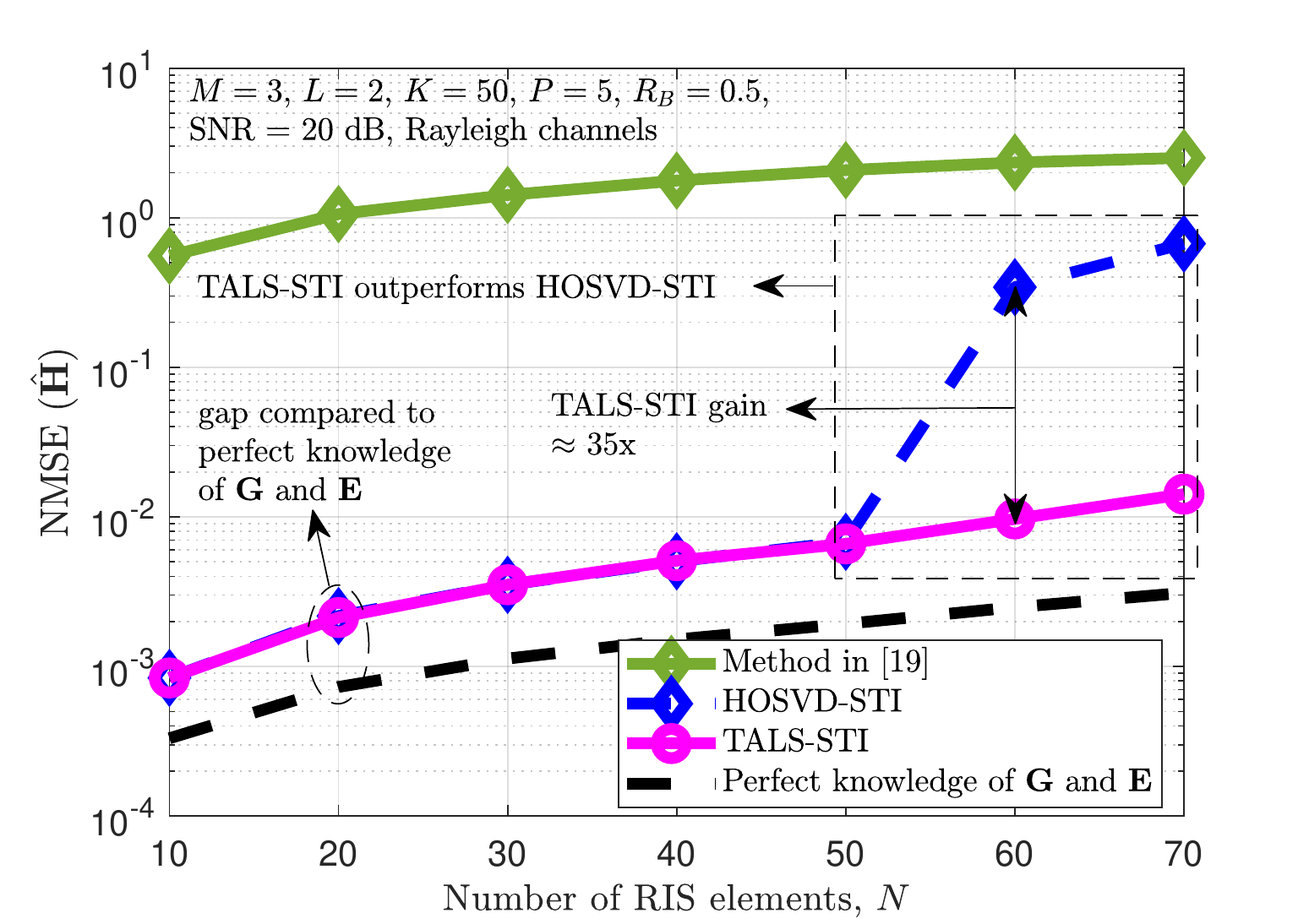}\label{result1}}\\
	\subfloat[\textcolor{black}{NMSE of $\hat{\mathbf{G}}$ \textit{versus} $N$.}]{\includegraphics[scale=0.57]{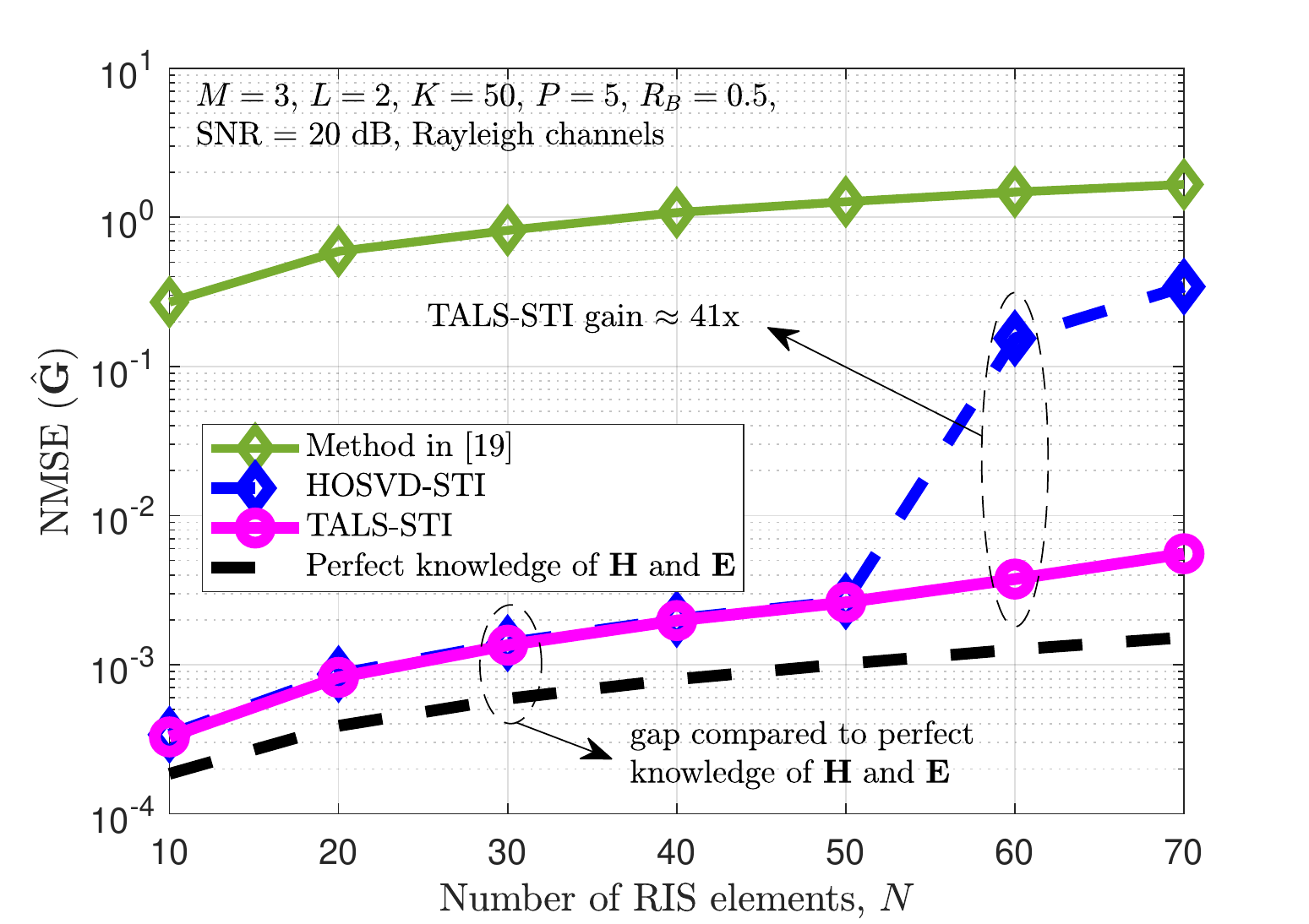}\label{result2}}\\
	\subfloat[\textcolor{black}{NMSE of $\hat{\mathbf{E}}$ \textit{versus} $N$.}]{\includegraphics[scale=0.57]{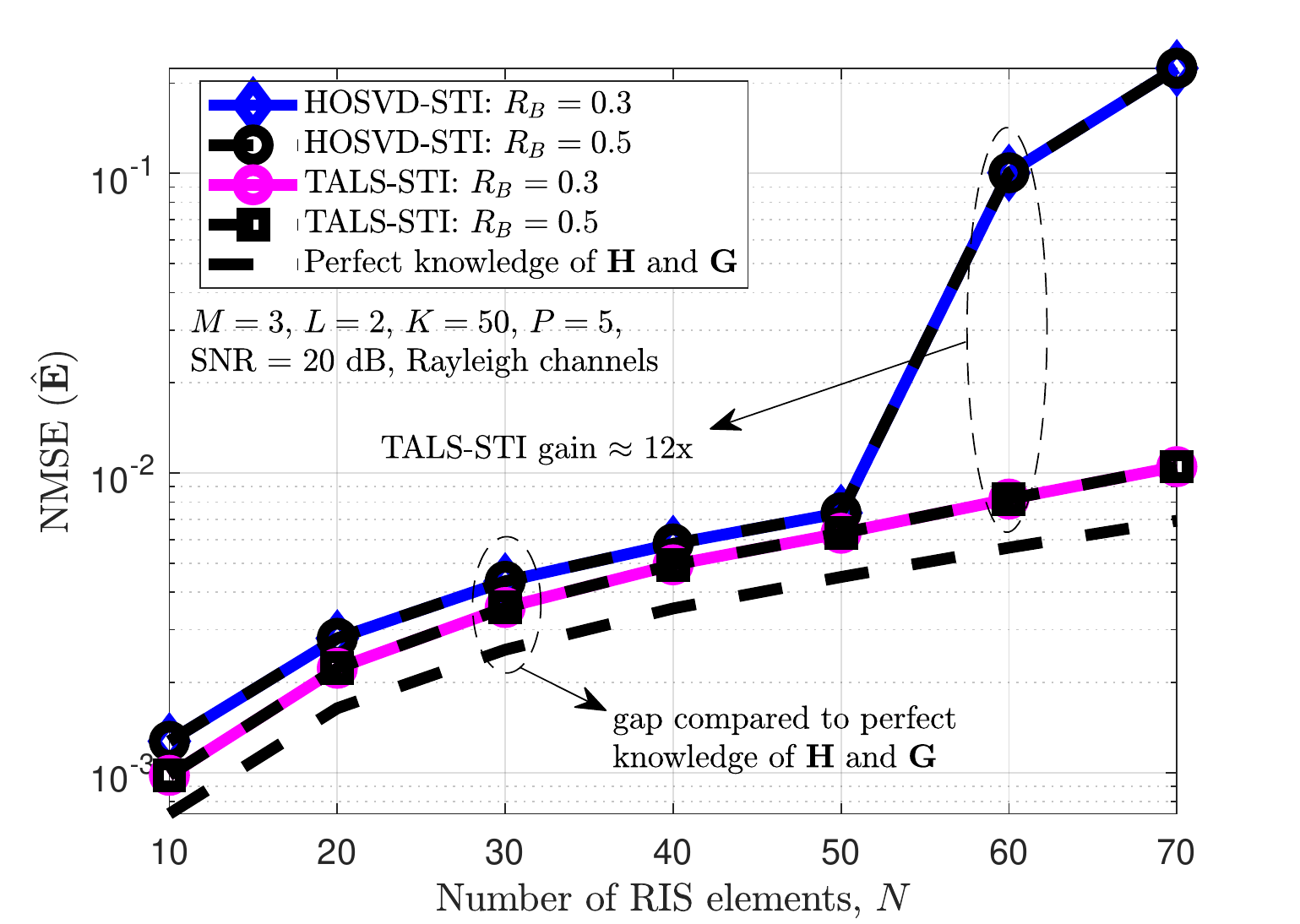}\label{result3}}
	\caption{\small{NMSE performance of the TALS-STI and HOSVD-STI algorithms \textit{versus} the number of the RIS elements $N$ for $M=3$ transmit antennas, $L=2$ receive antennas, $K=50$ time-blocks, $P=5$ frames, SNR $=$ 20 dB and different impairments occurrence probability $R_{B}$ assuming i.i.d. Rayleigh fading channels.}}
	\label{Figura10}
\end{figure}
\begin{figure}[!t]
	\centering
	\subfloat[\textcolor{black}{NMSE of $\hat{\mathbf{H}}$ \textit{versus} $R_{B}$.}]{\includegraphics[scale=0.57]{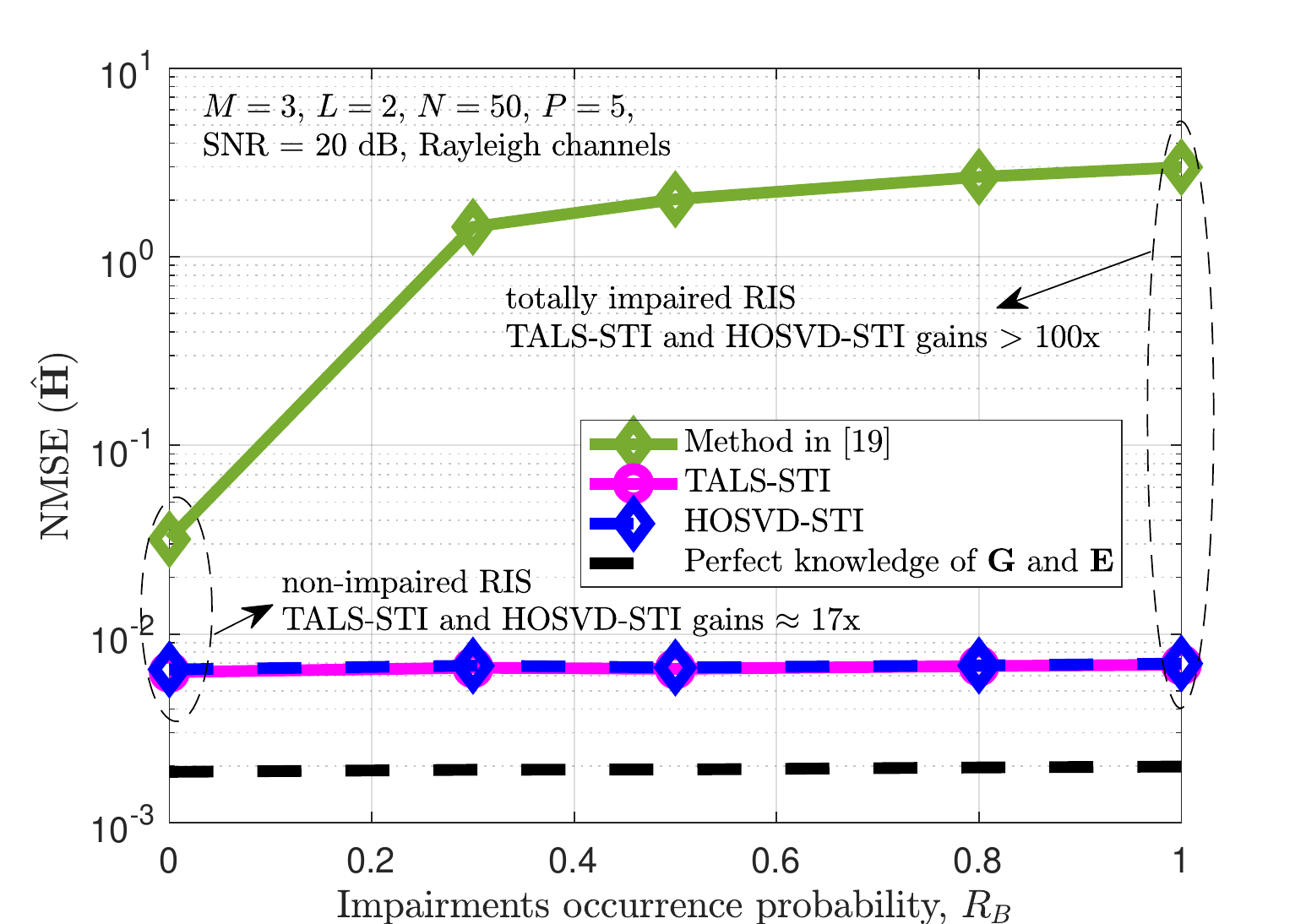}\label{result1}}\\
	\subfloat[\textcolor{black}{NMSE of $\hat{\mathbf{G}}$ \textit{versus} $R_{B}$.}]{\includegraphics[scale=0.57]{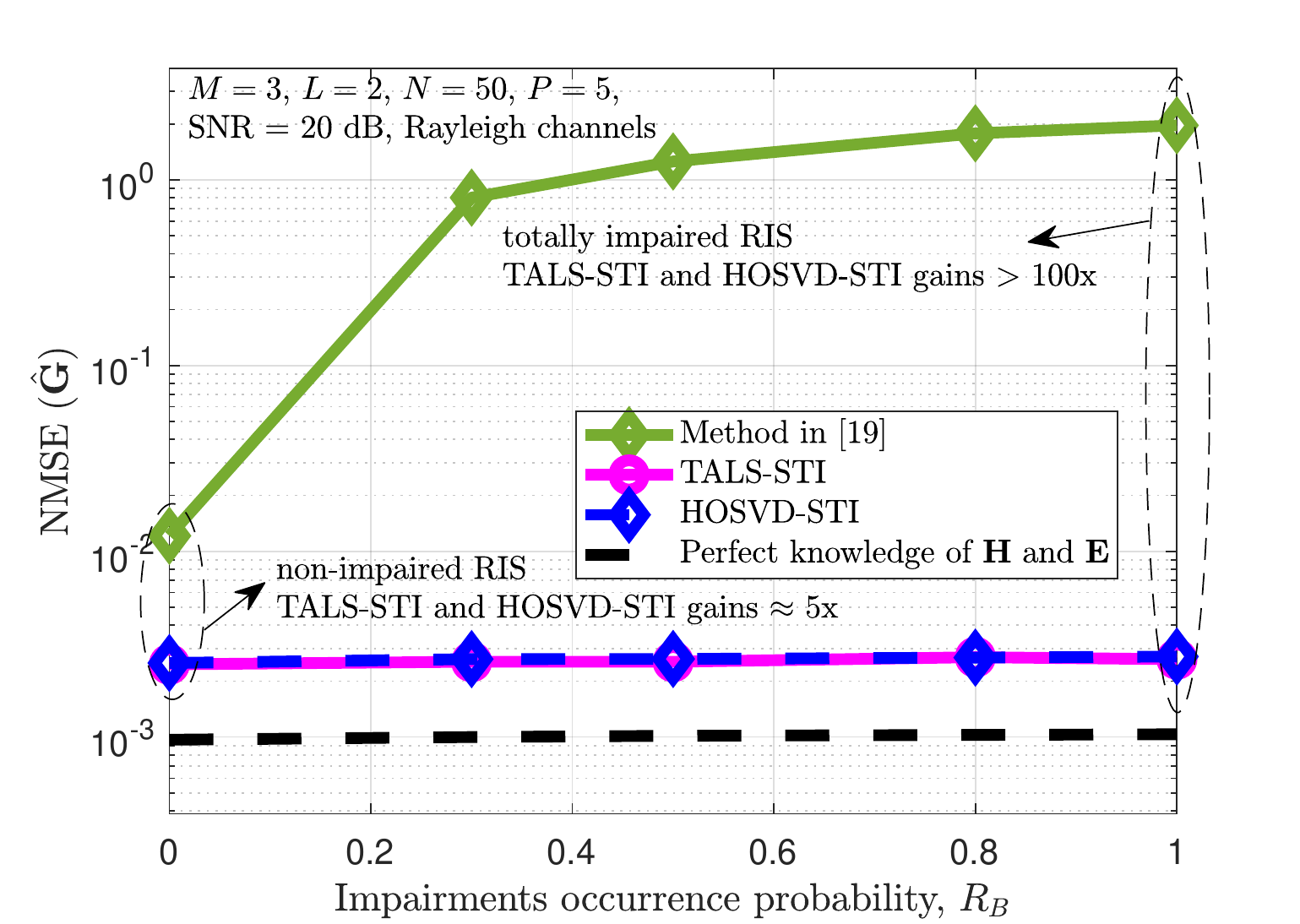}\label{result2}}\\
	\subfloat[\textcolor{black}{NMSE of $\hat{\mathbf{E}}$ \textit{versus} $R_{B}$.}]{\includegraphics[scale=0.57]{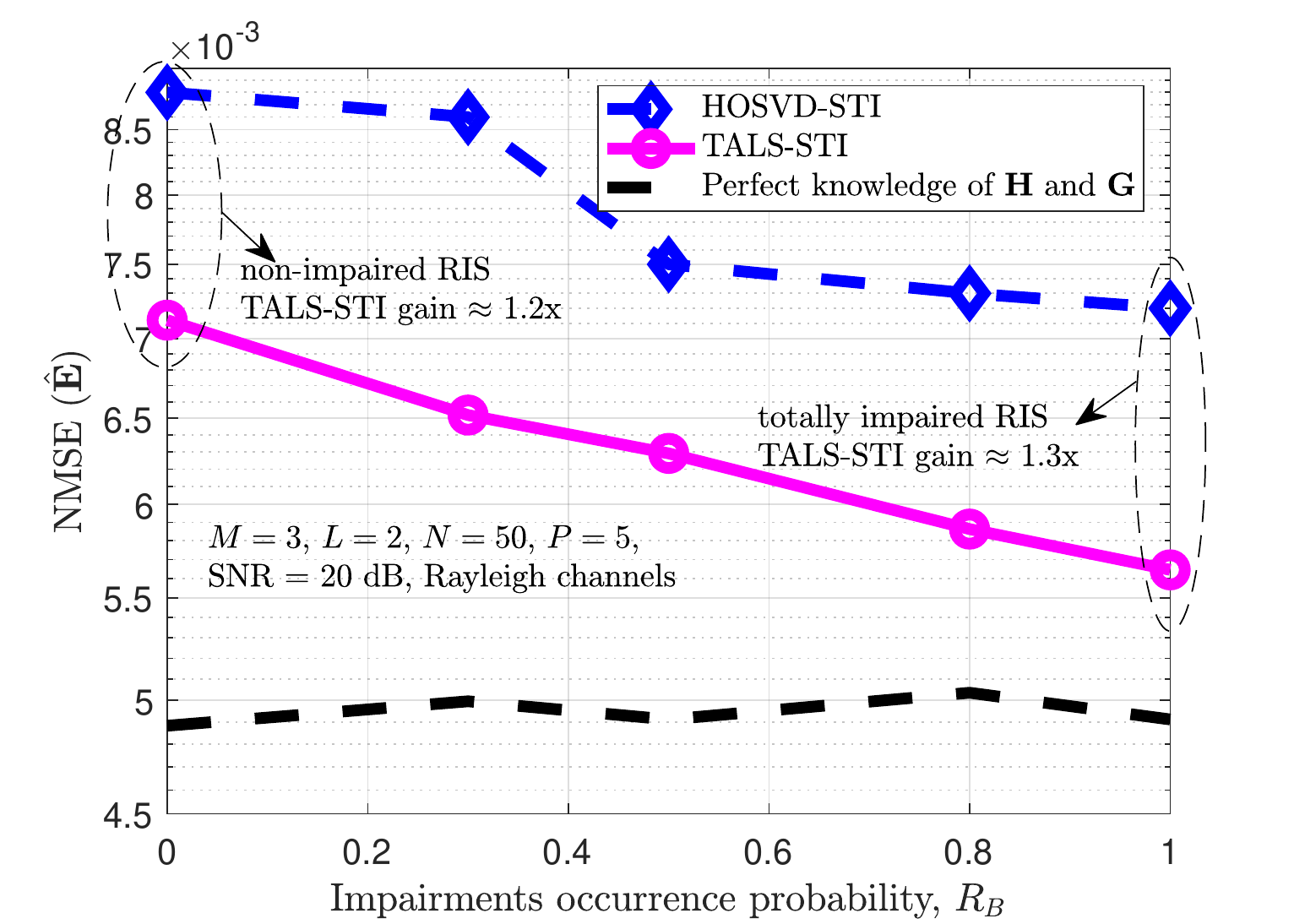}\label{result3}}
	\caption{\small{NMSE performance of the TALS-STI and HOSVD-STI algorithms \textit{versus} the impairments occurrence probability $R_{B}$ for $M=3$ transmit antennas, $L=2$ receive antennas, $N=50$ RIS elements, $K=50$ time-blocks, $P=5$ frames, SNR $=$ 20 dB assuming i.i.d. Rayleigh fading channels.}}
	\label{Figura11}
\end{figure}

In Fig. \ref{Figura9}, we evaluate the overall computational complexity of the TALS-STI and HOSVD-STI \textcolor{black}{algorithms} using as metric the average runtime (in seconds). This metric takes into account the number of iterations for convergence, making it possible to compare the complexity between iterative and closed-form solutions. It can be seen that the runtime grows when the number of RIS elements $N$ increases, \textcolor{black}{confirming} 
%thus corroborating with 
the results shown in Fig. \ref{Figura7}. The runtime required by HOSVD-STI is not sensitive to the SNR since it is a closed-form algorithm. On the other hand, the runtime of the TALS-STI \textcolor{black}{algorithm} decreases with the SNR \textcolor{black}{as} it is an iterative algorithm, achieving performance close to HOSVD-STI algorithm in the high SNR regime thanks to its rapid convergence \textcolor{black}{(Fig. \ref{Figura8}).}
%as shown in Fig. \ref{Figura8}. 

\textcolor{black}{We} can conclude that TALS-STI and HOSVD-STI \textcolor{black}{algorithms} outperform the method in \cite{GilJournal} in terms of both channel estimation accuracy and overall computational complexity. This is because due to the RIS imperfections, the method in \cite{GilJournal} does not estimate channels properly and has very slow convergence which directly increases its runtime. However, 
%as previously discussed in Section \ref{Iden}, 
the TALS-STI can operate under more flexible choices for the number of time-blocks $K$ for channel estimation compared to HOSVD-STI \textcolor{black}{(see Fig. \ref{Figura10}).}
%This is illustrated in Fig. \ref{Figura10}.
From this, we can note that when the condition in (\ref{hgf}) is not satisfied, the TALS-STI significantly outperforms the HOSVD-STI in terms of estimation accuracy. \textcolor{black}{Therefore, a tradeoff between overall computational complexity, estimation performance and operation conditions for the proposed solutions can be observed.}
%Therefore, a tradeoff between overall computational complexity, channel and imperfections estimation performance, and operation conditions for the proposed algorithms has been verified through our simulation results. 
Thus, the TALS-STI may be attractive when more flexible choices for the \textcolor{black}{number of time-blocks for channel estimation} $K$ are required, while the \textcolor{black}{HOSVD-STI is} preferred specially when low processing delay is \textcolor{black}{desired.} 
We can also observe that the estimation performance degrades as the number of RIS elements $N$ \textcolor{black}{increases,} which is an expected result since the number of channel and imperfections coefficients to be estimated also increases with $N$.

In our last experiment, shown in Fig. \ref{Figura11}, we evaluate the NMSE as a function of the impairments occurrence probability \textcolor{black}{$R_{B}$.} It can be seen \textcolor{black}{that,} in terms of channel estimation, for a broad range of parameter \textcolor{black}{settings,} our proposed approaches are not sensitive to the number of impaired elements at the RIS. This is because the imperfections matrix is estimated as an independent variable by both methods. However, in terms of imperfections estimation, the TALS-STI proves to be more accurate for all \textcolor{black}{considered values of} $R_{B}$. It is also important to note that the proposed TALS-STI and HOSVD-STI algorithms outperform the method in \cite{GilJournal}, even when ideal non-impaired RIS is assumed. However, the gains achieved by our methods become more evident in the challenging scenario, when the number of impaired elements at the RIS increases. \textcolor{black}{In this case, the method of \cite{GilJournal} is sensitive to $R_{B}$. Thus, if the number of impaired elements changes, it does not show a stable/predictable performance. This leads to the need for channel estimation correction and retransmissions. In contrast, our proposed methods are quite robust to $R_{B}$ variation which simplifies the system design.}
\vspace{-0.3cm}
\section{Conclusion}\label{Conc}

We have proposed \textcolor{black}{different efficient} tensor-based algorithms for channel estimation in RIS-assisted MIMO systems, in which the RIS elements are affected by real-world imperfections. We resort to the multidimensional structure of the received signal to solve these non-idealized channel estimation problems by means of trilinear and quadrilinear PARAFAC models. The proposed TALS-LTI algorithm solves the problem when static imperfections are assumed. In a generalized way, we have formulated the TALS-STI and HOSVD-STI algorithms for the more challenging scenario in which the behavior of the RIS imperfections is non-static with respect to the channel coherence time. The TALS-LTI and TALS-STI \textcolor{black}{algorithms} are iterative solutions that relax the system design requirements, operating under more flexible choices for the \textcolor{black}{training parameters.} In contrast, the HOSVD-STI algorithm is a closed-form solution that has a lower computational complexity compared to the competing ALS-based solutions, while affording parallel processing. Simulation results \textcolor{black}{illustrate} the high estimation performance of the proposed tensor-based algorithms for different kinds of imperfections, \textcolor{black}{channel models,} and system configurations. The TALS-STI and HOSVD-STI algorithms \textcolor{black}{present similar} channel estimation performances. However, the TALS-STI \textcolor{black}{is} preferable \textcolor{black}{for} the imperfections detection in the low SNR regime and when more flexible choices for \textcolor{black}{training parameters} is required, while the \textcolor{black}{HOSVD-STI is} preferred when low processing delay is \textcolor{black}{desired}.

\vspace{-0.1cm}
%\bibliographystyle{IEEEtran}
%\bibliography{refs}

\begin{thebibliography}{10}
	\providecommand{\url}[1]{#1}
	\csname url@samestyle\endcsname
	\providecommand{\newblock}{\relax}
	\providecommand{\bibinfo}[2]{#2}
	\providecommand{\BIBentrySTDinterwordspacing}{\spaceskip=0pt\relax}
	\providecommand{\BIBentryALTinterwordstretchfactor}{4}
	\providecommand{\BIBentryALTinterwordspacing}{\spaceskip=\fontdimen2\font plus
		\BIBentryALTinterwordstretchfactor\fontdimen3\font minus
		\fontdimen4\font\relax}
	\providecommand{\BIBforeignlanguage}[2]{{%
			\expandafter\ifx\csname l@#1\endcsname\relax
			\typeout{** WARNING: IEEEtran.bst: No hyphenation pattern has been}%
			\typeout{** loaded for the language `#1'. Using the pattern for}%
			\typeout{** the default language instead.}%
			\else
			\language=\csname l@#1\endcsname
			\fi
			#2}}
	\providecommand{\BIBdecl}{\relax}
	\BIBdecl
	
	\bibitem{ourGlobecom}
	\BIBentryALTinterwordspacing
	P.~R.~B. Gomes, G.~T. de~Araújo, B.~Sokal, A.~L.~F. de~Almeida, B.~Makki, and
	G.~Fodor, ``Tensor-based channel estimation for ris-assisted networks
	operating under imperfections,'' 2022. [Online]. Available:
	\url{https://arxiv.org/abs/2206.03557}
	\BIBentrySTDinterwordspacing
	
	\bibitem{SCisco}
	G.~Forecast, ``Cisco visual networking index: global mobile data traffic
	forecast update, 2017--2022,'' \emph{Update}, vol. 2017, p. 2022, 2019.
	
	\bibitem{RCITU}
	ITU, ``{IMT traffic estimates for the years 2020 to 2030 },'' {ITU-R M}, TR
	{2370-0}, Jan. 2015, {V.15.0.0}.
	
	\bibitem{SZhang2017}
	S.~Zhang, Q.~Wu, S.~Xu, and G.~Y. Li, ``Fundamental green tradeoffs:
	Progresses, challenges, and impacts on {5G} networks,'' \emph{IEEE Commun.
		Surveys Tuts.}, vol.~19, no.~1, pp. 33--56, Feb. 2016.
	
	\bibitem{CYou}
	C.~You, B.~Zheng, and R.~Zhang, ``Channel estimation and passive beamforming
	for intelligent reflecting surface: Discrete phase shift and progressive
	refinement,'' \emph{IEEE J. Sel. Areas Commun.}, vol.~38, no.~11, pp.
	2604--2620, Jul. 2020.
	
	\bibitem{QWu}
	Q.~Wu, G.~Y. Li, W.~Chen, D.~W.~K. Ng, and R.~Schober, ``An overview of
	sustainable green {5G} networks,'' \emph{IEEE Wireless Communications},
	vol.~24, no.~4, pp. 72--80, Aug. 2017.
	
	\bibitem{Zhou:22}
	G.~Zhou, C.~Pan, H.~Ren, P.~Popovski, and A.~L. Swindlehurst, ``Channel
	estimation for ris-aided multiuser millimeter-wave systems,'' \emph{IEEE
		Transactions on Signal Processing}, vol.~70, pp. 1478--1492, 2022.
	
	\bibitem{DiRenzo2019}
	M.~Di~Renzo, M.~Debbah, D.-T. Phan-Huy, A.~Zappone, M.-S. Alouini, C.~Yuen,
	V.~Sciancalepore, G.~C. Alexandropoulos, J.~Hoydis, H.~Gacanin \emph{et~al.},
	``Smart radio environments empowered by reconfigurable ai meta-surfaces: An
	idea whose time has come,'' \emph{EURASIP Journal on Wireless Communications
		and Networking}, vol. 2019, no.~1, pp. 1--20, May. 2019.
	
	\bibitem{YLiang2019}
	Y.-C. Liang, R.~Long, Q.~Zhang, J.~Chen, H.~V. Cheng, and H.~Guo, ``Large
	intelligent surface/antennas ({LISA}): Making reflective radios smart,''
	\emph{Journal of Communications and Information Networks}, vol.~4, no.~2, pp.
	40--50, Jun 2019.
	
	\bibitem{DiRenzo2020}
	M.~Di~Renzo, A.~Zappone, M.~Debbah, M.-S. Alouini, C.~Yuen, J.~de~Rosny, and
	S.~Tretyakov, ``Smart radio environments empowered by reconfigurable
	intelligent surfaces: How it works, state of research, and the road ahead,''
	\emph{IEEE J. Sel. Areas Commun.}, vol.~38, no.~11, pp. 2450--2525, Jul.
	2020.
	
	\bibitem{BehroozSug}
	N.~Rajatheva, I.~Atzeni, S.~Bicais, E.~Bjornson, A.~Bourdoux, S.~Buzzi,
	C.~D'Andrea, J.-B. Dore, S.~Erkucuk, M.~Fuentes, K.~Guan, Y.~Hu, X.~Huang,
	J.~Hulkkonen, J.~M. Jornet, M.~Katz, B.~Makki, R.~Nilsson, E.~Panayirci,
	K.~Rabie, N.~Rajapaksha, M.~Salehi, H.~Sarieddeen, S.~Shahabuddin,
	T.~Svensson, O.~Tervo, A.~Tolli, Q.~Wu, and W.~Xu, ``Scoring the terabit/s
	goal:broadband connectivity in {6G},'' Feb. 2021, arXiv:2008.07220v2.
	
	\bibitem{EBasar2019}
	E.~Basar, M.~Di~Renzo, J.~De~Rosny, M.~Debbah, M.-S. Alouini, and R.~Zhang,
	``Wireless communications through reconfigurable intelligent surfaces,''
	\emph{IEEE access}, vol.~7, pp. 116\,753--116\,773, Jun. 2019.
	
	\bibitem{QWuRZhang20192}
	Q.~Wu and R.~Zhang, ``Towards smart and reconfigurable environment: Intelligent
	reflecting surface aided wireless network,'' \emph{IEEE Communications
		Magazine}, vol.~58, no.~1, pp. 106--112, Nov. 2019.
	
	\bibitem{arxivBehrooz}
	\BIBentryALTinterwordspacing
	H.~Guo, B.~Makki, M.~Åström, M.-S. Alouini, and T.~Svensson, ``Dynamic
	blockage pre-avoidance using reconfigurable intelligent surfaces,'' 2022.
	[Online]. Available: \url{https://arxiv.org/abs/2201.06659}
	\BIBentrySTDinterwordspacing
	
	\bibitem{ivd01}
	3GPP, ``{RWS-210300} {NR} repeaters and reconfigurable intelligent surface,''
	in \emph{3GPP TSG RAN Rel-18 workshop}.\hskip 1em plus 0.5em minus
	0.4em\relax Electronic Meeting, June 28 - July 2, 2021.
	
	\bibitem{AZappone2021}
	A.~Zappone, M.~Di~Renzo, F.~Shams, X.~Qian, and M.~Debbah, ``Overhead-aware
	design of reconfigurable intelligent surfaces in smart radio environments,''
	\emph{IEEE Trans. Wireless Commun.}, vol.~20, no.~1, pp. 126--141, Jan. 2021.
	
	\bibitem{p20221}
	A.~L. Swindlehurst, G.~Zhou, R.~Liu, C.~Pan, and M.~Li, ``Channel estimation
	with reconfigurable intelligent surfaces--a general framework,''
	\emph{Proceedings of the IEEE}, pp. 1--27, 2022.
	
	\bibitem{GilWCL}
	G.~T. de~Ara\'{u}jo, P.~R.~B. Gomes, A.~L.~F. de~Almeida, G.~Fodor, and
	B.~Makki, ``Semi-blind joint channel and symbol estimation in irs-assisted
	multi-user mimo networks,'' \emph{IEEE Wireless Commun. Lett.}, pp. 1--1,
	2022.
	
	\bibitem{GilJournal}
	G.~T. de~Ara{\'u}jo, A.~L. de~Almeida, and R.~Boyer, ``Channel estimation for
	intelligent reflecting surface assisted {MIMO} systems: A tensor modeling
	approach,'' \emph{IEEE J. Sel. Topics Signal Process.}, vol.~15, no.~3, pp.
	789--802, Feb. 2021.
	
	\bibitem{YYang2020}
	Y.~Yang, B.~Zheng, S.~Zhang, and R.~Zhang, ``Intelligent reflecting surface
	meets {OFDM}: Protocol design and rate maximization,'' \emph{IEEE Trans.
		Commun.}, vol.~68, no.~7, pp. 4522--4535, Mar. 2020.
	
	\bibitem{Mishra2019}
	D.~Mishra and H.~Johansson, ``Channel estimation and low-complexity beamforming
	design for passive intelligent surface assisted {MISO} wireless energy
	transfer,'' in \emph{proc. ICASSP ``2019"}, Brighton, UK, Apr.
	
	\bibitem{ZQing}
	Z.-Q. He and X.~Yuan, ``Cascaded channel estimation for large intelligent
	metasurface assisted massive {MIMO},'' \emph{IEEE Wireless Commun. Lett.},
	vol.~9, no.~2, pp. 210--214, Feb. 2019.
	
	\bibitem{Taha2021}
	A.~Taha, M.~Alrabeiah, and A.~Alkhateeb, ``Enabling large intelligent surfaces
	with compressive sensing and deep learning,'' \emph{IEEE Access}, vol.~9, pp.
	44\,304--44\,321, Mar. 2021.
	
	\bibitem{JChen2019}
	J.~Chen, Y.-C. Liang, H.~V. Cheng, and W.~Yu, ``Channel estimation for
	reconfigurable intelligent surface aided multi-user {MIMO} systems,'' Dec.
	2019, arXiv:1912.03619v1 [eess.SP].
	
	\bibitem{JHe}
	J.~He, M.~Leinonen, H.~Wymeersch, and M.~Juntti, ``Channel estimation for
	{RIS}-aided mmwave {MIMO} systems,'' in \emph{proc. GLOBECOM ``2020"},
	Taipei, Taiwan, Feb.
	
	\bibitem{CDMA}
	N.~Sidiropoulos, G.~Giannakis, and R.~Bro, ``Blind {PARAFAC} receivers for
	{DS-CDMA} systems,'' \emph{IEEE Trans. Signal Process.}, vol.~48, no.~3, pp.
	810--823, Mar. 2000.
	
	\bibitem{tensorOverview}
	H.~Chen, F.~Ahmad, S.~Vorobyov, and F.~Porikli, ``Tensor decompositions in
	wireless communications and mimo radar,'' \emph{IEEE J.Sel. Topics in Signal
		Process.}, vol.~15, no.~3, pp. 438--453, Feb. 2021.
	
	\bibitem{sug01}
	N.~D. Sidiropoulos, L.~De~Lathauwer, X.~Fu, K.~Huang, E.~E. Papalexakis, and
	C.~Faloutsos, ``Tensor decomposition for signal processing and machine
	learning,'' \emph{IEEE Trans. Signal Process.}, vol.~65, no.~13, pp.
	3551--3582, Feb. 2017.
	
	\bibitem{RIS1}
	L.~Wei, C.~Huang, G.~C. Alexandropoulos, and C.~Yuen, ``Parallel factor
	decomposition channel estimation in {RIS}-assisted multi-user {MISO}
	communication,'' in \emph{proc. SAM ``2020"}, Hangzhou, China, Jun.
	
	\bibitem{gg1}
	G.~T. de~Ara{\'u}jo and A.~L. de~Almeida, ``{PARAFAC}-based channel estimation
	for intelligent reflective surface assisted {MIMO} system,'' in \emph{proc.
		SAM ``2020"}, Hangzhou, China, Jun.
	
	\bibitem{gg2}
	G.~T. de~Ara\'{u}jo and A.~L.~F. de~Almeida, ``Channel estimation for {MIMO}
	system assisted by intelligent reflective surface,'' in \emph{proc. SBrT
		``2020"}, Santa Catarina, Brazil, Nov.
	
	\bibitem{imp1}
	X.~Qian, M.~Di~Renzo, J.~Liu, A.~Kammoun, and M.-S. Alouini, ``Beamforming
	through reconfigurable intelligent surfaces in single-user {MIMO} systems:
	{SNR} distribution and scaling laws in the presence of channel fading and
	phase noise,'' \emph{IEEE Wireless Commun. Lett.}, vol.~10, no.~1, pp.
	77--81, Sep. 2021.
	
	\bibitem{imp3}
	M.-A. Badiu and J.~P. Coon, ``Communication through a large reflecting surface
	with phase errors,'' \emph{IEEE Wireless Commun. Lett.}, vol.~9, no.~2, pp.
	184--188, Feb. 2020.
	
	\bibitem{imp4}
	S.~Zhou, W.~Xu, K.~Wang, M.~Di~Renzo, and M.-S. Alouini, ``Spectral and energy
	efficiency of {IRS}-assisted {MISO} communication with hardware
	impairments,'' \emph{IEEE wireless commun. lett.}, vol.~9, no.~9, pp.
	1366--1369, Sep. 2020.
	
	\bibitem{imp5}
	K.~Zhi, C.~Pan, H.~Ren, and K.~Wang, ``Uplink achievable rate of intelligent
	reflecting surface-aided millimeter-wave communications with low-resolution
	{ADC} and phase noise,'' \emph{IEEE Wireless Commun. Lett.}, vol.~10, no.~3,
	pp. 654--658, Mar. 2021.
	
	\bibitem{mainblock}
	B.~Li, Z.~Zhang, Z.~Hu, and Y.~Chen, ``Joint array diagnosis and channel
	estimation for {RIS}-aided mmwave {MIMO} system,'' \emph{IEEE Access},
	vol.~8, pp. 193\,992--194\,006, Oct. 2020.
	
	\bibitem{BiLi02}
	B.~Li, Z.~Zhang, and Z.~Hu, ``Channel estimation for reconfigurable intelligent
	surface-assisted multiuser mmwave {MIMO} system in the presence of array
	blockage,'' \emph{Transactions on Emerging Telecommunications Technologies},
	vol.~32, no.~11, p. e4322, Jun. 2021.
	
	\bibitem{Kolda}
	T.~G. Kolda and B.~W. Bader, ``Tensor decompositions and applications,''
	\emph{SIAM review}, vol.~51, no.~3, pp. 455--500, Aug. 2009.
	
	\bibitem{hosvd}
	L.~De~Lathauwer, B.~De~Moor, and J.~Vandewalle, ``A multilinear singular value
	decomposition,'' \emph{SIAM Journal on Matrix Analysis and Applications},
	vol.~21, no.~4, pp. 1253--1278, Apr. 2000.
	
	\bibitem{var1}
	S.~Abeywickrama, R.~Zhang, Q.~Wu, and C.~Yuen, ``Intelligent reflecting
	surface: Practical phase shift model and beamforming optimization,''
	\emph{IEEE Trans. Commun.}, vol.~68, no.~9, pp. 5849--5863, Sep. 2020.
	
	\bibitem{Yigit2020}
	Z.~Yigit, E.~Basar, and I.~Altunbas, ``Low complexity adaptation for
	reconfigurable intelligent surface-based {MIMO} systems,'' \emph{IEEE Commun.
		Lett.}, vol.~24, no.~12, pp. 2946--2950, Dec. 2020.
	
	\bibitem{Sokal01}
	\BIBentryALTinterwordspacing
	B.~Sokal, P.~R.~B. Gomes, A.~L.~F. de~Almeida, B.~Makki, and G.~Fodor, ``{IRS}
	phase-shift feedback overhead-aware model based on rank-one tensor
	approximation,'' 2022. [Online]. Available:
	\url{https://arxiv.org/abs/2205.12024}
	\BIBentrySTDinterwordspacing
	
	\bibitem{Sokal02}
	\BIBentryALTinterwordspacing
	------, ``Reducing the control overhead of intelligent reconfigurable surfaces
	via a tensor-based low-rank factorization approach,'' 2022. [Online].
	Available: \url{https://arxiv.org/abs/2206.05341}
	\BIBentrySTDinterwordspacing
	
	\bibitem{Bro98}
	R.~Bro, ``Multi-way analysis in the food industry-models, algorithms, and
	applications,'' in \emph{MRI, EPG and EMA,” Proc ICSLP 2000}.\hskip 1em
	plus 0.5em minus 0.4em\relax Citeseer, 1998.
	
	\bibitem{Feng2011}
	X.-F. Gong and Q.-H. Lin, ``Spatially constrained parallel factor analysis for
	semi-blind beamforming,'' in \emph{2011 Seventh International Conference on
		Natural Computation}, vol.~1.\hskip 1em plus 0.5em minus 0.4em\relax IEEE,
	2011, pp. 416--420.
	
	\bibitem{Dong2018}
	Z.~Wang, C.~Cai, F.~Wen, and D.~Huang, ``A quadrilinear decomposition method
	for direction estimation in bistatic {MIMO} radar,'' \emph{IEEE Access},
	vol.~6, pp. 13\,766--13\,772, Mar. 2018.
	
	\bibitem{Optimal}
	T.~L. Jensen and E.~De~Carvalho, ``An optimal channel estimation scheme for
	intelligent reflecting surfaces based on a minimum variance unbiased
	estimator,'' in \emph{proc. ICASSP ``2020"}, Barcelona, Spain, May.
	
\end{thebibliography}


\end{document}